\documentclass[twocolumn,traditabstract,longauth]{aa}
\usepackage[english]{babel} 
\usepackage[utf8]{inputenc} 
\usepackage[T1]{fontenc}

\usepackage{graphicx}
\usepackage[breaklinks, colorlinks, citecolor=blue]{hyperref}
\usepackage{natbib}
\bibpunct{(}{)}{;}{a}{}{,}
\graphicspath{{figures/}}

\def\setsymbol#1#2{\expandafter\def\csname #1\endcsname{#2}}
\def\getsymbol#1{\csname #1\endcsname}

\def\Planck{{\it Planck\/}}

\def\HeJT{$^4$He-JT}

\def\allearlypapers{\nocite{planck2011-1.1, planck2011-1.3, planck2011-1.4, planck2011-1.5, planck2011-1.6, planck2011-1.7, planck2011-1.10, planck2011-1.10sup, planck2011-5.1a, planck2011-5.1b, planck2011-5.2a, planck2011-5.2b, planck2011-5.2c, planck2011-6.1, planck2011-6.2, planck2011-6.3a, planck2011-6.4a, planck2011-6.4b, planck2011-6.6, planck2011-7.0, planck2011-7.2, planck2011-7.3, planck2011-7.7a, planck2011-7.7b, planck2011-7.12, planck2011-7.13}}

\newbox\tablebox    \newdimen\tablewidth
\def\leaderfil{\leaders\hbox to 5pt{\hss.\hss}\hfil}
\def\tablenote#1 #2\par{\begingroup \parindent=0.8em
    \abovedisplayshortskip=0pt\belowdisplayshortskip=0pt
    \noindent
    $$\hss\vbox{\hsize\tablewidth \hangindent=\parindent \hangafter=1 \noindent
    \hbox to \parindent{\sup{\rm #1}\hss}\strut#2\strut\par}\hss$$
    \endgroup}

\def\L2{\ifmmode L_2\else $L_2$\fi}

\def\DeltaT{\ifmmode \Delta T\else $\Delta T$\fi}
\def\deltat{\ifmmode \Delta t\else $\Delta t$\fi}
\def\fknee{\ifmmode f_{\rm knee}\else $f_{\rm knee}$\fi}
\def\Fmax{\ifmmode F_{\rm max}\else $F_{\rm max}$\fi}
\def\solar{\ifmmode{\rm M}_{\mathord\odot}\else${\rm M}_{\mathord\odot}$\fi}

\def\inv{\ifmmode^{-1}\else$^{-1}$\fi}
\def\mo{\ifmmode^{-1}\else$^{-1}$\fi}
\def\sup#1{\ifmmode ^{\rm #1}\else $^{\rm #1}$\fi}
\def\expo#1{\ifmmode \times 10^{#1}\else $\times 10^{#1}$\fi}
\def\,{\thinspace}
\def\lsim{\mathrel{\raise .4ex\hbox{\rlap{$<$}\lower 1.2ex\hbox{$\sim$}}}}
\def\gsim{\mathrel{\raise .4ex\hbox{\rlap{$>$}\lower 1.2ex\hbox{$\sim$}}}}

\def\simprop{\mathrel{\raise .4ex\hbox{\rlap{$\propto$}\lower 1.2ex\hbox{$\sim$}}}}
\def\deg{\ifmmode^\circ\else$^\circ$\fi}
\def\pdeg{\ifmmode $\setbox0=\hbox{$^{\circ}$}\rlap{\hskip.11\wd0 .}$^{\circ}
          \else \setbox0=\hbox{$^{\circ}$}\rlap{\hskip.11\wd0 .}$^{\circ}$\fi}
\def\arcs{\ifmmode {^{\scriptstyle\prime\prime}}
          \else $^{\scriptstyle\prime\prime}$\fi}
\def\arcm{\ifmmode {^{\scriptstyle\prime}}
          \else $^{\scriptstyle\prime}$\fi}
\newdimen\sa  \newdimen\sb
\def\parcs{\sa=.07em \sb=.03em
     \ifmmode \hbox{\rlap{.}}^{\scriptstyle\prime\kern -\sb\prime}\hbox{\kern -\sa}
     \else \rlap{.}$^{\scriptstyle\prime\kern -\sb\prime}$\kern -\sa\fi}
\def\parcm{\sa=.08em \sb=.03em
     \ifmmode \hbox{\rlap{.}\kern\sa}^{\scriptstyle\prime}\hbox{\kern-\sb}
     \else \rlap{.}\kern\sa$^{\scriptstyle\prime}$\kern-\sb\fi}
\def\ra[#1 #2 #3.#4]{#1\sup{h}#2\sup{m}#3\sup{s}\llap.#4}
\def\dec[#1 #2 #3.#4]{#1\deg#2\arcm#3\arcs\llap.#4}
\def\deco[#1 #2 #3]{#1\deg#2\arcm#3\arcs}
\def\rra[#1 #2]{#1\sup{h}#2\sup{m}}

\def\dots{\relax\ifmmode \ldots\else $\ldots$\fi}
\def\WHzsr{\ifmmode $W\,Hz\mo\,sr\mo$\else W\,Hz\mo\,sr\mo\fi}
\def\mHz{\ifmmode $\,mHz$\else \,mHz\fi}
\def\GHz{\ifmmode $\,GHz$\else \,GHz\fi}
\def\mKs{\ifmmode $\,mK\,s$^{1/2}\else \,mK\,s$^{1/2}$\fi}
\def\muKs{\ifmmode \,\mu$K\,s$^{1/2}\else \,$\mu$K\,s$^{1/2}$\fi}
\def\muKRJs{\ifmmode \,\mu$K$_{\rm RJ}$\,s$^{1/2}\else \,$\mu$K$_{\rm RJ}$\,s$^{1/2}$\fi}
\def\muKHz{\ifmmode \,\mu$K\,Hz$^{-1/2}\else \,$\mu$K\,Hz$^{-1/2}$\fi}
\def\MJysr{\ifmmode \,$MJy\,sr\mo$\else \,MJy\,sr\mo\fi}
\def\MJysrmK{\ifmmode \,$MJy\,sr\mo$\,mK$_{\rm CMB}\mo\else \,MJy\,sr\mo\,mK$_{\rm CMB}\mo$\fi}
\def\microns{\ifmmode \,\mu$m$\else \,$\mu$m\fi}

\def\muK{\ifmmode \,\mu$K$\else \,$\mu$\hbox{K}\fi}
\def\microK{\ifmmode \,\mu$K$\else \,$\mu$\hbox{K}\fi}
\def\muW{\ifmmode \,\mu$W$\else \,$\mu$\hbox{W}\fi}
\def\kms{\ifmmode $\,km\,s$^{-1}\else \,km\,s$^{-1}$\fi}
\def\kmsMpc{\ifmmode $\,\kms\,Mpc\mo$\else \,\kms\,Mpc\mo\fi}
\setsymbol{LFI:center:frequency:70GHz:units}{70.3\,GHz}
\setsymbol{LFI:center:frequency:44GHz:units}{44.1\,GHz}
\setsymbol{LFI:center:frequency:30GHz:units}{28.5\,GHz}

\setsymbol{LFI:center:frequency:70GHz}{70.3}
\setsymbol{LFI:center:frequency:44GHz}{44.1}
\setsymbol{LFI:center:frequency:30GHz}{28.5}

\setsymbol{LFI:center:frequency:LFI18:Rad:M:units}{71.7\GHz}
\setsymbol{LFI:center:frequency:LFI19:Rad:M:units}{67.5\GHz}
\setsymbol{LFI:center:frequency:LFI20:Rad:M:units}{69.2\GHz}
\setsymbol{LFI:center:frequency:LFI21:Rad:M:units}{70.4\GHz}
\setsymbol{LFI:center:frequency:LFI22:Rad:M:units}{71.5\GHz}
\setsymbol{LFI:center:frequency:LFI23:Rad:M:units}{70.8\GHz}
\setsymbol{LFI:center:frequency:LFI24:Rad:M:units}{44.4\GHz}
\setsymbol{LFI:center:frequency:LFI25:Rad:M:units}{44.0\GHz}
\setsymbol{LFI:center:frequency:LFI26:Rad:M:units}{43.9\GHz}
\setsymbol{LFI:center:frequency:LFI27:Rad:M:units}{28.3\GHz}
\setsymbol{LFI:center:frequency:LFI28:Rad:M:units}{28.8\GHz}
\setsymbol{LFI:center:frequency:LFI18:Rad:S:units}{70.1\GHz}
\setsymbol{LFI:center:frequency:LFI19:Rad:S:units}{69.6\GHz}
\setsymbol{LFI:center:frequency:LFI20:Rad:S:units}{69.5\GHz}
\setsymbol{LFI:center:frequency:LFI21:Rad:S:units}{69.5\GHz}
\setsymbol{LFI:center:frequency:LFI22:Rad:S:units}{72.8\GHz}
\setsymbol{LFI:center:frequency:LFI23:Rad:S:units}{71.3\GHz}
\setsymbol{LFI:center:frequency:LFI24:Rad:S:units}{44.1\GHz}
\setsymbol{LFI:center:frequency:LFI25:Rad:S:units}{44.1\GHz}
\setsymbol{LFI:center:frequency:LFI26:Rad:S:units}{44.1\GHz}
\setsymbol{LFI:center:frequency:LFI27:Rad:S:units}{28.5\GHz}
\setsymbol{LFI:center:frequency:LFI28:Rad:S:units}{28.2\GHz}

\setsymbol{LFI:center:frequency:LFI18:Rad:M}{71.7}
\setsymbol{LFI:center:frequency:LFI19:Rad:M}{67.5}
\setsymbol{LFI:center:frequency:LFI20:Rad:M}{69.2}
\setsymbol{LFI:center:frequency:LFI21:Rad:M}{70.4}
\setsymbol{LFI:center:frequency:LFI22:Rad:M}{71.5}
\setsymbol{LFI:center:frequency:LFI23:Rad:M}{70.8}
\setsymbol{LFI:center:frequency:LFI24:Rad:M}{44.4}
\setsymbol{LFI:center:frequency:LFI25:Rad:M}{44.0}
\setsymbol{LFI:center:frequency:LFI26:Rad:M}{43.9}
\setsymbol{LFI:center:frequency:LFI27:Rad:M}{28.3}
\setsymbol{LFI:center:frequency:LFI28:Rad:M}{28.8}
\setsymbol{LFI:center:frequency:LFI18:Rad:S}{70.1}
\setsymbol{LFI:center:frequency:LFI19:Rad:S}{69.6}
\setsymbol{LFI:center:frequency:LFI20:Rad:S}{69.5}
\setsymbol{LFI:center:frequency:LFI21:Rad:S}{69.5}
\setsymbol{LFI:center:frequency:LFI22:Rad:S}{72.8}
\setsymbol{LFI:center:frequency:LFI23:Rad:S}{71.3}
\setsymbol{LFI:center:frequency:LFI24:Rad:S}{44.1}
\setsymbol{LFI:center:frequency:LFI25:Rad:S}{44.1}
\setsymbol{LFI:center:frequency:LFI26:Rad:S}{44.1}
\setsymbol{LFI:center:frequency:LFI27:Rad:S}{28.5}
\setsymbol{LFI:center:frequency:LFI28:Rad:S}{28.2}

\setsymbol{LFI:white:noise:sensitivity:70GHz:units}{152.6\muKs}
\setsymbol{LFI:white:noise:sensitivity:44GHz:units}{173.1\muKs}
\setsymbol{LFI:white:noise:sensitivity:30GHz:units}{146.8\muKs}

\setsymbol{LFI:white:noise:sensitivity:70GHz}{152.6}
\setsymbol{LFI:white:noise:sensitivity:44GHz}{173.1}
\setsymbol{LFI:white:noise:sensitivity:30GHz}{146.8}

\setsymbol{LFI:white:noise:sensitivity:LFI18:Rad:M:units}{512.0\muKs}
\setsymbol{LFI:white:noise:sensitivity:LFI19:Rad:M:units}{581.4\muKs}
\setsymbol{LFI:white:noise:sensitivity:LFI20:Rad:M:units}{590.8\muKs}
\setsymbol{LFI:white:noise:sensitivity:LFI21:Rad:M:units}{455.2\muKs}
\setsymbol{LFI:white:noise:sensitivity:LFI22:Rad:M:units}{492.0\muKs}
\setsymbol{LFI:white:noise:sensitivity:LFI23:Rad:M:units}{507.7\muKs}
\setsymbol{LFI:white:noise:sensitivity:LFI24:Rad:M:units}{462.2\muKs}
\setsymbol{LFI:white:noise:sensitivity:LFI25:Rad:M:units}{413.6\muKs}
\setsymbol{LFI:white:noise:sensitivity:LFI26:Rad:M:units}{478.6\muKs}
\setsymbol{LFI:white:noise:sensitivity:LFI27:Rad:M:units}{277.7\muKs}
\setsymbol{LFI:white:noise:sensitivity:LFI28:Rad:M:units}{312.3\muKs}
\setsymbol{LFI:white:noise:sensitivity:LFI18:Rad:S:units}{465.7\muKs}
\setsymbol{LFI:white:noise:sensitivity:LFI19:Rad:S:units}{555.6\muKs}
\setsymbol{LFI:white:noise:sensitivity:LFI20:Rad:S:units}{623.2\muKs}
\setsymbol{LFI:white:noise:sensitivity:LFI21:Rad:S:units}{564.1\muKs}
\setsymbol{LFI:white:noise:sensitivity:LFI22:Rad:S:units}{534.4\muKs}
\setsymbol{LFI:white:noise:sensitivity:LFI23:Rad:S:units}{542.4\muKs}
\setsymbol{LFI:white:noise:sensitivity:LFI24:Rad:S:units}{399.2\muKs}
\setsymbol{LFI:white:noise:sensitivity:LFI25:Rad:S:units}{392.6\muKs}
\setsymbol{LFI:white:noise:sensitivity:LFI26:Rad:S:units}{418.6\muKs}
\setsymbol{LFI:white:noise:sensitivity:LFI27:Rad:S:units}{302.9\muKs}
\setsymbol{LFI:white:noise:sensitivity:LFI28:Rad:S:units}{285.3\muKs}

\setsymbol{LFI:white:noise:sensitivity:LFI18:Rad:M}{512.0}
\setsymbol{LFI:white:noise:sensitivity:LFI19:Rad:M}{581.4}
\setsymbol{LFI:white:noise:sensitivity:LFI20:Rad:M}{590.8}
\setsymbol{LFI:white:noise:sensitivity:LFI21:Rad:M}{455.2}
\setsymbol{LFI:white:noise:sensitivity:LFI22:Rad:M}{492.0}
\setsymbol{LFI:white:noise:sensitivity:LFI23:Rad:M}{507.7}
\setsymbol{LFI:white:noise:sensitivity:LFI24:Rad:M}{462.2}
\setsymbol{LFI:white:noise:sensitivity:LFI25:Rad:M}{413.6}
\setsymbol{LFI:white:noise:sensitivity:LFI26:Rad:M}{478.6}
\setsymbol{LFI:white:noise:sensitivity:LFI27:Rad:M}{277.7}
\setsymbol{LFI:white:noise:sensitivity:LFI28:Rad:M}{312.3}
\setsymbol{LFI:white:noise:sensitivity:LFI18:Rad:S}{465.7}
\setsymbol{LFI:white:noise:sensitivity:LFI19:Rad:S}{555.6}
\setsymbol{LFI:white:noise:sensitivity:LFI20:Rad:S}{623.2}
\setsymbol{LFI:white:noise:sensitivity:LFI21:Rad:S}{564.1}
\setsymbol{LFI:white:noise:sensitivity:LFI22:Rad:S}{534.4}
\setsymbol{LFI:white:noise:sensitivity:LFI23:Rad:S}{542.4}
\setsymbol{LFI:white:noise:sensitivity:LFI24:Rad:S}{399.2}
\setsymbol{LFI:white:noise:sensitivity:LFI25:Rad:S}{392.6}
\setsymbol{LFI:white:noise:sensitivity:LFI26:Rad:S}{418.6}
\setsymbol{LFI:white:noise:sensitivity:LFI27:Rad:S}{302.9}
\setsymbol{LFI:white:noise:sensitivity:LFI28:Rad:S}{285.3}

\setsymbol{LFI:knee:frequency:70GHz:units}{29.5\mHz}
\setsymbol{LFI:knee:frequency:44GHz:units}{56.2\mHz}
\setsymbol{LFI:knee:frequency:30GHz:units}{113.7\mHz}

\setsymbol{LFI:knee:frequency:70GHz}{29.5}
\setsymbol{LFI:knee:frequency:44GHz}{56.2}
\setsymbol{LFI:knee:frequency:30GHz}{113.7}

\setsymbol{LFI:knee:frequency:LFI18:Rad:M:units}{16.3\mHz}
\setsymbol{LFI:knee:frequency:LFI19:Rad:M:units}{15.1\mHz}
\setsymbol{LFI:knee:frequency:LFI20:Rad:M:units}{18.7\mHz}
\setsymbol{LFI:knee:frequency:LFI21:Rad:M:units}{37.2\mHz}
\setsymbol{LFI:knee:frequency:LFI22:Rad:M:units}{12.7\mHz}
\setsymbol{LFI:knee:frequency:LFI23:Rad:M:units}{34.6\mHz}
\setsymbol{LFI:knee:frequency:LFI24:Rad:M:units}{46.2\mHz}
\setsymbol{LFI:knee:frequency:LFI25:Rad:M:units}{24.9\mHz}
\setsymbol{LFI:knee:frequency:LFI26:Rad:M:units}{67.6\mHz}
\setsymbol{LFI:knee:frequency:LFI27:Rad:M:units}{187.4\mHz}
\setsymbol{LFI:knee:frequency:LFI28:Rad:M:units}{122.2\mHz}
\setsymbol{LFI:knee:frequency:LFI18:Rad:S:units}{17.7\mHz}
\setsymbol{LFI:knee:frequency:LFI19:Rad:S:units}{22.0\mHz}
\setsymbol{LFI:knee:frequency:LFI20:Rad:S:units}{8.7\mHz}
\setsymbol{LFI:knee:frequency:LFI21:Rad:S:units}{25.9\mHz}
\setsymbol{LFI:knee:frequency:LFI22:Rad:S:units}{15.8\mHz}
\setsymbol{LFI:knee:frequency:LFI23:Rad:S:units}{129.8\mHz}
\setsymbol{LFI:knee:frequency:LFI24:Rad:S:units}{100.9\mHz}
\setsymbol{LFI:knee:frequency:LFI25:Rad:S:units}{38.9\mHz}
\setsymbol{LFI:knee:frequency:LFI26:Rad:S:units}{58.9\mHz}
\setsymbol{LFI:knee:frequency:LFI27:Rad:S:units}{104.4\mHz}
\setsymbol{LFI:knee:frequency:LFI28:Rad:S:units}{40.7\mHz}

\setsymbol{LFI:knee:frequency:LFI18:Rad:M}{16.3}
\setsymbol{LFI:knee:frequency:LFI19:Rad:M}{15.1}
\setsymbol{LFI:knee:frequency:LFI20:Rad:M}{18.7}
\setsymbol{LFI:knee:frequency:LFI21:Rad:M}{37.2}
\setsymbol{LFI:knee:frequency:LFI22:Rad:M}{12.7}
\setsymbol{LFI:knee:frequency:LFI23:Rad:M}{34.6}
\setsymbol{LFI:knee:frequency:LFI24:Rad:M}{46.2}
\setsymbol{LFI:knee:frequency:LFI25:Rad:M}{24.9}
\setsymbol{LFI:knee:frequency:LFI26:Rad:M}{67.6}
\setsymbol{LFI:knee:frequency:LFI27:Rad:M}{187.4}
\setsymbol{LFI:knee:frequency:LFI28:Rad:M}{122.2}
\setsymbol{LFI:knee:frequency:LFI18:Rad:S}{17.7}
\setsymbol{LFI:knee:frequency:LFI19:Rad:S}{22.0}
\setsymbol{LFI:knee:frequency:LFI20:Rad:S}{8.7}
\setsymbol{LFI:knee:frequency:LFI21:Rad:S}{25.9}
\setsymbol{LFI:knee:frequency:LFI22:Rad:S}{15.8}
\setsymbol{LFI:knee:frequency:LFI23:Rad:S}{129.8}
\setsymbol{LFI:knee:frequency:LFI24:Rad:S}{100.9}
\setsymbol{LFI:knee:frequency:LFI25:Rad:S}{38.9}
\setsymbol{LFI:knee:frequency:LFI26:Rad:S}{58.9}
\setsymbol{LFI:knee:frequency:LFI27:Rad:S}{104.4}
\setsymbol{LFI:knee:frequency:LFI28:Rad:S}{40.7}

\setsymbol{LFI:slope:70GHz:units}{$-1.03$\mHz}
\setsymbol{LFI:slope:44GHz:units}{$-0.89$\mHz}
\setsymbol{LFI:slope:30GHz:units}{$-0.87$\mHz}

\setsymbol{LFI:slope:70GHz}{$-1.03$}
\setsymbol{LFI:slope:44GHz}{$-0.89$}
\setsymbol{LFI:slope:30GHz}{$-0.87$}

\setsymbol{LFI:slope:LFI18:Rad:M:units}{$-1.04$\mHz}
\setsymbol{LFI:slope:LFI19:Rad:M:units}{$-1.09$\mHz}
\setsymbol{LFI:slope:LFI20:Rad:M:units}{$-0.69$\mHz}
\setsymbol{LFI:slope:LFI21:Rad:M:units}{$-1.56$\mHz}
\setsymbol{LFI:slope:LFI22:Rad:M:units}{$-1.01$\mHz}
\setsymbol{LFI:slope:LFI23:Rad:M:units}{$-0.96$\mHz}
\setsymbol{LFI:slope:LFI24:Rad:M:units}{$-0.83$\mHz}
\setsymbol{LFI:slope:LFI25:Rad:M:units}{$-0.91$\mHz}
\setsymbol{LFI:slope:LFI26:Rad:M:units}{$-0.95$\mHz}
\setsymbol{LFI:slope:LFI27:Rad:M:units}{$-0.87$\mHz}
\setsymbol{LFI:slope:LFI28:Rad:M:units}{$-0.88$\mHz}
\setsymbol{LFI:slope:LFI18:Rad:S:units}{$-1.15$\mHz}
\setsymbol{LFI:slope:LFI19:Rad:S:units}{$-1.00$\mHz}
\setsymbol{LFI:slope:LFI20:Rad:S:units}{$-0.95$\mHz}
\setsymbol{LFI:slope:LFI21:Rad:S:units}{$-0.92$\mHz}
\setsymbol{LFI:slope:LFI22:Rad:S:units}{$-1.01$\mHz}
\setsymbol{LFI:slope:LFI23:Rad:S:units}{$-0.95$\mHz}
\setsymbol{LFI:slope:LFI24:Rad:S:units}{$-0.73$\mHz}
\setsymbol{LFI:slope:LFI25:Rad:S:units}{$-1.16$\mHz}
\setsymbol{LFI:slope:LFI26:Rad:S:units}{$-0.79$\mHz}
\setsymbol{LFI:slope:LFI27:Rad:S:units}{$-0.82$\mHz}
\setsymbol{LFI:slope:LFI28:Rad:S:units}{$-0.91$\mHz}

\setsymbol{LFI:slope:LFI18:Rad:M}{$-1.04$}
\setsymbol{LFI:slope:LFI19:Rad:M}{$-1.09$}
\setsymbol{LFI:slope:LFI20:Rad:M}{$-0.69$}
\setsymbol{LFI:slope:LFI21:Rad:M}{$-1.56$}
\setsymbol{LFI:slope:LFI22:Rad:M}{$-1.01$}
\setsymbol{LFI:slope:LFI23:Rad:M}{$-0.96$}
\setsymbol{LFI:slope:LFI24:Rad:M}{$-0.83$}
\setsymbol{LFI:slope:LFI25:Rad:M}{$-0.91$}
\setsymbol{LFI:slope:LFI26:Rad:M}{$-0.95$}
\setsymbol{LFI:slope:LFI27:Rad:M}{$-0.87$}
\setsymbol{LFI:slope:LFI28:Rad:M}{$-0.88$}
\setsymbol{LFI:slope:LFI18:Rad:S}{$-1.15$}
\setsymbol{LFI:slope:LFI19:Rad:S}{$-1.00$}
\setsymbol{LFI:slope:LFI20:Rad:S}{$-0.95$}
\setsymbol{LFI:slope:LFI21:Rad:S}{$-0.92$}
\setsymbol{LFI:slope:LFI22:Rad:S}{$-1.01$}
\setsymbol{LFI:slope:LFI23:Rad:S}{$-0.95$}
\setsymbol{LFI:slope:LFI24:Rad:S}{$-0.73$}
\setsymbol{LFI:slope:LFI25:Rad:S}{$-1.16$}
\setsymbol{LFI:slope:LFI26:Rad:S}{$-0.79$}
\setsymbol{LFI:slope:LFI27:Rad:S}{$-0.82$}
\setsymbol{LFI:slope:LFI28:Rad:S}{$-0.91$}

\setsymbol{LFI:FWHM:70GHz:units}{13\parcm01}
\setsymbol{LFI:FWHM:44GHz:units}{27\parcm92}
\setsymbol{LFI:FWHM:30GHz:units}{32\parcm65}

\setsymbol{LFI:FWHM:70GHz}{13.01}
\setsymbol{LFI:FWHM:44GHz}{27.92}
\setsymbol{LFI:FWHM:30GHz}{32.65}

\setsymbol{LFI:FWHM:LFI18:units}{13\parcm39}
\setsymbol{LFI:FWHM:LFI19:units}{13\parcm01}
\setsymbol{LFI:FWHM:LFI20:units}{12\parcm75}
\setsymbol{LFI:FWHM:LFI21:units}{12\parcm74}
\setsymbol{LFI:FWHM:LFI22:units}{12\parcm87}
\setsymbol{LFI:FWHM:LFI23:units}{13\parcm27}
\setsymbol{LFI:FWHM:LFI24:units}{22\parcm98}
\setsymbol{LFI:FWHM:LFI25:units}{30\parcm46}
\setsymbol{LFI:FWHM:LFI26:units}{30\parcm31}
\setsymbol{LFI:FWHM:LFI27:units}{32\parcm65}
\setsymbol{LFI:FWHM:LFI28:units}{32\parcm66}

\setsymbol{LFI:FWHM:LFI18}{13.39}
\setsymbol{LFI:FWHM:LFI19}{13.01}
\setsymbol{LFI:FWHM:LFI20}{12.75}
\setsymbol{LFI:FWHM:LFI21}{12.74}
\setsymbol{LFI:FWHM:LFI22}{12.87}
\setsymbol{LFI:FWHM:LFI23}{13.27}
\setsymbol{LFI:FWHM:LFI24}{22.98}
\setsymbol{LFI:FWHM:LFI25}{30.46}
\setsymbol{LFI:FWHM:LFI26}{30.31}
\setsymbol{LFI:FWHM:LFI27}{32.65}
\setsymbol{LFI:FWHM:LFI28}{32.66}

\setsymbol{LFI:FWHM:uncertainty:LFI18:units}{0.170\arcm}
\setsymbol{LFI:FWHM:uncertainty:LFI19:units}{0.174\arcm}
\setsymbol{LFI:FWHM:uncertainty:LFI20:units}{0.170\arcm}
\setsymbol{LFI:FWHM:uncertainty:LFI21:units}{0.156\arcm}
\setsymbol{LFI:FWHM:uncertainty:LFI22:units}{0.164\arcm}
\setsymbol{LFI:FWHM:uncertainty:LFI23:units}{0.171\arcm}
\setsymbol{LFI:FWHM:uncertainty:LFI24:units}{0.652\arcm}
\setsymbol{LFI:FWHM:uncertainty:LFI25:units}{1.075\arcm}
\setsymbol{LFI:FWHM:uncertainty:LFI26:units}{1.131\arcm}
\setsymbol{LFI:FWHM:uncertainty:LFI27:units}{1.266\arcm}
\setsymbol{LFI:FWHM:uncertainty:LFI28:units}{1.287\arcm}

\setsymbol{LFI:FWHM:uncertainty:LFI18}{0.170}
\setsymbol{LFI:FWHM:uncertainty:LFI19}{0.174}
\setsymbol{LFI:FWHM:uncertainty:LFI20}{0.170}
\setsymbol{LFI:FWHM:uncertainty:LFI21}{0.156}
\setsymbol{LFI:FWHM:uncertainty:LFI22}{0.164}
\setsymbol{LFI:FWHM:uncertainty:LFI23}{0.171}
\setsymbol{LFI:FWHM:uncertainty:LFI24}{0.652}
\setsymbol{LFI:FWHM:uncertainty:LFI25}{1.075}
\setsymbol{LFI:FWHM:uncertainty:LFI26}{1.131}
\setsymbol{LFI:FWHM:uncertainty:LFI27}{1.266}
\setsymbol{LFI:FWHM:uncertainty:LFI28}{1.287}

\setsymbol{HFI:center:frequency:100GHz:units}{100\,GHz}
\setsymbol{HFI:center:frequency:143GHz:units}{143\,GHz}
\setsymbol{HFI:center:frequency:217GHz:units}{217\,GHz}
\setsymbol{HFI:center:frequency:353GHz:units}{353\,GHz}
\setsymbol{HFI:center:frequency:545GHz:units}{545\,GHz}
\setsymbol{HFI:center:frequency:857GHz:units}{857\,GHz}

\setsymbol{HFI:center:frequency:100GHz}{100}
\setsymbol{HFI:center:frequency:143GHz}{143}
\setsymbol{HFI:center:frequency:217GHz}{217}
\setsymbol{HFI:center:frequency:353GHz}{353}
\setsymbol{HFI:center:frequency:545GHz}{545}
\setsymbol{HFI:center:frequency:857GHz}{857}

\setsymbol{HFI:Ndetectors:100GHz}{8}
\setsymbol{HFI:Ndetectors:143GHz}{11}
\setsymbol{HFI:Ndetectors:217GHz}{12}
\setsymbol{HFI:Ndetectors:353GHz}{12}
\setsymbol{HFI:Ndetectors:545GHz}{3}
\setsymbol{HFI:Ndetectors:857GHz}{4}

\setsymbol{HFI:FWHM:Maps:100GHz:units}{9\parcm88}
\setsymbol{HFI:FWHM:Maps:143GHz:units}{7\parcm18}
\setsymbol{HFI:FWHM:Maps:217GHz:units}{4\parcm87}
\setsymbol{HFI:FWHM:Maps:353GHz:units}{4\parcm65}
\setsymbol{HFI:FWHM:Maps:545GHz:units}{4\parcm72}
\setsymbol{HFI:FWHM:Maps:857GHz:units}{4\parcm39}
\setsymbol{HFI:FWHM:Maps:100GHz}{9.88}
\setsymbol{HFI:FWHM:Maps:143GHz}{7.18}
\setsymbol{HFI:FWHM:Maps:217GHz}{4.87}
\setsymbol{HFI:FWHM:Maps:353GHz}{4.65}
\setsymbol{HFI:FWHM:Maps:545GHz}{4.72}
\setsymbol{HFI:FWHM:Maps:857GHz}{4.39}

\setsymbol{HFI:beam:ellipticity:Maps:100GHz}{1.15}
\setsymbol{HFI:beam:ellipticity:Maps:143GHz}{1.01}
\setsymbol{HFI:beam:ellipticity:Maps:217GHz}{1.06}
\setsymbol{HFI:beam:ellipticity:Maps:353GHz}{1.05}
\setsymbol{HFI:beam:ellipticity:Maps:545GHz}{1.14}
\setsymbol{HFI:beam:ellipticity:Maps:857GHz}{1.19}

\setsymbol{HFI:FWHM:Mars:100GHz:units}{9\parcm37}
\setsymbol{HFI:FWHM:Mars:143GHz:units}{7\parcm04}
\setsymbol{HFI:FWHM:Mars:217GHz:units}{4\parcm68}
\setsymbol{HFI:FWHM:Mars:353GHz:units}{4\parcm43}
\setsymbol{HFI:FWHM:Mars:545GHz:units}{3\parcm80}
\setsymbol{HFI:FWHM:Mars:857GHz:units}{3\parcm67}

\setsymbol{HFI:FWHM:Mars:100GHz}{9.37}
\setsymbol{HFI:FWHM:Mars:143GHz}{7.04}
\setsymbol{HFI:FWHM:Mars:217GHz}{4.68}
\setsymbol{HFI:FWHM:Mars:353GHz}{4.43}
\setsymbol{HFI:FWHM:Mars:545GHz}{3.80}
\setsymbol{HFI:FWHM:Mars:857GHz}{3.67}

\setsymbol{HFI:beam:ellipticity:Mars:100GHz}{1.18}
\setsymbol{HFI:beam:ellipticity:Mars:143GHz}{1.03}
\setsymbol{HFI:beam:ellipticity:Mars:217GHz}{1.14}
\setsymbol{HFI:beam:ellipticity:Mars:353GHz}{1.09}
\setsymbol{HFI:beam:ellipticity:Mars:545GHz}{1.25}
\setsymbol{HFI:beam:ellipticity:Mars:857GHz}{1.03}

\setsymbol{HFI:CMB:relative:calibration:100GHz}{$\lsim 1\%$}
\setsymbol{HFI:CMB:relative:calibration:143GHz}{$\lsim 1\%$}
\setsymbol{HFI:CMB:relative:calibration:217GHz}{$\lsim 1\%$}
\setsymbol{HFI:CMB:relative:calibration:353GHz}{$\lsim 1\%$}
\setsymbol{HFI:CMB:relative:calibration:545GHz}{}
\setsymbol{HFI:CMB:relative:calibration:857GHz}{}

\setsymbol{HFI:CMB:absolute:calibration:100GHz}{$\lsim 2\%$}
\setsymbol{HFI:CMB:absolute:calibration:143GHz}{$\lsim 2\%$}
\setsymbol{HFI:CMB:absolute:calibration:217GHz}{$\lsim 2\%$}
\setsymbol{HFI:CMB:absolute:calibration:353GHz}{$\lsim 2\%$}
\setsymbol{HFI:CMB:absolute:calibration:545GHz}{}
\setsymbol{HFI:CMB:absolute:calibration:857GHz}{}

\setsymbol{HFI:FIRAS:gain:calibration:accuracy:statistical:100GHz}{}
\setsymbol{HFI:FIRAS:gain:calibration:accuracy:statistical:143GHz}{}
\setsymbol{HFI:FIRAS:gain:calibration:accuracy:statistical:217GHz}{}
\setsymbol{HFI:FIRAS:gain:calibration:accuracy:statistical:353GHz}{2.5\%}
\setsymbol{HFI:FIRAS:gain:calibration:accuracy:statistical:545GHz}{1\%}
\setsymbol{HFI:FIRAS:gain:calibration:accuracy:statistical:857GHz}{0.5\%}

\setsymbol{HFI:FIRAS:gain:calibration:accuracy:systematic:100GHz}{}
\setsymbol{HFI:FIRAS:gain:calibration:accuracy:systematic:143GHz}{}
\setsymbol{HFI:FIRAS:gain:calibration:accuracy:systematic:217GHz}{}
\setsymbol{HFI:FIRAS:gain:calibration:accuracy:systematic:353GHz}{}
\setsymbol{HFI:FIRAS:gain:calibration:accuracy:systematic:545GHz}{7\%}
\setsymbol{HFI:FIRAS:gain:calibration:accuracy:systematic:857GHz}{7\%}

\setsymbol{HFI:FIRAS:zero:point:accuracy:100GHz:units}{0.8\MJysr}
\setsymbol{HFI:FIRAS:zero:point:accuracy:143GHz:units}{}
\setsymbol{HFI:FIRAS:zero:point:accuracy:217GHz:units}{}
\setsymbol{HFI:FIRAS:zero:point:accuracy:353GHz:units}{1.4\MJysr}
\setsymbol{HFI:FIRAS:zero:point:accuracy:545GHz:units}{2.2\MJysr}
\setsymbol{HFI:FIRAS:zero:point:accuracy:857GHz:units}{1.7\MJysr}

\setsymbol{HFI:FIRAS:zero:point:accuracy:100GHz}{0.8}
\setsymbol{HFI:FIRAS:zero:point:accuracy:143GHz}{}
\setsymbol{HFI:FIRAS:zero:point:accuracy:217GHz}{}
\setsymbol{HFI:FIRAS:zero:point:accuracy:353GHz}{1.4}
\setsymbol{HFI:FIRAS:zero:point:accuracy:545GHz}{2.2}
\setsymbol{HFI:FIRAS:zero:point:accuracy:857GHz}{1.7}

\setsymbol{HFI:unit:conversion:100GHz:units}{0.2415\MJysrmK}
\setsymbol{HFI:unit:conversion:143GHz:units}{0.3694\MJysrmK}
\setsymbol{HFI:unit:conversion:217GHz:units}{0.4811\MJysrmK}
\setsymbol{HFI:unit:conversion:353GHz:units}{0.2883\MJysrmK}
\setsymbol{HFI:unit:conversion:545GHz:units}{0.05826\MJysrmK}
\setsymbol{HFI:unit:conversion:857GHz:units}{0.002238\MJysrmK}

\setsymbol{HFI:unit:conversion:100GHz}{0.2415}
\setsymbol{HFI:unit:conversion:143GHz}{0.3694}
\setsymbol{HFI:unit:conversion:217GHz}{0.4811}
\setsymbol{HFI:unit:conversion:353GHz}{0.2883}
\setsymbol{HFI:unit:conversion:545GHz}{0.05826}
\setsymbol{HFI:unit:conversion:857GHz}{0.002238}

\setsymbol{HFI:colour:correction:alpha=-2:V1.01:100GHz}{0.9893}
\setsymbol{HFI:colour:correction:alpha=-2:V1.01:143GHz}{0.9759}
\setsymbol{HFI:colour:correction:alpha=-2:V1.01:217GHz}{1.0007}
\setsymbol{HFI:colour:correction:alpha=-2:V1.01:353GHz}{1.0028}
\setsymbol{HFI:colour:correction:alpha=-2:V1.01:545GHz}{1.0019}
\setsymbol{HFI:colour:correction:alpha=-2:V1.01:857GHz}{0.9889}

\setsymbol{HFI:colour:correction:alpha=0:V1.01:100GHz}{1.0008}
\setsymbol{HFI:colour:correction:alpha=0:V1.01:143GHz}{1.0148}
\setsymbol{HFI:colour:correction:alpha=0:V1.01:217GHz}{0.9909}
\setsymbol{HFI:colour:correction:alpha=0:V1.01:353GHz}{0.9888}
\setsymbol{HFI:colour:correction:alpha=0:V1.01:545GHz}{0.9878}
\setsymbol{HFI:colour:correction:alpha=0:V1.01:857GHz}{1.0014}

\begin{document}

\title{ \textit{Planck} early results: first assessment \\
of the High Frequency Instrument in-flight performance}

\authorrunning{The \Planck\ HFI core team}
\titlerunning{\textit{Planck} early results: first assessment of
the HFI in-flight performance}

\author{\small
Planck HFI Core Team:
P.~A.~R.~Ade\inst{47}
\and
N.~Aghanim\inst{25}
\and
R.~Ansari\inst{39}
\and
M.~Arnaud\inst{35}
\and
M.~Ashdown\inst{33, 54}
\and
J.~Aumont\inst{25}
\and
A.~J.~Banday\inst{52, 5, 41}
\and
M.~Bartelmann\inst{51, 41}
\and
J.~G.~Bartlett\inst{1, 31}
\and
E.~Battaner\inst{57}
% ...
}
\institute{\small
Astroparticule et Cosmologie, CNRS (UMR7164), Universit\'{e} Denis Diderot Paris 7, B\^{a}timent Condorcet, 10 rue A. Domon et L\'{e}onie Duquet, Paris, France\\
\and
Atacama Large Millimeter/submillimeter Array, ALMA Santiago Central Offices Alonso de Cordova 3107, Vitacura, Casilla 763 0355, Santiago, Chile\\
\and
CITA, University of Toronto, 60 St. George St., Toronto, ON M5S 3H8, Canada\\
\and
CNES, 18 avenue Edouard Belin, 31401 Toulouse Cedex 9, France\\
\and
CNRS, IRAP, 9 Av. colonel Roche, BP 44346, F-31028 Toulouse cedex 4, France\\
\and
California Institute of Technology, Pasadena, California, U.S.A.\\
\and
DAMTP, Centre for Mathematical Sciences, Wilberforce Road, Cambridge CB3 0WA, U.K.\\
\and
DSM/Irfu/SPP, CEA-Saclay, F-91191 Gif-sur-Yvette Cedex, France\\
\and
DTU Space, National Space Institute, Juliane Mariesvej 30, Copenhagen, Denmark\\
\and
Department of Astronomy and Astrophysics, University of Toronto, 50 Saint George Street, Toronto, Ontario, Canada\\
\and
Department of Physics, Princeton University, Princeton, New Jersey, U.S.A.\\
\and
Department of Physics, University of California, One Shields Avenue, Davis, California, U.S.A.\\
\and
Department of Physics, University of Illinois at Urbana-Champaign, 1110 West Green Street, Urbana, Illinois, U.S.A.\\
\and
Dipartimento di Fisica, Universit\`{a} La Sapienza, P. le A. Moro 2, Roma, Italy\\
\and
Dipartimento di Fisica, Universit\`{a} degli Studi di Milano, Via Celoria, 16, Milano, Italy\\
\and
European Southern Observatory, ESO Vitacura, Alonso de Cordova 3107, Vitacura, Casilla 19001, Santiago, Chile\\
\and
European Space Agency, ESTEC, Keplerlaan 1, 2201 AZ Noordwijk, The Netherlands\\
\and
INAF - Osservatorio Astronomico di Trieste, Via G.B. Tiepolo 11, Trieste, Italy\\
\and
INAF/IASF Bologna, Via Gobetti 101, Bologna, Italy\\
\and
INAF/IASF Milano, Via E. Bassini 15, Milano, Italy\\
\and
INSU, Institut des sciences de l'univers, CNRS, 3, rue Michel-Ange, 75794 Paris Cedex 16, France\\
\and
IPAG: Institut de Plan\'{e}tologie et d'Astrophysique de Grenoble, Universit\'{e} Joseph Fourier, Grenoble 1 / CNRS-INSU, UMR 5274, Grenoble, F-38041, France\\
\and
Imperial College London, Astrophysics group, Blackett Laboratory, Prince Consort Road, London, SW7 2AZ, U.K.\\
\and
Infrared Processing and Analysis Center, California Institute of Technology, Pasadena, CA 91125, U.S.A.\\
\and
Institut d'Astrophysique Spatiale, CNRS (UMR8617) Universit\'{e} Paris-Sud 11, B\^{a}timent 121, Orsay, France\\
\and
Institut d'Astrophysique de Paris, CNRS UMR7095, Universit\'{e} Pierre \& Marie Curie, 98 bis boulevard Arago, Paris, France\\
\and
Institut de Ci\`{e}ncies de l'Espai, CSIC/IEEC, Facultat de Ci\`{e}ncies, Campus UAB, Torre C5 par-2, Bellaterra 08193, Spain\\
\and
Institut de Radioastronomie Millim\'{e}trique (IRAM), Avenida Divina Pastora 7, Local 20, 18012 Granada, Spain\\
\and
Institut de Radioastronomie Millim\'{e}trique (IRAM), Domaine Universitaire de Grenoble, 300 rue de la Piscine, 38406, Grenoble, France\\
\and
Instituto de Astrof\'{\i}sica de Canarias, C/V\'{\i}a L\'{a}ctea s/n, La Laguna, Tenerife, Spain\\
\and
Jet Propulsion Laboratory, California Institute of Technology, 4800 Oak Grove Drive, Pasadena, California, U.S.A.\\
\and
Jodrell Bank Centre for Astrophysics, Alan Turing Building, School of Physics and Astronomy, The University of Manchester, Oxford Road, Manchester, M13 9PL, U.K.\\
\and
Kavli Institute for Cosmology Cambridge, Madingley Road, Cambridge, CB3 0HA, U.K.\\
\and
LERMA, CNRS, Observatoire de Paris, 61 Avenue de l'Observatoire, Paris, France\\
\and
Laboratoire AIM, IRFU/Service d'Astrophysique - CEA/DSM - CNRS - Universit\'{e} Paris Diderot, B\^{a}t. 709, CEA-Saclay, F-91191 Gif-sur-Yvette Cedex, France\\
\and
Laboratoire Traitement et Communication de l'Information, CNRS (UMR 5141) and T\'{e}l\'{e}com ParisTech, 46 rue Barrault F-75634 Paris Cedex 13, France\\
\and
Laboratoire d'Astrophysique de Marseille, 38 rue Fr\'{e}d\'{e}ric Joliot-Curie, 13388, Marseille Cedex 13, France\\
\and
Laboratoire de Physique Subatomique et de Cosmologie, CNRS, Universit\'{e} Joseph Fourier Grenoble I, 53 rue des Martyrs, Grenoble, France\\
\and
Laboratoire de l'Acc\'{e}l\'{e}rateur Lin\'{e}aire, Universit\'{e} Paris-Sud 11, CNRS/IN2P3, Orsay, France\\
\and
Lawrence Berkeley National Laboratory, Berkeley, California, U.S.A.\\
\and
Max-Planck-Institut f\"{u}r Astrophysik, Karl-Schwarzschild-Str. 1, 85741 Garching, Germany\\
\and
National University of Ireland, Department of Experimental Physics, Maynooth, Co. Kildare, Ireland\\
\and
Observational Cosmology, Mail Stop 367-17, California Institute of Technology, Pasadena, CA, 91125, U.S.A.\\
\and
Optical Science Laboratory, University College London, Gower Street, London, U.K.\\
\and
Rutherford Appleton Laboratory, Chilton, Didcot, U.K.\\
\and
SUPA, Institute for Astronomy, University of Edinburgh, Royal Observatory, Blackford Hill, Edinburgh EH9 3HJ, U.K.\\
\and
School of Physics and Astronomy, Cardiff University, Queens Buildings, The Parade, Cardiff, CF24 3AA, U.K.\\
\and
Space Research Institute (IKI), Russian Academy of Sciences, Profsoyuznaya Str, 84/32, Moscow, 117997, Russia\\
\and
Space Sciences Laboratory, University of California, Berkeley, California, U.S.A.\\
\and
Stanford University, Dept of Physics, Varian Physics Bldg, 382 Via Pueblo Mall, Stanford, California, U.S.A.\\
\and
Universit\"{a}t Heidelberg, Institut f\"{u}r Theoretische Astrophysik, Albert-\"{U}berle-Str. 2, 69120, Heidelberg, Germany\\
\and
Universit\'{e} de Toulouse, UPS-OMP, IRAP, F-31028 Toulouse cedex 4, France\\
\and
Universities Space Research Association, Stratospheric Observatory for Infrared Astronomy, MS 211-3, Moffett Field, CA 94035, U.S.A.\\
\and
University of Cambridge, Cavendish Laboratory, Astrophysics group, J J Thomson Avenue, Cambridge, U.K.\\
\and
University of Cambridge, Institute of Astronomy, Madingley Road, Cambridge, U.K.\\
\and
University of Cambridge/Astrophysics, Group Madingley, Road CB3 0HE, Cambridge, U.K.\\
\and
University of Granada, Departamento de F\'{\i}sica Te\'{o}rica y del Cosmos, Facultad de Ciencias, Granada, Spain\\
% ...
}

\date{\today}

\abstract{The \Planck\ High Frequency Instrument (HFI) is designed to
  measure the temperature and polarization anisotropies of the Cosmic
  Microwave Background and galactic foregrounds in six wide bands
  centered at 100, 143, 217, 353, 545 and 857\,GHz at an angular
  resolution of 10\arcm\ (100\,GHz), 7\arcm\ (143\,GHz), and
  5\arcm\ (217\,GHz and higher). HFI has been operating flawlessly
  since launch on 14 May 2009. The bolometers cooled to 100\,mK as
  planned.  The settings of the readout electronics, such as the
  bolometer bias current, that optimize HFI's noise performance on
  orbit are nearly the same as the ones chosen during ground testing.
  Observations of Mars, Jupiter, and Saturn verified both the optical
  system and the time response of the detection chains. The optical
  beams are close to predictions from physical optics modeling. The
  time response of the detection chains is close to pre-launch
  measurements.
The detectors suffer from an unexpected high flux of cosmic rays
related to low solar activity. Due to the redundancy of Planck's
observation strategy, the removal of a few percent of data
contaminated by glitches does not significantly affect the
sensitivity. The cosmic rays heat up the bolometer plate
and the modulation on periods of days to months of the heat load
creates a common drift of all bolometer signals which do not affect
the scientific capabilities.  Only the high energy cosmic ray showers
induce inhomogeneous heating which is a probable source of low
frequency noise.
The removal of systematic effects in the time ordered data provides a
signal with an average level of noise less than 70\% of our goal
values in the 0.6--2.5\,Hz range. This is slightly higher than the
pre-launch measurements but better than predicted in the early phases
of the project.  This is attributed to the low level of photon noise
resulting from an optimized optical and thermal design.

~
}


\maketitle

\allearlypapers

\section{Introduction}
\label{PlanckIPF1.5:Sect1}

\Planck\footnote{\Planck\ (http://www.esa.int/\Planck) is a project
of the European Space Agency (ESA) with instruments provided by two
scientific consortia funded by ESA member states (in particular the
lead countries France and Italy), with contributions from NASA (USA)
and telescope reflectors provided by a collaboration between ESA and a
scientific consortium led and funded by Denmark.} \citep{tauber2010a,
planck2011-1.1}
is the third-generation space mission to measure the
anisotropy of the cosmic microwave background (CMB).  It observes the
sky in nine frequency bands covering 30--857\,GHz with high
sensitivity and angular resolution from 31\arcm\ to 5\arcm.  The Low
Frequency Instrument (LFI; \citealt{mandolesi2010, bersanelli2010,
planck2011-1.4}) covers the 30, 44, and 70\,GHz bands with amplifiers
cooled to 20\,\hbox{K}.  The High Frequency Instrument (HFI;
\citealt{lamarre2010, planck2011-1.5}) covers the 100, 143, 217, 353,
545, and 857\,GHz bands with bolometers cooled to 0.1\,\hbox{K}.
Polarization is measured in all but the highest two bands
\citep{leahy2010, rosset2010}.  A combination of radiative cooling and
three mechanical coolers provides the temperatures needed for the
detectors and optics \citep{planck2011-1.3}.  Two data processing
centres (DPCs) check and calibrate the data and make maps of the sky
\citep{planck2011-1.7, planck2011-1.6}.  \Planck's sensitivity,
angular resolution, and frequency coverage make it a powerful
instrument for galactic and extragalactic astrophysics as well as
cosmology.  Early astrophysics results are given in Planck
Collaboration (2011h--x).

The goal of this paper is to describe the in-flight performance of the
HFI in space and after the challenging launch conditions. It does
not attempt to duplicate the content of the \Planck\ pre-launch status
papers \citep{lamarre2010,pajot2010}, but rather presents the
operational status from an instrumental viewpoint.  These results
propagate to scientific products through the data processing reported
in the companion paper \citep{planck2011-1.7} which describes the
instrumental properties as they appear in the maps used by the
``\Planck\ early results'' companion papers.  This paper focuses on
the ability of the HFI to measure intensity without any description of
its performance in measuring polarization, which will be reported
later.

Section \ref{PlanckIPF1.5:Sect2} summarizes the instrument design.
Section \ref{PlanckIPF1.5:Sect3} focuses on early in-flight
operations, the verification phase and the setting of the parameters
that have to be tuned in flight. Section \ref{PlanckIPF1.5:Sect4}
addresses the measurement of the beams on planets and the
disentangling of time response effects from the beam shape. It also
presents the best current knowledge of the physical beams resulting
from this work. The effective beam obtained after data processing are
to be found in \cite{planck2011-1.7}. Sections
\ref{PlanckIPF1.5:Sect5}, \ref{PlanckIPF1.5:Sect6} and
\ref{PlanckIPF1.5:Sect7} are dedicated to noise, systematic effects
and instrument stability respectively. A summary of the HFI in-flight
performance and a comparison with pre-launch expectations are
presented in section \ref{PlanckIPF1.5:Sect8}.

\section{The HFI instrument}
\label{PlanckIPF1.5:Sect2}

\begin{table*}[!tb]
\caption{The HFI receivers. P stands for polarisation sensitive bolometers.}
\label{PlanckIPF1.5:tab:Channels}
\centering
\begin{tabular}{l*{10}{c}}
\hline \hline
Channel &  & 100P & 143P &143 &217P &217 & 353P & 353 & 545 & 857 \\
\hline
Central frequency & (GHz) & 100 & 143 & 143 & 217 & 217 & 353 & 353 & 545 & 857 \\
Bandwidth & (\%)          & 33  & 32  & 30  & 29  & 33  & 29  & 28  & 31  & 30 \\
Number of bolometers & &8 &8 &4 &8 &4 &8 &4 &4 &4 \\
\hline
\end{tabular}
\end{table*}

\subsection{Design}

The High Frequency Instrument (HFI) was proposed to ESA in response to
the announcement of opportunity for instruments for the Planck mission
in 1995. It is designed to measure the sky in six bands
(Tab. \ref{PlanckIPF1.5:tab:Channels}) with bolometer sensitivity close to the
fundamental limit set by photon noise. The lower four frequency bands
include the measurement of the polarization. This sensitivity is
obtained through a combination of technological breakthroughs in each
of the critical components needed for bolometric detection:
\begin{itemize}
\item Spider web bolometers \citep{Bock1995,Holmes2008} and
 polarization sensitive bolometers \citep{Jones2003} which can reach
 the photon noise limit with sufficient bandwidth to enable scanning
 great circles on the sky at roughly 1\,rpm. They offer a very low
  cross-section to cosmic rays that proves to be essential in this
 environment and with this sensitivity.
\item A space qualified 100\,mK dilution cooler \citep{Benoit1997}
 associated with a 
high precision temperature
 control system.
\item An active cooler for 4\,K \citep{Bradshaw1997} using vibration
controlled mechanical compressors to prevent excessive warming of the
100\,mK stage and minimize parasitic effects on bolometers.
\item AC biased readout electronics that
 extend high sensitivity to very slow signals \citep{Gaertner1997}.
\item A thermo-optical design consisting, for each optical channel, of
 three corrugated horns and a set of compact reflective filters and
 lenses at cryogenic temperatures \citep{Church1996}. These include
 high throughput (multimoded) corrugated horns for the 545 and
 857\,GHz channels \citep{Murphy2002}.
\end{itemize}

The angular resolution was chosen to extend the measurement of the
small scale features in the CMB, while keeping the level of stray
light to extremely low levels.  At the same time, at this sensitivity,
the measurement and removal of foregrounds requires a large number of
bands extending on both sides of the foreground minimum. This is
achieved with the six bands of the HFI (Table~\ref{PlanckIPF1.5:tab:Channels})
and the three bands of the Low Frequency Instrument (LFI;
\citealt{planck2011-1.4}).

The instrument uses a $\sim20$\,K sorption cooler common to the HFI
and the LFI \citep{planck2011-1.3,Bhandari2000,Bhandari2004}. The HFI
focal plane unit (FPU) is integrated inside the mechanical structure
of the LFI, on axis of the focal plane of a common telescope
\citep{tauber2010a}.

The ability to achieve background limited sensitivity was demonstrated
by the ARCHEOPS balloon-borne experiment \citep{Benoit2003a,
Benoit2003b}, an adaptation of the HFI designed for operation in the
environment of a stratospheric balloon. Similarly, the method of
polarimetry employed by the HFI was demonstrated by the Boomerang
experiment \citep{Montroy2006,Piacentini2006, Jones2006}. The HFI
itself was extensively tested on the ground during the calibration
campaigns \citep{pajot2010} at IAS in Orsay and CSL at Liège. However,
the fully integrated instrument was never characterized in an
operational environment like that of the second Earth-Sun Lagrange
point (L2).  In addition to thermal and gravitational environmental
conditions, the spectrum and flux of cosmic rays at L2 is vastly
different than that during the pre-flight testing. Finally, due to the
operational constraints of the cryogenic receiver, the end to end
optical assembly could not be tested on the ground with the focal
plane instruments.

The instrument design and development are described in
\cite{lamarre2010}. The calibration of the instrument is described in
\cite{pajot2010}. The overall thermal and cryogenic design and the
Planck payload performance are critical aspects of the mission.
Detailed system-level aspects are described in \citet{planck2011-1.1}
and \citet{planck2011-1.3}.

\begin{figure}
\includegraphics[width=\columnwidth,keepaspectratio]{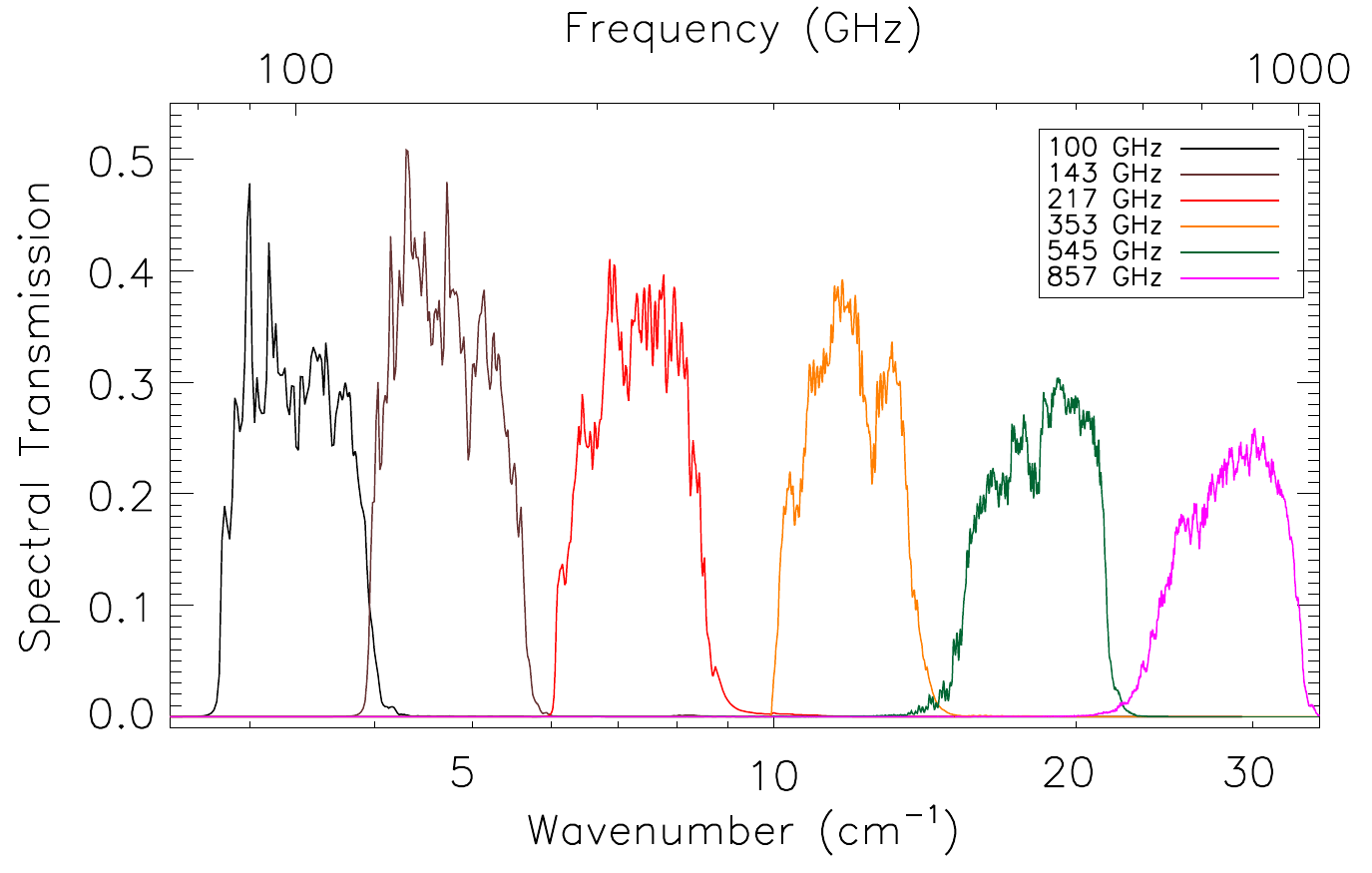}
\caption{HFI spectral transmission}
\label{PlanckIPF1.5:fig:spectral}  
\end{figure}

\subsection{Spectral transmission}

The spectral calibration is described in \citet{pajot2010} and consists
of pre-launch data, in the passband and around, combined with
component level data to determine the out of band rejection over an
extended frequency range (radio--UV).  Analysis of the in-flight data
shows that the contribution of CO rotational transitions to the HFI
measurements is important.  An evaluation of this contribution for the
$J=1\rightarrow0$ (100 and 143 GHz), $J=2\rightarrow1$ (217 GHz) and
$J=3\rightarrow2$ (353 GHz) transitions of CO is presented in
\citet{planck2011-1.7}.

\section{Early HFI operation}
\label{PlanckIPF1.5:Sect3}

\subsection{ HFI Cool down and cryogenic operating point}
\label{PlanckIPF1.5:SS:cooldown}

The \Planck\ satellite cooldown is described in \cite{planck2011-1.3}.

The first two weeks after launch were used for passive outgassing,
which ended on 2 June 2009.  During this period, gas was circulated
through the \HeJT\ cooler and the dilution cooler to prevent clogging
by condensable gases. The sorption cooler thermal interface with HFI
reached a temperature of 17.2\,K on 13~June. The $^4$He-JT cooler was
only operated at its nominal stroke amplitude of 3.5\,mm on 24~June to
leave time for the LFI to carry out a specific calibration with their
reference loads around 20\,\hbox{K}.  The operating temperature was
reached on 27~June, with the thermal interface with the focal plane
unit at 4.37\,K.

The dilution cooler cold head reached 93\,mK on 3 July 2009. Taking
into account the specific LFI calibration requirement that slowed down
the cooldown, the system behaved as expected within a few days,
according to the thermal models adjusted to the full system cryogenic
tests in the summer 2008 at CSL (Li\`ege).

The regulated operating temperature point of the 4\,K stage was set at
4.8\,K for the 4\,K feed horns on the \hbox{FPU}.  The other stages
were set to 1.395\,K for the so called 1.4\,K stage, 100.4\,mK for the
regulated dilution plate, and 103\,mK for the bolometer regulated
plate.

These numbers were very close to the planned operating point. As the
whole system worked nominally, margins on the cooling chain for
interface temperatures and heat lift are large. The \Planck\ active
cooling chain was one of the great technological challenges of this
mission and is fully successful.  A full description of the
performance of the cryogenic chain and its system aspects can be found
in \cite{planck2011-1.3}. The parameters of the operating points of
the 4\,K, 1.4\,K and 100\,mK stages are summarised in
Table~\ref{PlanckIPF1.5:tab:cryopoint}.

The temperature stability of the regulated stages has a direct impact
on the scientific performance of the \hbox{HFI}.  These stabilities
are discussed in detail in \cite{planck2011-1.3}.  Their impact on the
power received by the detectors is given in
Sect.~\ref{PlanckIPF1.5:SSS:Stability}.

\begin{table*}
\caption{Main operation and interface parameters of the cooling chain}
\label{PlanckIPF1.5:tab:cryopoint}
\begin{center}
\vspace{-0.5cm}
\begin{tabular}{lcc}
\hline \hline
Interface Sorption cooler-$^4$He-JT cooler (4\,K gas pre-cooling temperature)&17.2~K\\
Interface $^4$He-JT cooler-dilution cooler (dilution gas pre-cooling temperature)&4.37~K\\
Interface  1.4\,K cooler-dilution gas precooling&1.34~K\\
Temperature of dilution plate (after regulation)&100.4~mK\\
Temperature of bolometer plate (after regulation)&103~mK\\
Temperature of  1.4\,K plate (after regulation)&1.395~K\\
Temperature of 4\,K plate (after regulation)&4.80~K\\
Dilution plate PID power&24.3--30.7~nW\\
Bolometer plate PID power&5.1--7.4~nW\\
1.4\,K PID power&270~$\mu$W\\
4\,K PID power&1.7~mW\\
$^4$He-JT cooler stroke amplitude&3450~$\mu$m\\
Dilution cooler $^4$He flow rate&16.19--16.65~$\mu$mole/s\\
Dilution cooler $^3$He flow rate&5.92--6.00~$\mu$mole/s\\
Present survey life time (started 6 August 2009) &29.4~months\\
 \hline
\end{tabular}
\end{center}
\end{table*}

\subsection{Calibration and performance verification phase}

\subsubsection{Overview}

The calibration and performance verification (CPV) phase of the HFI
consisted of activities during the initial cooldown to 100\,mK and
during a period of about six weeks before the start of the survey.
The cooldown phase is summarized in Sect.~\ref{PlanckIPF1.5:SS:cooldown}.
The pre-launch value of the $^4$He-JT cooler operating frequency was
used (see Sect.~\ref{PlanckIPF1.5:SSS:freq4K}).  Activities related to the
optimization of the detection chain settings were performed first
during the cooldown of the JFET amplifiers, and again when the
bolometers were at their operating temperature.  Most of the operating
conditions were pre-determined during the ground calibration. The main
unknown was the in-flight background on the detectors. The detection
chain settings are presented in Sect.~\ref{PlanckIPF1.5:SSS:detchain}.
Other CPV activities performed are:
\begin{itemize}
\item determination of the detection chain time response under the
flight background
\item determination of the detection chain channel-to-channel
crosstalk under the flight background
\item characterization of the bolometer response to the 4\,K and
1.4\,K optical stages, and to the bolometer plate temperature
variations
\item checking the immunity of the instrument to the satellite
transponder
\item optimization of the numerical compression parameters for the
actual sky signal and high energy particle glitch rate
\item various ring-to-ring slew angles (1\parcm7, 2\parcm0 [nominal],
2\parcm5)
\item checking the effect of the scan angle with respect to the Sun
\item checking the effect of the satellite spin rate around its
nominal value of 1\,rpm.
\end{itemize}

On 5 August 2009, an unexpected shutdown of the $^4$He-JT cooler was
triggered by its current regulator unit (CRU). Despite investigations
into this event, its origin is still unexplained. A procedure for a
quick restart was developed and implemented in case the problem
recurred, but it has not. Six days were required to re-cool the
instrument to its operating point. The two-week first light survey
(FLS) followed this recovery, starting on 15 August 2009.  The FLS
allowed assessment of the quality of the instrument settings,
readiness of the data processing chain, and satellite scanning before
the start of science operations. The complete instrument and satellite
settings were validated and kept, and science operations began.  All
activities performed during the CPV phase confirmed the pre-launch
estimates of the instrument settings and operating mode.  We will
detail in the following paragraphs the most significant ones.

\subsubsection{$^4$He-JT cooler operating frequency setting}
\label{PlanckIPF1.5:SSS:freq4K}

The $^4$He-JT cooler operating frequency was set to the nominal value
of 40.08\,Hz determined during ground tests. Once the cryochain
stabilized, the in-flight behaviour of the cooler was found to be very
similar to that observed during ground tests. The lines observed in
the signal due to known electromagnetic interference (EMI) from the
$^4$He-JT cooler drive electronics have the same very narrow
width. The long term evolution of the $^4$He-JT cooler parasitic lines
is discussed in Sect.~\ref{PlanckIPF1.5:Sect6}.

\subsubsection{Detection chain parameters setting}
\label{PlanckIPF1.5:SSS:detchain}

The JFET preamplifiers are operated at the temperature which minimizes
their noise. This setting was checked when the bolometers were still
warm (above 100\,K) during the cooldown, since the bolometer 
Johnson noise was then much lower than the JFET noise. Optimum noise
performance of the JFETs was found close at 130 K, in agreement 
with the ground calibration.

\begin{figure*}
\includegraphics[width=\textwidth, keepaspectratio]{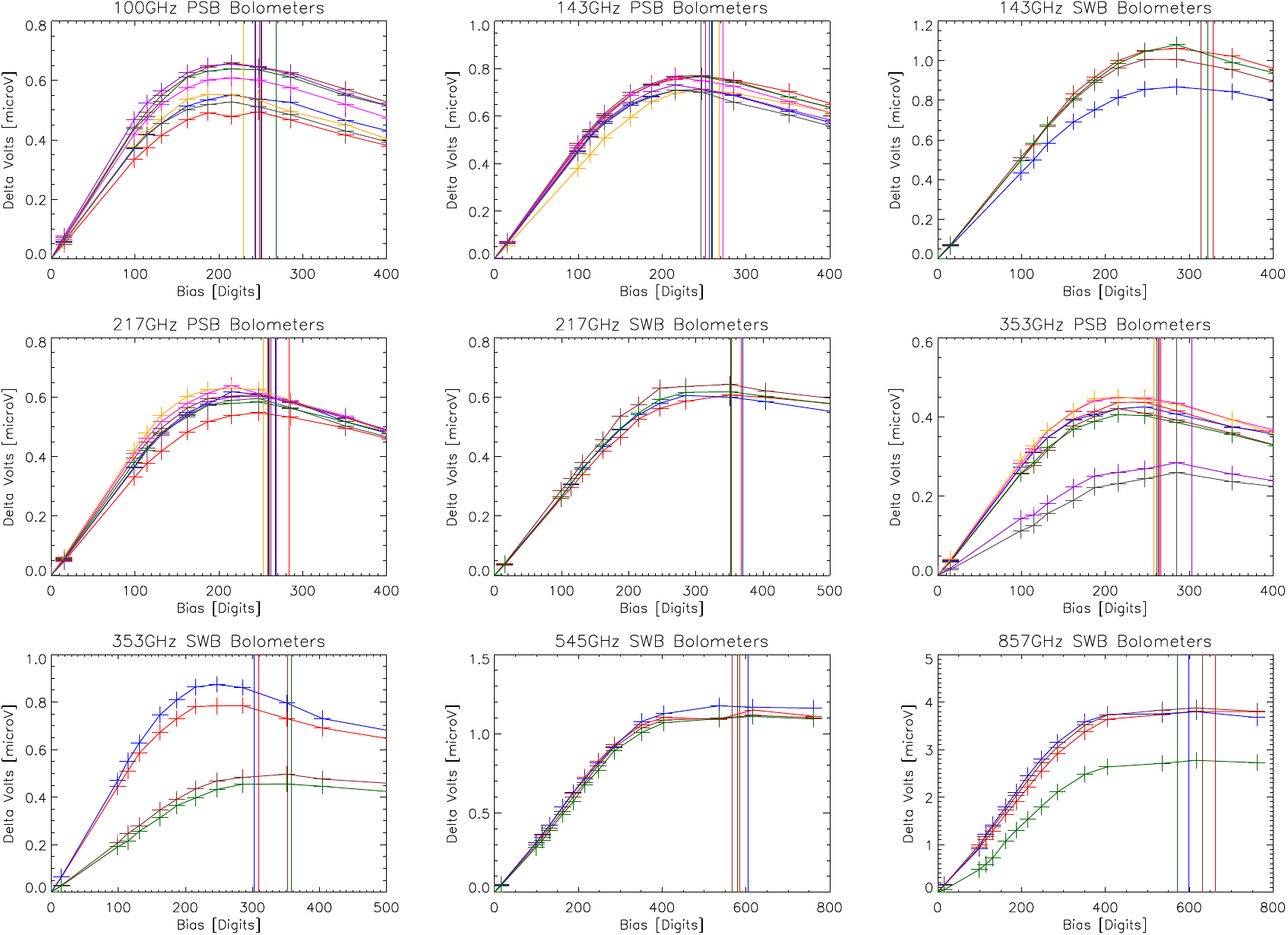}
\caption{Optimization of the bolometer bias currents. Vertical lines
   indicate the final bias value setting. These values are shifted
   with respect to the maximum because a dynamic response correction
   has been taken into account. A bias value of 100\,digits
   corresponds approximately to 0.1\,nA.}
\label{PlanckIPF1.5:fig:Resp_all}
\end{figure*}

After ground calibration, the only parameters of the REU remaining to
optimize in-flight were the bolometer bias current and the phase of
the lock-in detection, which slightly depends on the bolometer
impedance.  Fig.~\ref{PlanckIPF1.5:fig:Resp_all} shows the bolometer
responses for a set of bias current values measured while \Planck\ was
scanning the sky. For this sequence, the satellite rotation axis was
fixed.  For each bias value, the total detection chain noise was
computed after subtraction of the sky signal.  Ground measurements
have shown that the minimum NEP and the maximum responsivity bias
currents differ by less than 1\%.  Because of its higher
signal-to-noise ratio, we use the responsivity to optimize the bias
currents \citep{CatalanoACDC2010}. The optimum in-flight bias current
values correspond to the pre-launch estimates within 5\%. Therefore
the pre-launch settings, for which extensive ground characterizations
were performed, were kept (Fig.~\ref{PlanckIPF1.5:fig:Resp_all}). In a
similar way, the lock-in phase was explored and optimized, and again
the pre-launch settings were kept.

The optical background power on the bolometers is on the low end of
our rather conservative range of predictions, even lower than expected
from the ground measurements. This is attributed to a low telescope
temperature and no detectable contamination of the telescope surface
by dust during launch. This should result in a level of photon noise
lower than initially expected and an improved sensitivity.

\subsubsection{Numerical data-compression tuning}

The output of the readout electronic unit (REU) consists of one number
for each of the 72~science channels for each half-period of modulation
\citep{lamarre2010}. This number, $S_{\rm REU}$, is the exact sum of
the 40 16-bit ADC signal values obtained within the given
half-period. The data processor unit (DPU) performs a lossy
quantization of $S_{\rm REU}$.

First, 254 $S_{\rm REU}$ values corresponding to about 1.4\,s of
observation for each detector, covering a strip of sky about 8\deg\
long, are processed.  These 254 values are called a {\it compression
slice}.  The mean $<S_{\rm REU}>$ of the data within each compression
slice is computed, and data are demodulated using this mean:
\begin{equation}
  S_{{\rm demod},i}= (S_{{\rm REU},i}-<S_{\rm REU}>)*(-1)^i 
\end{equation}
where $1<i<254$ is the running index within the compression slice.

Then the mean $<S_{\rm demod}>$ of the demodulated data $S_{{\rm
demod},i}$ is computed and subtracted. The resulting data slice is
quantized according to a step $Q$ fixed per detector:
\begin{equation}
  S_{{\rm DPU},i}= \hbox{round}((S_{{\rm demod},i}-<S_{\rm demod}>)/Q)
\end{equation}
This is the lossy part of the algorithm: the required compression
factor, obtained through the tuning of the quantization step $Q$, adds
some extra noise to the data. For $\sigma/Q = 2$, where $\sigma$ is
the standard deviation of Gaussian white noise, the quantization adds
1\% to the noise \citep{pajot2010,pratt1978}. In flight, the value of
$\sigma$ was determined at the end of the CPV phase after subtraction
of the signal from the timeline.

The two means $<S_{\rm REU}>$ and $<S_{\rm demod}>$ computed as 32-bit
words are sent through the telemetry, together with the $S_{{\rm
DPU},i}$ values.  A variable length encoding of the $S_{{\rm DPU},i}$
values is performed on board, and the inverse decoding is applied on
ground.  This provides a lossless transmission of the quantized
values. A load limitation mechanism inhibits the data transmission,
first at the compression slice level (compression errors), and second
at the ring level \citep{lamarre2010}.

\begin{figure}
\centering
\includegraphics[width=0.5\textwidth,totalheight=.38\textwidth]{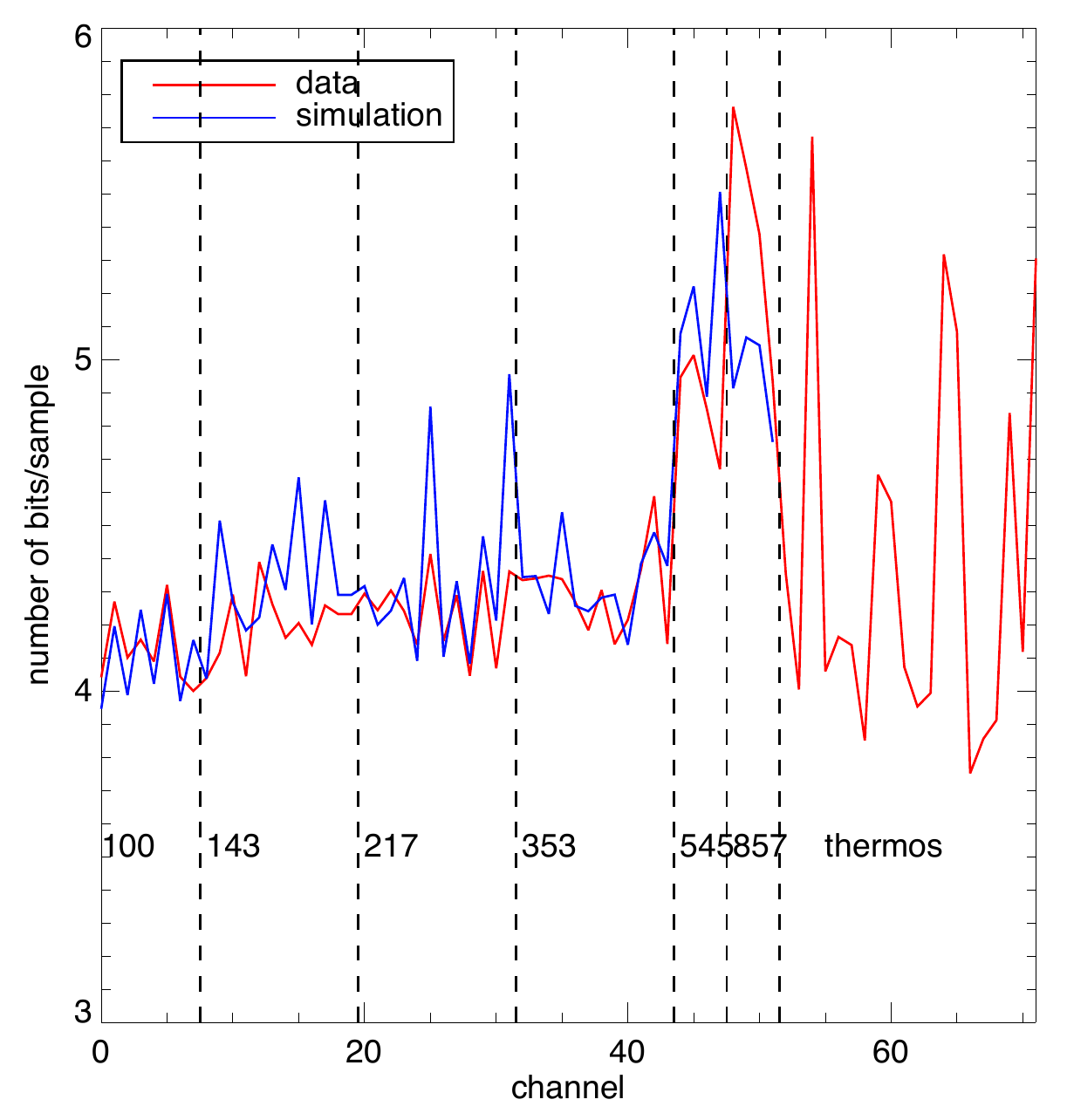}
\caption{Load measured for each HFI channel on 16 July 2010. Simulated
  data for the same patch of the sky are shown for
  bolometers. Channels \#54 and higher corrrespond to the fine
  thermometers on the optical stages of the instruments, plus a fixed
  resistor (\#60) and a capacitor (\#61) on the bolometer plate.}
\label{PlanckIPF1.5:fig:compload}
\end{figure}

\begin{figure}
\centering
\includegraphics[width=\columnwidth]{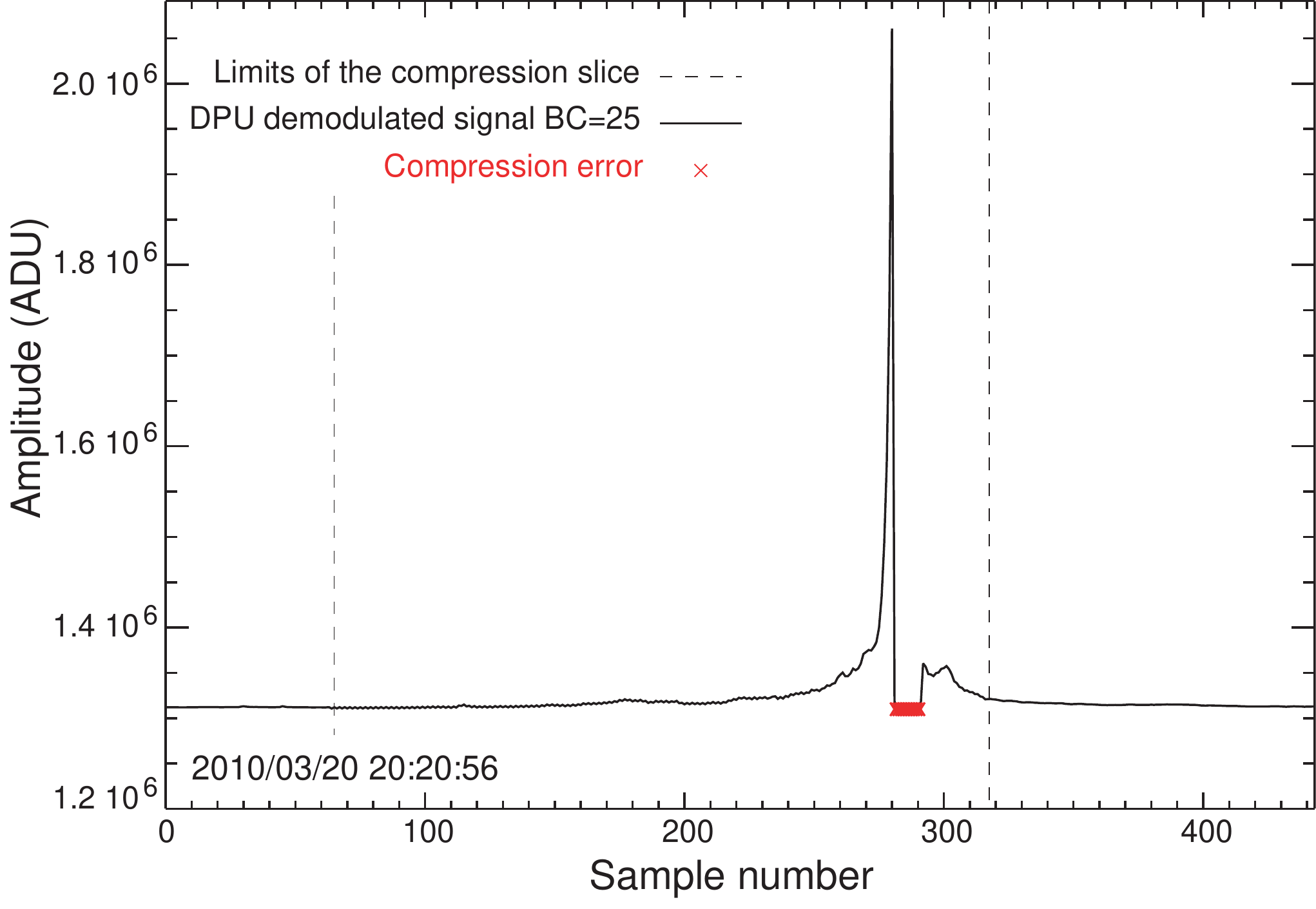}
\caption{Example of loss in one compression slice of data on bolometer
    857-1. Note the large signal-to-noise ratio while scanning through
    the galactic center.}
\label{PlanckIPF1.5:fig:comperror}
\end{figure}

For a given $Q$ value, the load on each channel depends on the dynamic
range of the signal above the level of the noise. This dynamic range
is largest for the high frequency bolometers because of the galactic
signal. The large rate of glitches due to high energy particle
interactions also contributes to the load of each channel.  Optimal
use of the bandpass available for the downlink (75\,kb\,s\mo\ average
for HFI science) was obtained by using initially a value of $Q =
\sigma/2.5$ for all bolometer signals, therefore including a margin
with respect to the requirement of $\sigma/Q = 2$. The load on each
HFI channel is shown and compared to simulated data in
Fig.~\ref{PlanckIPF1.5:fig:compload}. The increase of signal gradients while
scanning through the galactic center in September 2009 triggered the
load limitation mechanism (compression error) and up to 80000 samples
were lost for each of the 857\,GHz band bolometers. Therefore a new
value of $Q = \sigma/2$ was set for those bolometers from 21 December
2009 onward, reducing the number of samples lost to less than 200
during the following scan through the galactic center in March 2010.
An illustration of a compression error loss is shown in
Fig.~\ref{PlanckIPF1.5:fig:comperror}. Thanks to the redundancy of the \Planck\
scan strategy and the irregular distribution of the few remaining
compression errors, no pixels are missing in the maps of the high
signal-to-noise ratio galactic center regions. Periodic checks of the
noise value $\sigma$ are done for each channel, but no deviation
requiring a change in the quantization step $Q$ has been encountered
so far.

\subsubsection{Instrument readiness at the end of the CPV phase}

The overall readiness of the instrument was assessed during the
\hbox{FLS}. This end-to-end test was completely successful, from both
the instrument setting and the satellite scanning points of view. The
part of the sky covered during the FLS was included in the first all
sky survey.

\subsection{Response}

\subsubsection{Variation of the signal with background and with the bolometer plate temperature}
\label{PlanckIPF1.5:SSS:Stability}

The optical background on the bolometers originates from the sky, the
telescope, and from the HFI itself.  The operating point of the
bolometers is constrained by this total optical background, and the
fluctuations of this background have a direct impact on the stability
of the HFI measurements.

The power spectral density of each contribution to the background is
compared to 30\% of the total noise measured in-flight (NEP$_1$ column
of Table~\ref{PlanckIPF1.5:tab:noiseJML}). This specification corresponds to a
quadratic contribution smaller than 5\% on the total noise.

\begin{figure*}
\centering
\includegraphics[width=\columnwidth,keepaspectratio]{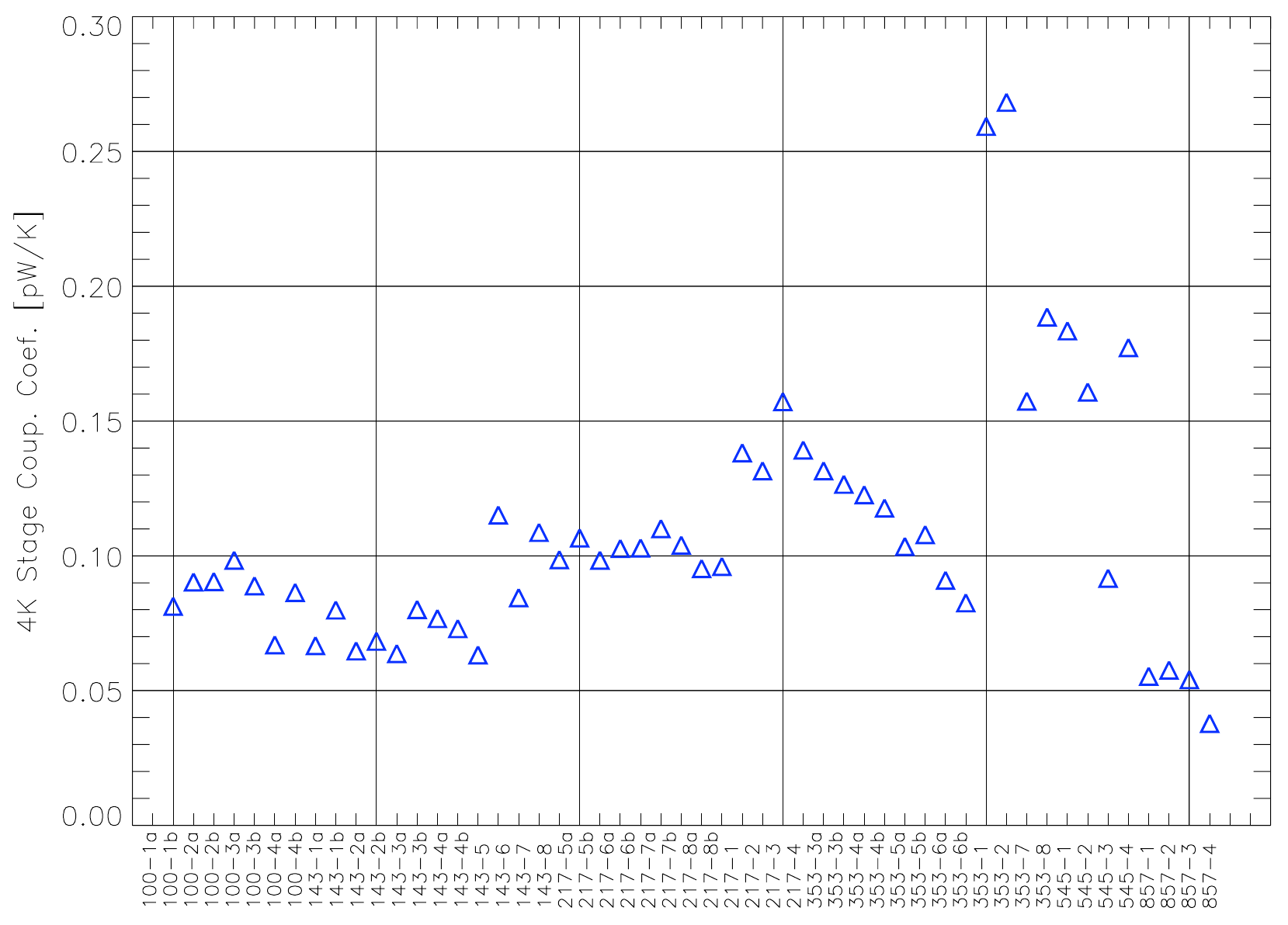}
\includegraphics[width=\columnwidth,keepaspectratio]{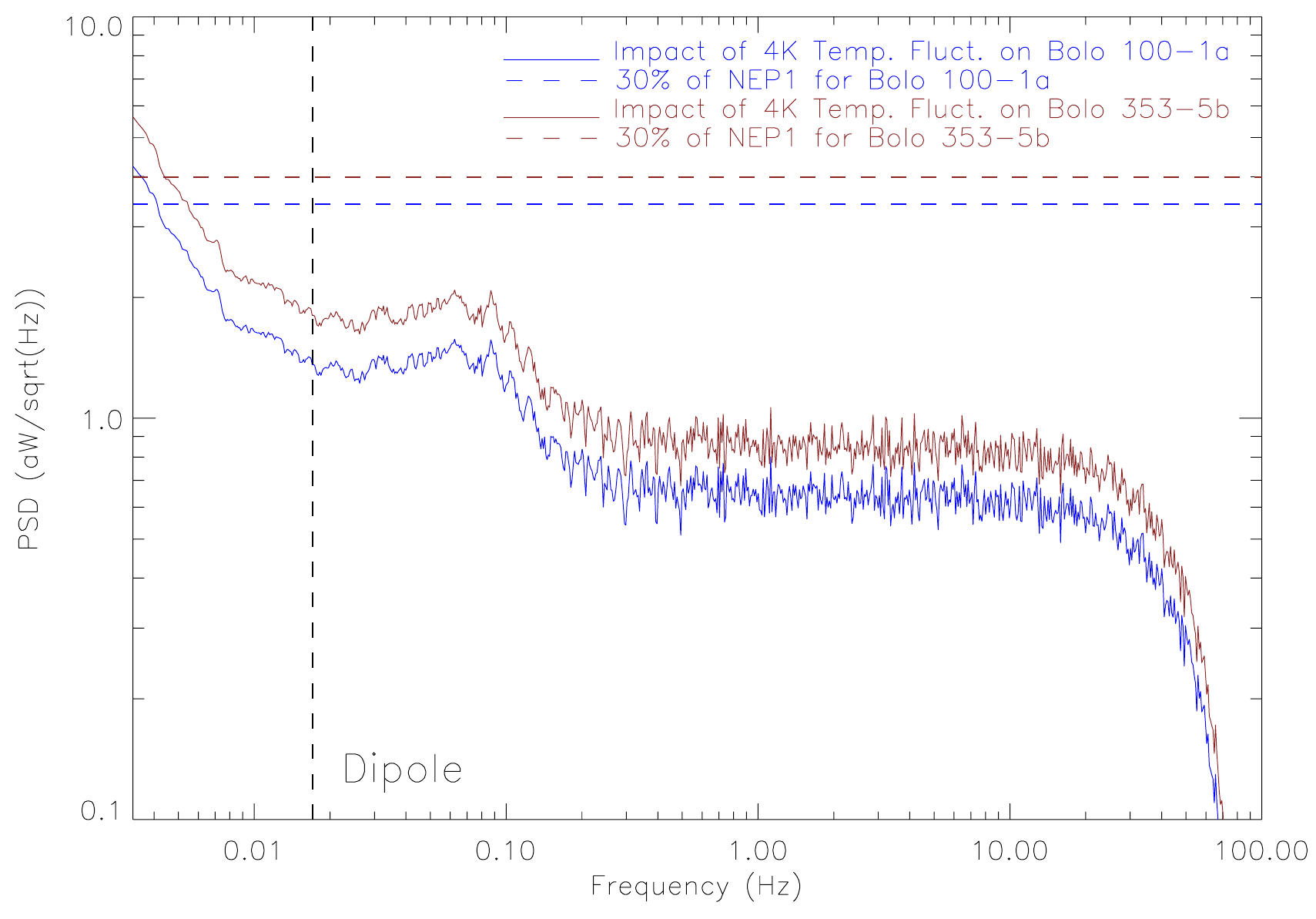}
\caption{Left: coupling coefficients of the 4\,K stage. Right: scaled
  power spectral density (PSD) of the 4\,K stage thermal fluctuations
  for the 100-1a and 353-5a-7a bolometers.}
\label{PlanckIPF1.5:fig:4k}
\end{figure*}

\begin{figure*}
\centering
\includegraphics[width=\columnwidth,keepaspectratio]{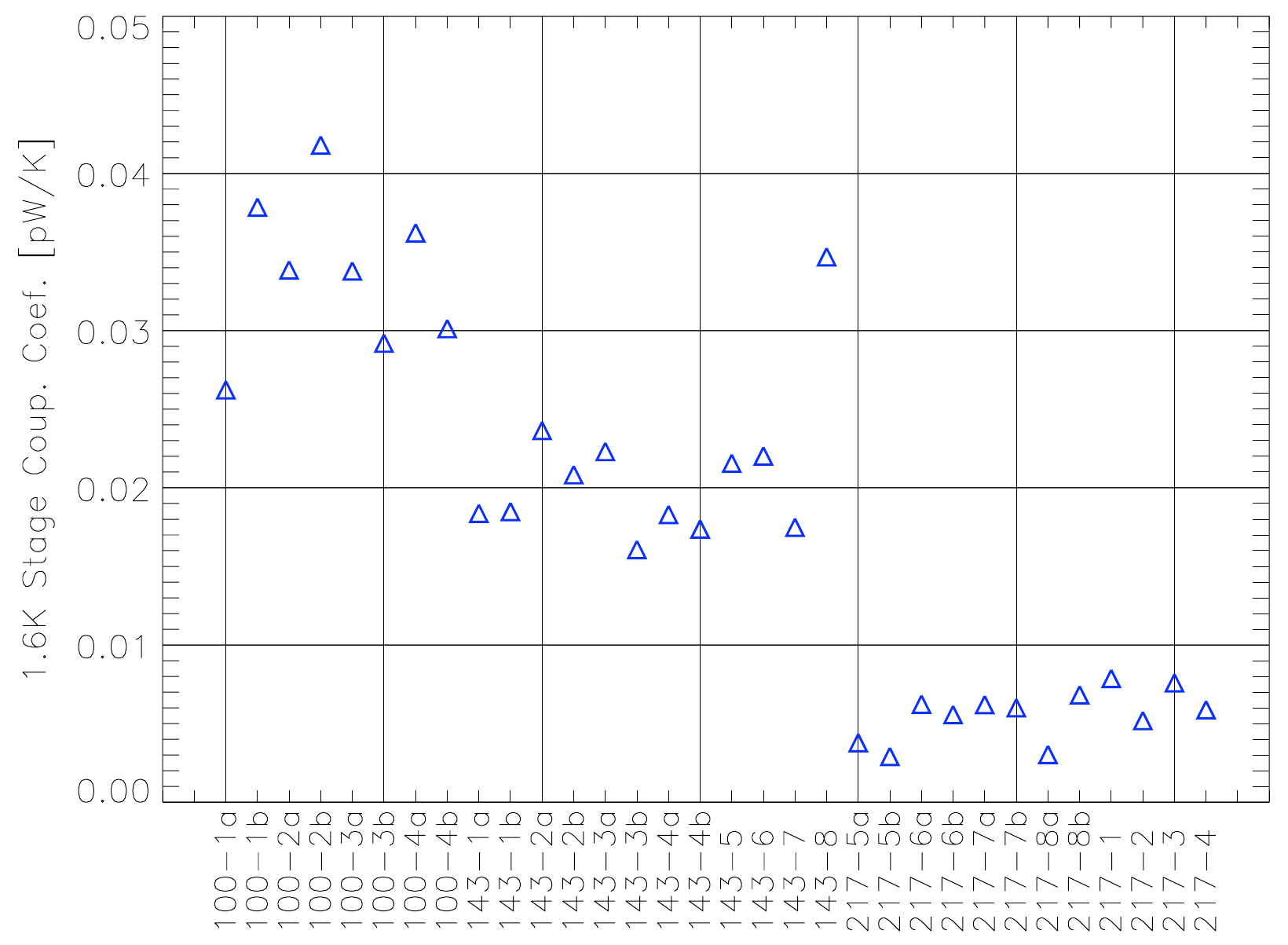}
\includegraphics[width=\columnwidth,keepaspectratio]{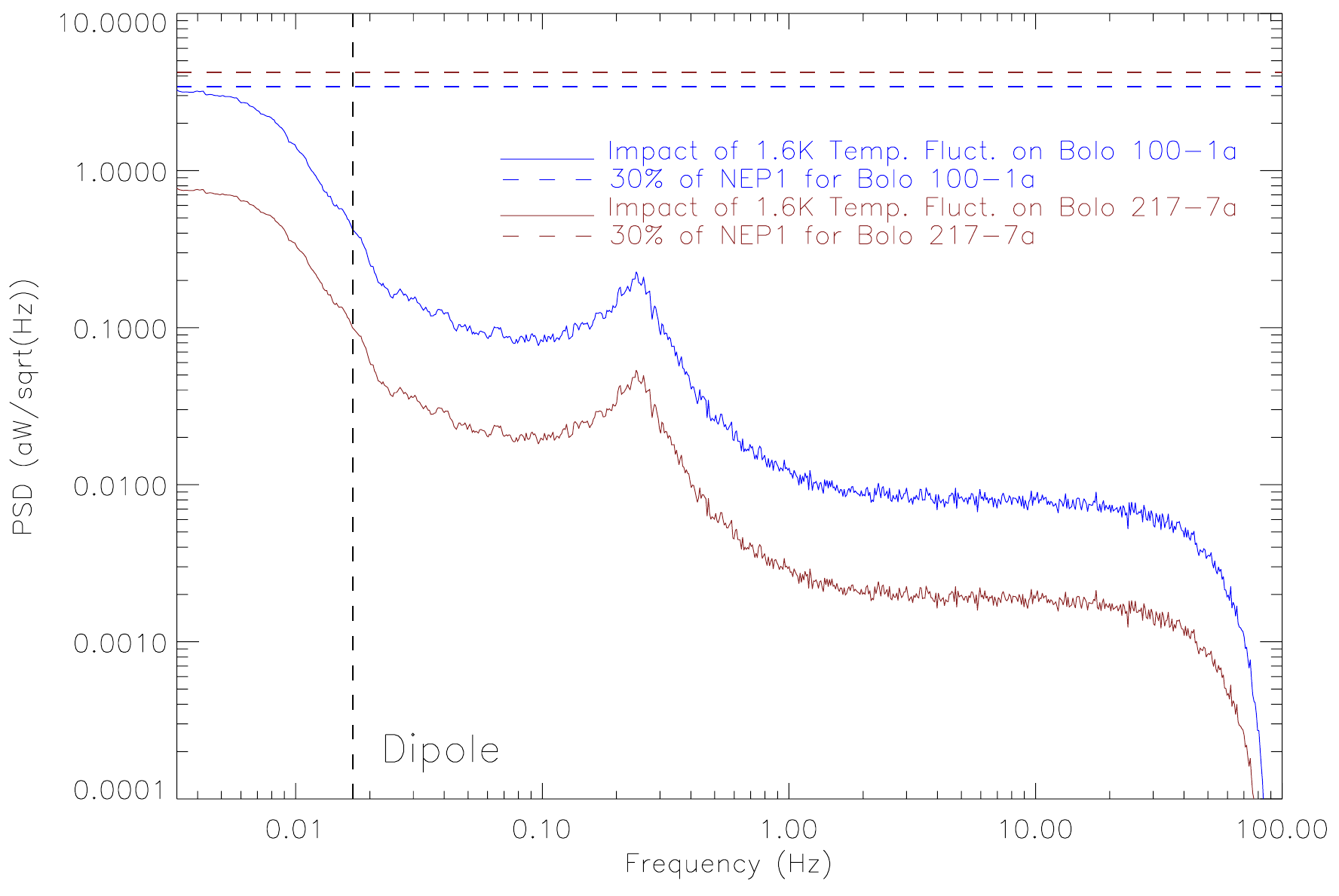}
\caption{Left: coupling coefficients of the 1.4\,K stage. The thermal
    emission in high frequency bands becomes too small to be
    measured. Right: scaled PSD of the 1.4\,K stage thermal
    fluctuations for the 100-1a and 353-5a-7a bolometers.}
\label{PlanckIPF1.5:fig:16k}
\end{figure*}

\begin{figure}
\centering
\includegraphics[width=\columnwidth,keepaspectratio]{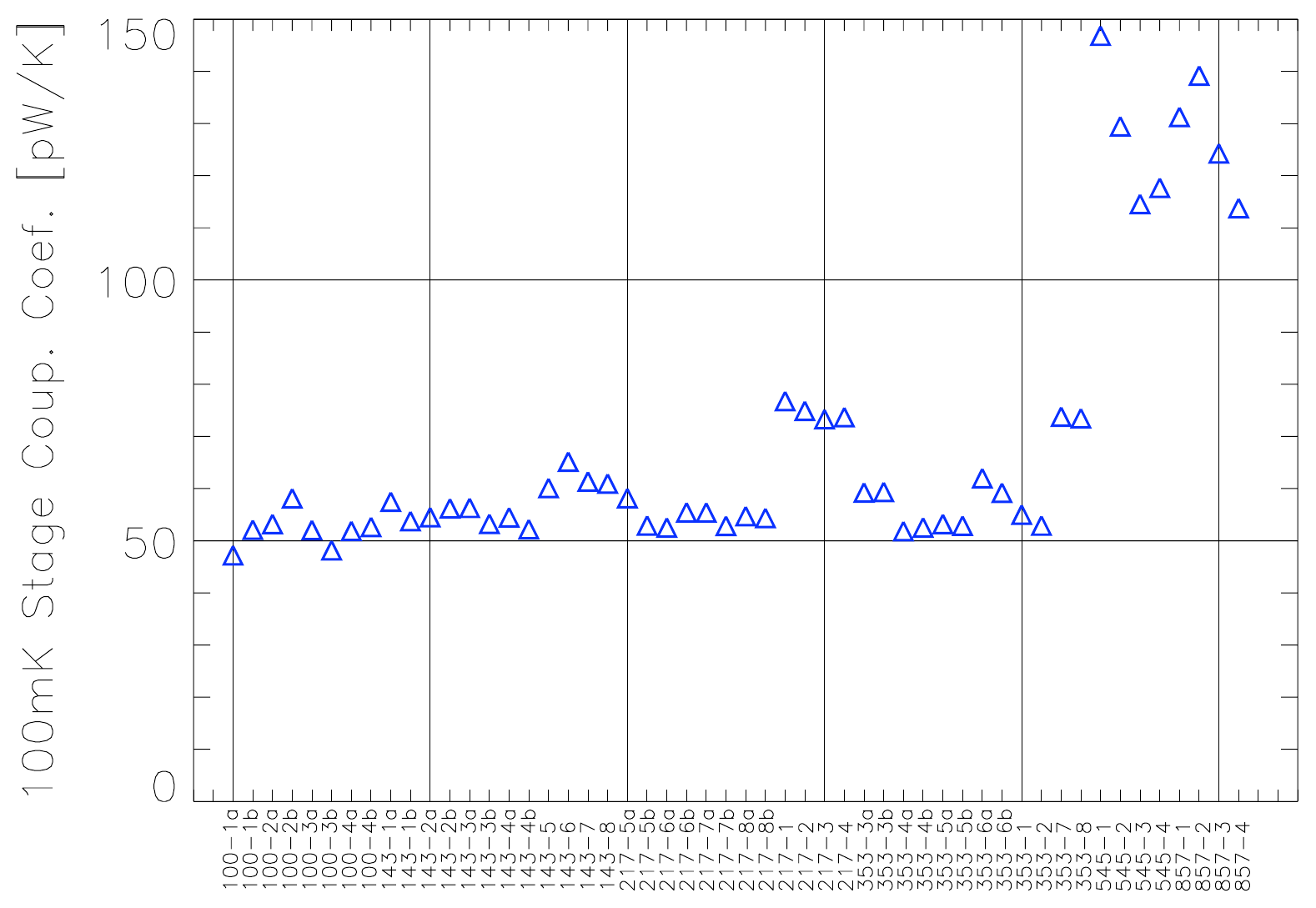}
\caption{Bolometer signal coupling coefficients to the 100mK bolometer plate.}
\label{PlanckIPF1.5:fig:100mk}
\end{figure}

The in-flight temperature stability of the HFI cryogenic stages is
discussed in \cite{planck2011-1.3}. The optical coupling of the HFI
bolometers to each cryogenic stage is shown in the left panels of
Figs. \ref{PlanckIPF1.5:fig:4k} and \ref{PlanckIPF1.5:fig:16k} and in
Fig. \ref{PlanckIPF1.5:fig:100mk}.  (The fact that the 100\,mK
couplings all agree with pre-launch measurements shows that no
bolometers were damaged during launch.)  These couplings are used to
calculate the effect of the fluctuations of each cryogenic stage on
the bolometer signals. The right panels of
Figs.~\ref{PlanckIPF1.5:fig:4k} and \ref{PlanckIPF1.5:fig:16k} show
the power spectral density (PSD) of the respective thermometers scaled
by the optical coupling factors for the most extreme bolometers.  The
scaled PSDs of the thermal fluctuations of the 4\,K and 1.4\,K stages
are below the line corresponding to 30\% of the total noise of the
corresponding bolometer for all frequencies above the spacecraft spin
frequency.

The bolometer plate thermometers have a large cosmic particle hit rate
\citep{planck2011-1.3} because of the large size of their sensors
compared to that of the bolometers.  Cosmic ray hits detection and
removal do not allow us to reach the thermometer nominal sensitivity,
therefore they cannot be used to remove the effect of bolometer plate
temperature fluctuations on the bolometer signal.  Instead, the data
processing pipeline \citep{planck2011-1.7} uses blind bolometers
located on the bolometer plate. The bolometer noise components are 
discussed in Sect.~\ref{PlanckIPF1.5:Sect6}.

\subsubsection{Linearity}
\label{PlanckIPF1.5:SSS:Linearity}

The way a bolometer transforms absorbed optical power into a voltage
is not a linear process because both the conductance between the
bolometer and the heat sink, and the bolometer impedance have a
non-linear dependence on the temperature (see
e.g. \cite{Catalano2008Thesis,Sudiwala2000}).

\begin{table}
\caption{Relative response deviation (in \%) from linearity for the
CMB dipole, the galactic center (GC) and planets. Saturation (Sat.)
occurs for the Jupiter measurements at high frequency.}
\label{PlanckIPF1.5:tab:nonlinearity}
\vspace{-0.25cm}
\begin{tabular}{lcccccc}
\hline\hline
& Dipole & GC &Mars & Saturn & Jupiter\\
\hline
100 GHz &$3.8~10^{-4}$ &0.001 & 0.01& 0.13 &  0.8\\
143 GHz &$10^{-3}$&0.0017 & 0.02 &               0.18 &  1.0   \\
217 GHz&    $8~10^{-4}$ &0.003 &   0.05&              0.53 & 3.2     \\
353 GHz&    $6.4~10^{-4}$ &0.007 &   0.06&                
0.8 & 4.5      \\
545 GHz &$<10^{-4}$ &0.01 &  0.08&               0.8  & Sat.      \\
857 GHz& $<10^{-4}$  &0.1 &   0.06&                   0.8  &  Sat.     \\
\hline
\end{tabular}
\end{table}

The characterization of the linearity of the HFI detectors has a
direct impact on the calibration of the instrument: strong
non-linearity takes place during the galaxy crossing for high
frequency bolometers and during planet crossings. An accurate absolute
calibration is also necessary for the CMB dipole. Finally, the energy
scale of large glitches can be corrected. The static response has been
characterized during ground calibration showing a small deviation from
linearity around a tenth of one percent for fainter sources (few
hundreds of attowatts) and around a few percent for brighter sources
like planets \citep{pajot2010}. The static response measured during
the CPV phase agrees with the ground estimate to better than
1\%. Nevertheless, the use of the static non-linearity determination
does not represent the true bolometric non-linear behaviour when
scanning through bright point sources like planets. The linearization
of the response done by multiplying the signal times the gain
(depending on the amplitude of the signal) and convolving it with the
temporal transfer function normalized to 1 at the lowest frequency, is
valid in the case of small signals.  However this is not the case for
bright point sources for which the estimate of non-linearity using the
static response may be incorrect by up to 40\% in the extreme case of
Jupiter. For these sources we use a model to correct the static
results. The use of fainter planets like Mars to characterize the
beams minimizes this effect.

Table~\ref{PlanckIPF1.5:tab:nonlinearity} gives the deviation from linearity for
various sources at the center of the beam for the bolometers at each
frequency.

\subsection{Electrical crosstalk on HFI detectors}

The electrical coupling of the signal of one bolometer into the
readout chain of another, or \emph{electrical crosstalk}, was measured
to be less than $-60$\,dB for all pairs of channels during ground-based
tests \citep{pajot2010}.  We performed two tests in flight to verify
this result, described below. 

\subsubsection{CPV crosstalk measurements}

During the CPV phase, we switched off each readout channel one at a
time for ten minutes, and observed the impact on all other channels.
For each bolometer we collected about 660\,minutes of data.

\begin{figure}
\includegraphics[width=\columnwidth]{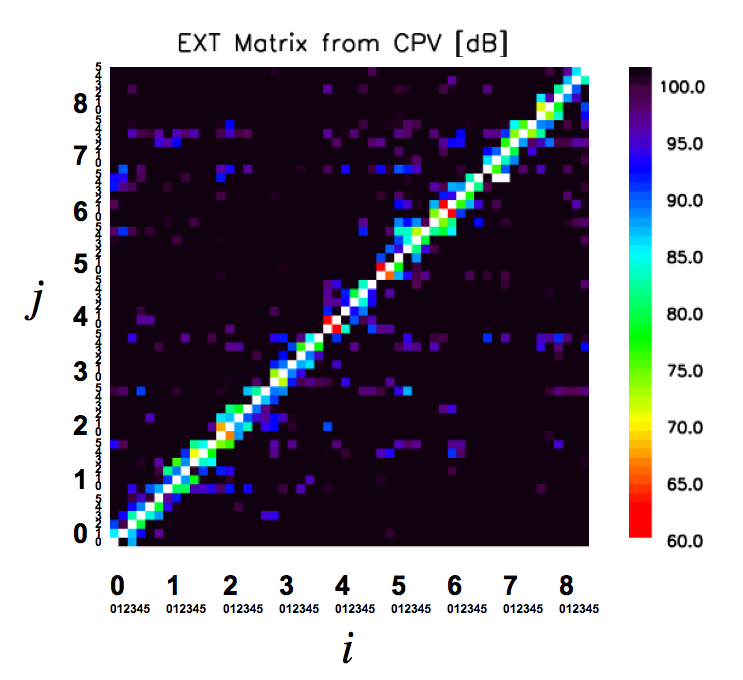}
\includegraphics[width=\columnwidth]{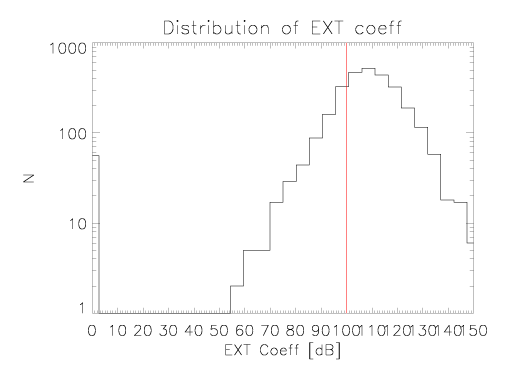}
\caption{Top: Electrical crosstalk matrix {$C_{ij}$} 54x54 for all
bolometers, coefficients are in \hbox{dB}.  Bottom: Distribution of
electrical crosstalk coefficients in dB.}
\label{PlanckIPF1.5:fig:CPV_EXT_matrix}
\end{figure}

The crosstalk coefficient between channels $i$ and $j$ is expressed as:
\begin{equation}
C_{ij} = \Delta{\tilde{V}_j} / \Delta{\tilde{V}_i},
\end{equation}
where {$\tilde{V}_i$} and {$\tilde{V}_j$} are the channel $i$ and $j$
voltages, corrected for thermal drift.  The crosstalk matrix and a
histogram of crosstalk levels are shown in
Fig.~\ref{PlanckIPF1.5:fig:CPV_EXT_matrix}.  The crosstalk is mostly
confined to nearest neighbours in the belt, channels whose wiring is
physically close.  The measured crosstalk level is in good agreement
with ground measurements, typically {$< -70$}\,dB, and thus meets the
requirement.  A few of the polarization sensitive bolometer pairs show
a crosstalk around $-60$\,dB.

In the next section we see that crosstalk measurement from glitches
shows a much lower level of crosstalk. These results suggest that the
CPV test has measured electrical crosstalk in current which is
unrelated to the scientific signal.

\subsubsection{Measurements using glitches}

We used high energy glitches in one channel to study the impact on the
signal of surrounding channels. Thousands of glitch events are
collected for one channel, and the signals of all other channels for
the same time period are stacked.  The crosstalk in volts for
individual glitches is defined as:
\begin{equation}
  c_{ij}^V = \Delta{V_j} / \Delta{V_i}
\end{equation}
where {$V_i$} is the glitch amplitude in volts in the channel hit by a
cosmic ray, and {$V_j$} the response amplitude of another channel $j$.
Then, for a pair of channels $i$ and $j$, the global voltage crosstalk
coefficient is
\begin{equation}
   C_{ij}^V = \hbox{median}(c_{ij}^V)
\end{equation}

For SWB channels, in contrast with the CPV previous results, no
evidence of crosstalk is seen, with an upper limit of $-100$\,\hbox{dB}.
There are outliers in galactic channels because of incorrect glitch 
flagging.  A second analysis using planet crossing data instead of 
glitches gave the same results.

Concerning the coupling between PSB pairs, we see crosstalk around
$-60$\,dB, in agreement with the CPV tests; however, this is likely an
upper limit because it includes the effects of coincident cosmic ray
glitches which produce a similar effect but are not crosstalk.

\section{Beams and time response}
\label{PlanckIPF1.5:Sect4}

\subsection{Measurement of Time Response}

\subsubsection{Introduction}

The \emph{time response} of HFI describes the shift, in amplitude and
phase, between the optical signal incident to each detector and the
output of the readout electronics.  The response can be approximated
by a linear complex transfer function in the frequency domain.  The
signal band of HFI extends from the spin frequency of the spacecraft
($f_{\rm spin} \simeq 16.7$\,mHz) to a cutoff defined by the angular
size of the beam (14--70\,Hz; see Table~4 from
\citet{lamarre2010}). For the channels at 100, 143, 217, and 353\,GHz,
the dipole calibration normalizes the time response at the spin
frequency. To properly measure the sky signal at small scales, the
time response must be characterized to high precision across the
entire signal band, spanning four decades from 16.7\,mHz to $\sim
100$\,Hz.

The time response of bolometers typically is nearly flat over a signal
band from zero frequency to a frequency defined by the bolometer's
thermal time constant, and then drops sharply at higher frequencies.
For the HFI bolometers, the thermal frequency is 20--50\,Hz
\citep{lamarre2010,Holmes2008}, as noted in \citet{lamarre2010} and
\citet{pajot2010}, however, the time response of HFI is not flat at
very low frequencies, but exhibits a low frequency excess response
(LFER).

We define the {\em optical beam} as the instantaneous directional
response to a point source.  Any sky signal is convolved with this
function, which is completely determined by the optical systems of HFI
and \Planck.

Since \Planck\ is rotating at a nearly constant rate and around the
same direction, the data are the convolution of the signal with both
the beam and the time response of \hbox{HFI}.  We separate the
two effects and deconvolve the time response from the time ordered
data. This deconvolution results in a flat signal response, but
necessarily amplifies any components of the system noise that are not
rolled off by the bolometric response.  This amplified noise is
supressed by a low-pass filter \citep{planck2011-1.7}.

\subsubsection{TF10 model}
\label{PlanckIPF1.5:SSS:TF10_model}

The main ingredients of the time response are: (i)~heat propagation
within the bolometer; (ii)~signal modulation at a frequency of $f_{\rm
mod}= 90.188$\,Hz performed by reversing the bolometer bias current;
(iii)~the effect of parasitic capacitance along the high impedance
wiring between the bolometer and the first electronics stage (JFETs);
(iv)~band-pass filtering, to reject the low frequency and high
frequency white noise in the electronics; (v)~signal averaging and
sampling; and (vi)~demodulation.

Because of the complexity of this sequence, a phenomenological
approach was chosen to build the time response model. The time
response is written as the product of three factors:
\begin{equation}
H_{\rm 10} (f) = H_{\rm bolo} \times H_{\rm res} \times H_{\rm filter}
\end{equation}

Schematically, the first factor takes into account step
(i), the second factor describes a resonance effect
that results from the combination of steps (ii) and (iii) , while the
purpose of $H_{\rm filter}$ is to account for step (iv).

\begin{figure}
\begin{center}
\includegraphics[width=1\columnwidth]{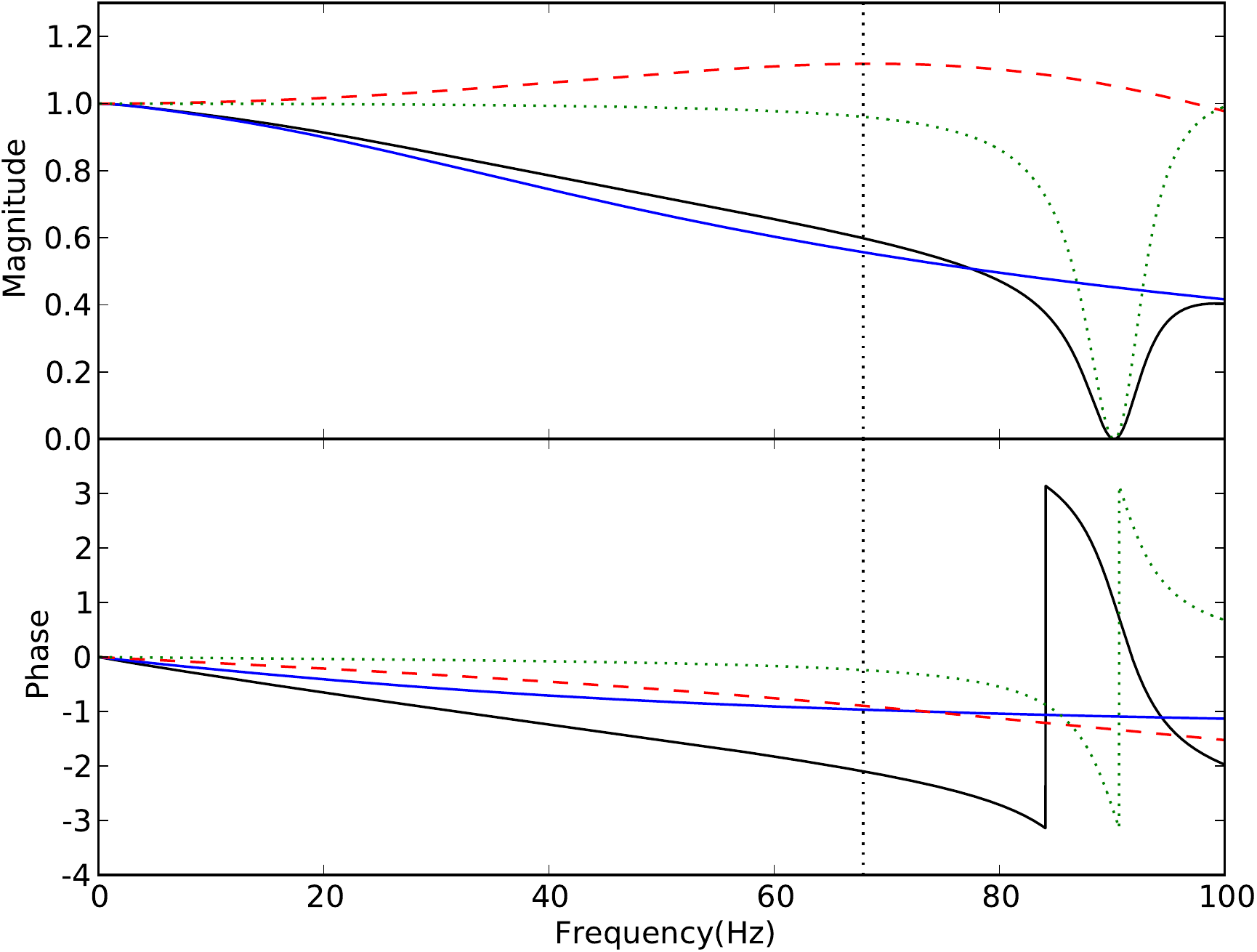}
\caption{The amplitude and phase (in radians) of the three components
   of the TF10 model of the time response.  The solid blue line is
   $H_{\rm bolo} (f)$, the dotted green line shows $H_{\rm filter}
   (f)$ and the dashed red line shows $H_{\rm res}$.  The solid black
   line is $H_{\rm 10} (f)$, the product of the three components.  The
   vertical dotted black line shows the signal frequency where the
   beam of a 217\,GHz channel cuts the signal power by half.}
\label{PlanckIPF1.5:fig:TF10_example}
\end{center}

\end{figure}

Detailed analysis and measurements of heat propagation within the
bolometer have shown that $H_{\rm bolo}$
is given by the algebraic sum of three single pole low pass filters.
Explicitly:
\begin{equation}
H_{\rm bolo} = \sum_{i = 1,3} \frac{a_i}{1 + j 2 \pi f \tau_i}
\end{equation}
with 6 parameters ($a_1$,\,$a_2$,\,$a_3$,\,$\tau_1$,\,$\tau_2$,\,$\tau_3$).
\begin{equation}
H_{\rm res} = {{1 + p_7 (2 \pi f)^2}\over{1 -p_8 (2 \pi f)^2 + j p_9 
(2 \pi f)}}
\end{equation}
with 3 free parameters ($p_7$,\,$p_8$,\,$p_9$),
\begin{equation}
H_{\rm filter} = { {1 - (f / F_{\rm mod})^2} \over
{1 - p_{10}(2 \pi f)^2 + j (f/F_{\rm filter})^2} }
\end{equation}
with one free parameter ($p_{10}$).  A total of 10 free parameters
describe this model, as indicated by its name.  See
Fig. \ref{PlanckIPF1.5:fig:TF10_example} for an illustration of the three
components of the time response model TF10 for a typical 217 GHz
channel.

The parameter $F_{\rm filter}$ characterizes the rejection filter width
and is kept fixed to 6\,Hz in the fitting process.
Besides the fact that this phenomenological model
is physically motivated, this parameterization:
\begin{itemize}
\item ensures causality
\item satisfies $H(-f) = H^\ast{}(f)$
\item goes to 1 when $f$ goes to zero (because we define $a_1+a_2+a_3
  = 1$), while it goes to 0 when $f$ goes to infinity
\item includes enough parameters to provide the necessary
  flexibility to fit the time response data of all 52 bolometers.
\end{itemize}

\subsubsection{Fitting the TF10 Model to Ground Data}
\label{PlanckIPF1.5:SSS:TF10_GroundData}

To obtain the 10 $\times$ 52 parameter values, we used three sets of
pre-launch measurements. (i) The bolometer response was measured at 10
different frequencies by illuminating all 52 bolometers with a chopped
light source. (ii) Other measurements were done using carbon fibers as
light sources; the latter were alternately turned on and off at a
variable frequency. (iii) The bolometer bias currents were
periodically stepped up and lowered by a small amount.  By adding a
square wave to the DC current, temperature steps are induced,
simulating turning on and off a light source (the analysis of these
data requires bolometer modelling).

None of these measurements was absolutely normalized; all
compared the relative response to inputs of
various frequencies. While
measurement (i) only provided the amplitude of the time response,
measurements (ii) and (iii) provided both the amplitude and the
phase. Note that for the phase analysis, because of the lack of
precise knowledge of the time origin $t_0$ of the light/current
pulses, a fourth factor $\exp(j 2 \pi \Delta t_0 f)$ is
introduced in the expression of $H_{10} (f)$, where the additional 
parameter $\Delta t_0$ represents
the uncertainty in time.

Among the three sets of measurements, the carbon fiber set covered the
largest frequency range. Thus it was the best for building the transfer
function model described above and for investigating its main
features.  However, it involves uncertainties
that could be resolved only with a very detailed simulation of the set-up,
therefore it was not used to calculate the final set of parameter
values.  The final values were calculated from data sets (i)
and (iii), whose frequency ranges are complementary, 2--140\,Hz and 
0.0167--10\,Hz, respectively.
Since no absolute normalization was available, the two datasets were matched
in the overlap frequency range.

The fitting of the analytic expression given above to the merged data
was done in the range between 16.7\,mHz and 120\,Hz. The 52 fits have
a $\chi^2$/DoF distribution whose mean value is 1.13, indicating that
the model is adequate to describe the data.  The numerical values of
$p_{10}$ displayed a small spread, $\sigma_{\rm mean} <
6\times10^{-4}$.  This parameter was set at its mean value to
calculate the 52 covariance matrices of the nine remaining parameters,
which were useful in propagating the statistical errors.

As described below, the time response thus obtained was further tuned
and checked using in-flight observations, in particular signals
produced by planets and by cosmic rays (glitches).

An alternative model has also been defined, based on the analytical
expression of the steps (ii) to (vi) of
section~\ref{PlanckIPF1.5:SSS:TF10_model}. Based on a closer analysis
of the electronics stages, this model is more physically motivated
than the 10-parameter model. It requires only 8 parameters and
provides better results near the modulation frequency.  Nevertheless
the model has not been used in the current data release.  It is only
used as a benchmark, to check possible systematic effects in the
current release.  Most of the effects of the difference between the
models disappear when the data are low-pass filtered.

\subsubsection{Fitting TF10 to Flight Data}
\label{PlanckIPF1.5:SSS:TF10_FlightData}

The planets Mars, Jupiter, and Saturn are bright, compact sources that
are suitable for measuring the beam and provide a near-delta-function
stimulus to the system that can be used to constrain the time
response.  During the first sky survey, \Planck\ observed Mars twice
and Jupiter and Saturn once \citep{planck2011-1.7}.  During a planet
observation, the spacecraft scans in its usual observing mode
\citep{planck2011-1.1}, shifting the spin axis in 2\arcm\ steps along
a cycloidal path on the sky.  Since planets are close to the ecliptic
plane, the coverage in the cross-scan direction is not as fine as in
the scan direction.  In the case of Jupiter and Saturn, each channel
observes the planet once per rotation for a period of approximately 6
hours (9 periods of stationary pointing, or "rings"). Because Mars has
a large proper motion, the first observation lasted 12 hours (or 18
rings).

We use the forward-sense time domain approach \citep{Huffenberger2010}
to simultaneously fit for Gaussian beam parameters and TF10 time
response parameters. A custom processing pipeline avoids filtering the
data. We extract the raw bolometer signal and demodulate it using the
parity bit. We use the flags created by the 
time ordered information (TOI) processing pipeline to
exclude data samples contaminated by cosmic rays, and we additionally
flag all data samples where the nonlinear gain correction is more than
0.1\%. We use Horizons\footnote{\url{ssd.jpl.nasa.gov/?horizons}}
emphemerides to compute the pointing of each horn relative to the
planet center.

The time domain signal from the planet is modeled as an elliptical
Gaussian convolved with the TF10 time response as follows:
\begin{equation}
d(t) = H_{10} \star A (t) G\left[\vec{x}(t); \vec{x}_0,\epsilon, 
\theta_{FWHM}, \psi \right]
\end{equation}
where the Gaussian optical beam model $G$ is parameterized as in
Eqs.~9--11 of \cite{Huffenberger2010}, except the planet amplitude is
parameterized with a disk temperature rather than a single amplitude:
\begin{equation}
A (t) = T_{disk} \frac{\Omega_{p} (t)}{\Omega_{b}},
\end{equation}
where $T_{disk}$ is the whole-disk temperature of the planet,
$\Omega_p$ is the solid angle of the planet, which can vary significantly
during the observation, and $\Omega_b$ is the solid angle of the 
beam. $\Omega_p$
is computed using Horizons, which is programmed with \Planck's orbit.

The free parameters of the fit are the six parameters of the time
response corresponding to $H_{\rm bolo}$, the two components of the
centroid of the beam $\vec{x}_0$, the mean FWHM $\theta_{\rm FWHM}$, the
ellipticity $\epsilon$, the ellipse orientation angle $\psi$, and the
planetary disk temperature $T_{\rm disk}$. The four parameters describing
the electronics are somewhat degenerate with the bolometer part of the
time response, and we fix them at the ground-based values.

Because of the large nonlinear response and highly non-Gaussian beams
at 545 and 857\,GHz, we do not perform fits to the planet data at
these frequencies.  Instead we rely on pre-launch fits for the time
response.

By taking the Fourier transform of the time response function derived
on planets, one obtains the system response to a Dirac impulse. This
response can be compared to the glitches generated by cosmic rays that
deposit energy in the sensor grids.

The glitches detected by HFI are sampled with time steps $1/(2 F_{\rm
mod})$. However, the glitches can be superresolved in time by
normalizing, phasing, and stacking single glitch events
\citep{Crill2003}.  This gives glitch templates for each channel
(Alexandre Sauv\'e, private communication) that are effectively
sampled at a much higher frequency.

\begin{figure}
\includegraphics[width=1\columnwidth]{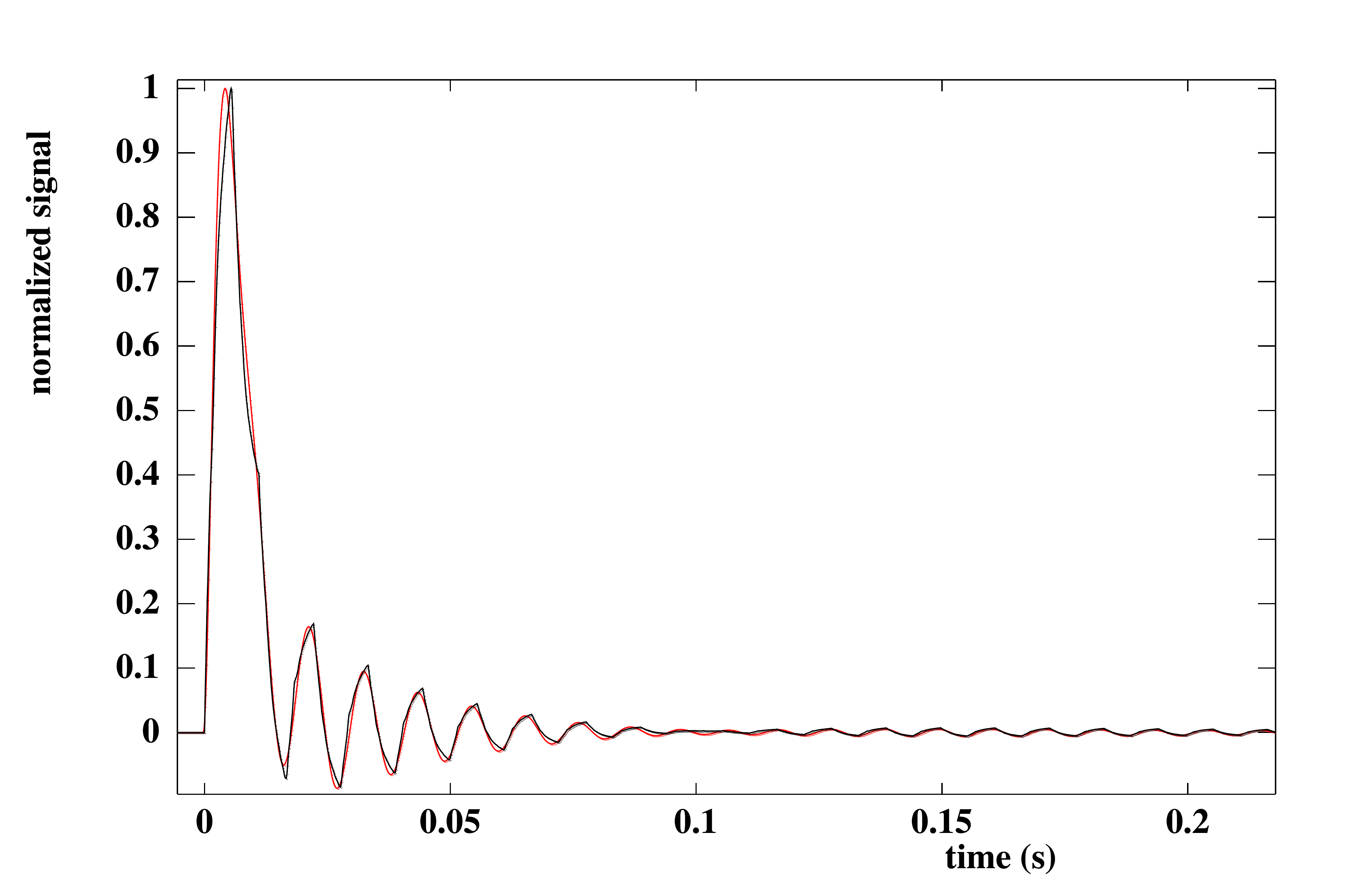}
\caption{Comparison of the impulse response of channel 143-2a (red
   curve) and a template made from stacking glitch events (black
   curve). Noise begins to dominate further in the timeline.  The
   ringing observed at the modulation frequency is generated by the
   electronics rejection filter.}
\label{PlanckIPF1.5:fig:glitch_vs_tf}
\end{figure}

Figure \ref{PlanckIPF1.5:fig:glitch_vs_tf} shows the comparison between a
superresolved glitch template and the corresponding calculated
response. There is good agreement in general, but there are
discrepancies at high frequency ($f > 100$\,Hz).  The physical model
for the electronics transfer functions briefly described at the end of
section~\ref{PlanckIPF1.5:SSS:TF10_GroundData} suppresses this discrepancy at
high frequency.

Each planet observation suffers unique systematic effects, so a
comparison of the time response recovered on each gives a good assessment
of the effects of these different systematics.  Mars has a large
proper motion, giving excellent sampling in the cross-scan
direction; however, there is a known diurnal variability in its
brightness temperature \citep{Swinyard2010}.  Jupiter has a large
angular diameter (48\arcs) relative to the HFI beam size, and
Saturn and Jupiter are so bright that the HFI detectors are
driven significantly nonlinear (see Table \ref{PlanckIPF1.5:tab:nonlinearity}).
Nonetheless, we find that the time response is consistent to 0.1-0.5\%
when recovered from each of the planets individually as well as from
all planets simultaneously.

A further cross-check is done by stacking planet scans to build
a superresolved planet timeline.  Time response parameters are fit to
the superresolved planet using the assumption of a near-Gaussian beam 
profile, and are consistent with the first approach.

The in-flight time response differs from the ground-based
time-response by at worst 1.5\% between 1\,Hz and 40\,Hz.  We do not
include this difference in the final error budget, because it is
likely that the time response has changed due to differences in
background conditions.

\subsubsection{Low Frequency Excess Response}

The HFI bolometers show low frequency excess response (LFER),
~\citep{lamarre2010}.  Though the planets are bright, the short impulse
they provide is close to a delta function and the energy is spread
evenly across nearly all harmonic components. In combination with
low frequency noise, the measurements are not sensitive to frequencies below
$\sim0.5$\,Hz; so with planet observations alone we cannot constrain
an excess response at very low frequency.  We maintain the ground
measurements as our best estimation of the \hbox{LFER}.  In the
ground-based measurements the bias step vs.~carbon fiber differs by at
most 1.5\% at low frequency, so we assign a systematic error of 1.5\%
for frequencies below 0.5\,Hz.

For future data releases, we will use the difference of sky signal
between surveys to constrain the \hbox{LFER}.

\subsubsection{Summary of Errors in the Time Response}

As noted above, the data represent the combined effect of the time
response and the optical beam.  The time response, however, is not
degenerate with a Gaussian parameterization of the beam; the true
beams deviate from a Gaussian shape at the several percent level near
the main lobe, while time response effects tend to give the beam an
extended tail following the planet in the scan direction.  The
Gaussian assumption could slightly bias the recovered time response;
however, any residual bias is captured in the measurement of the
post-deconvolution scanning beam \citep{planck2011-1.7}.

Because of the high signal-to-noise ratio of the planet data,
statistical errors in the fit are small, so we assess the systematic
errors in the resulting time response by checking the consistency of
various methods of recovering the time response. We fit to different
combinations of planet data: Mars, Jupiter, and Saturn data separately
and all of the data simultaneously to check for systematics resulting
from various planets. Additionally we compare the planet-fitted time
response with ground-based data and with the impulse response from
cosmic ray glitches.

Our final error budget is as follows:
\begin{itemize}
\item Low frequency ($f<0.5$\,Hz): the errors are dominated by the
   possibility of a low frequency excess response below 0.5\,Hz at a
   level of 1.5\%.
\item Middle frequency ($0.5\,{\rm Hz}<f<50$\,Hz): We set an error bar
   between 0.1\% and 0.5\% depending on the channel. This error bar is
   set by the consistency in results from different sets of planet
   data.
\item High frequency: ($f>50$\,Hz) Our empirical model of the
   electronics in the TF10 model does not describe the system very well
   at these frequencies, as shown by some disagreement between the
   glitches and the TF10 impulse response. However, for this data
   release, the low-pass filter applied to the data and the beam cutoff
   reduce the importance of this frequency band.
\end{itemize}

The \Planck\ scan strategy is such that the same region of the sky
is observed scanning in nearly opposite directions six months apart.  An
error in the time response is highlighted in the difference of maps
obtained from the first six months and the second six months of the
survey.  This difference map shows some level of contamination, in
particular near the Galactic plane, where the signal is higher.  The
same level of contamination is observed in simulations in which the
data are generated with a transfer function, and analysed with a
different one, in order to mimic the uncertainties described
above. With this technique, we validate the error budget.

\subsection{Optical Beams}

The {\em optical beam} is defined (Sect.~4.1.1) as the instantaneous
directional response to a point source.  For HFI, the optical beams
for each channel are determined by the telescope, the horn antennas in
the focal plane and, for the polarized channels, by the orientations
of their respective polarization sensitive bolometers (PSBs)
\citep{Maffei2010}.  Model calculations of the beams are essential,
since it was possible to measure only a limited number of beams in the
telescope far-field before launch. The 545 and 857\,GHz channels,
which employ multimoded corrugated horns and waveguides, were not
included in this campaign.  (The optical beam is related to, but is
not the same as the \emph{scanning beam} defined in
\cite{planck2011-1.7} and used for data analysis purposes.)
\cite{tauber2010b} reported the best pre-launch expectations for the
optical beams, obtained using physical optics calculations with
CORRUG\footnote{SMT Consultancies
Ltd. \url{www.smtconsultancies.co.uk}} and GRASP\footnote{TICRA,
\url{www.ticra.com}}.  Table~\ref{PlanckIPF1.5:tab:optical_beam_parameters}
compares the calculated and measured (Sect.~\ref{PlanckIPF1.5:SSS:TF10_model})
beams for the single-moded channels (up to 353\,GHz).

\begin{table*}
\caption{Comparison of pre-launch calculations and measured parameters
for the HFI optical beams (band averages). Standard deviations
$\sigma$ are computed as the dispersion between the Saturn, Jupiter,
and Mars data for each given channel.}
\label{PlanckIPF1.5:tab:optical_beam_parameters}
\centering
\begin{tabular}{l c c c c c c}
\hline\hline
Band & Expected & Mars & Mars & Expected & Mars & Mars \\
      & FWHM & FWHM & $\sigma_{\rm FWHM}$ & ellipticity & ellipticity 
& $\sigma_{\rm ellip}$ \\
& [\arcm] & [\arcm] & [\arcm] & & & \\
[0.5ex]
\hline
100P & 9.58 & 9.37 & 0.06 & 1.17 & 1.18 & 0.006\\
143P & 6.93 & 6.97 & 0.10 & 1.06 & 1.02 & 0.004\\
143 & 7.11 & 7.24 & 0.10 & 1.03 & 1.04 & 0.005\\
217P & 4.63 & 4.70 & 0.06 & 1.12 & 1.13 & 0.006\\
217 & 4.62 & 4.63 & 0.06 & 1.10 & 1.15 & 0.010\\
353P & 4.52 & 4.41 & 0.06 & 1.08 & 1.07 & 0.009\\
353 & 4.59 & 4.48 & 0.04 & 1.23 & 1.14 & 0.007\\
545  & 4.09 & 3.80 & --   & 1.03 & 1.25 & -- \\
857  & 3.93 & 3.67 & --   & 1.04 & 1.03 & -- \\
[1ex]
\hline
\end{tabular}
\end{table*}

\begin{figure}
\rotatebox{180}{%
\includegraphics[width=.95\columnwidth]{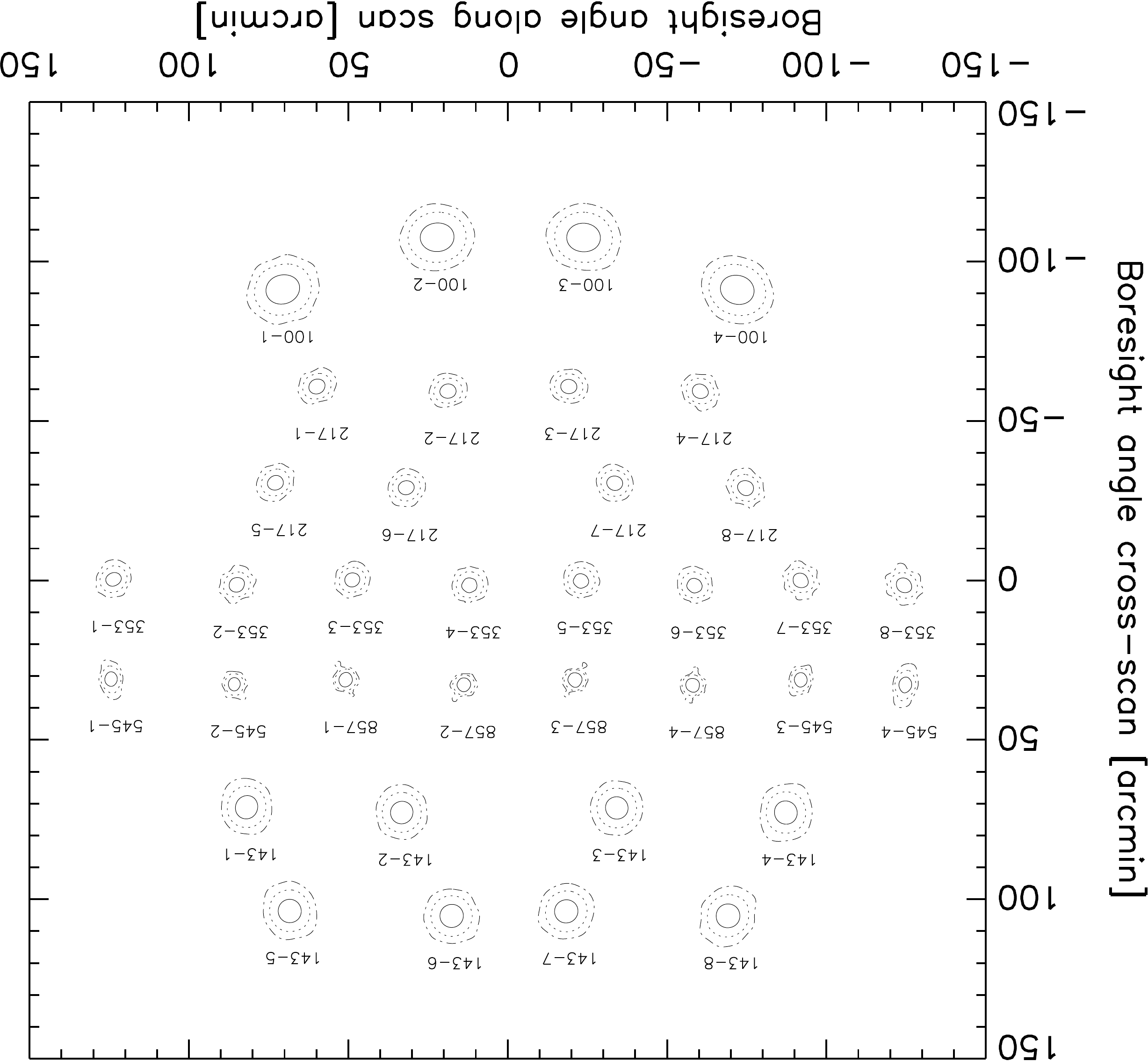}}
\caption{The distribution of the HFI beams on the sky relative to the
   telescope boresight as viewed from infinity.  Contours of the
   Gauss-Hermite decomposition of the Mars data at 1\%, 10\%, and 50\%
   power levels from the peak. For the photometers containing a pair
   of PSBs, the average beam of the two PSBs is shown.}
\label{PlanckIPF1.5:fig:focal_plane_layout}
\end{figure}

For these channels the pre-launch calculations of FWHM and ellipticity
and measured Mars values agree to within a few percent.  These
differences are contained within $2.7\sigma$ of the data errors. The
main source of discrepancy could be a slight misalignment of the
pre-launch telescope model with respect to the actual in-flight
telescope geometry, which is currently being investigated
\citep{Jensen2010}.

\begin{table}[!t]
\caption{Geometric mean of the asymmetric gaussian FWHM.}
\label{PlanckIPF1.5:tab:SummaryBeams}
\centering
\begin{tabular}{l c c c }
\hline\hline
Bolometer & beam & spectral Band & spectral Band \\
           & \ \ FWHM\ \ & cut-on & cut-off \\
Name      & [\arcm] & GHz & GHz \\
\hline
  100-1a & 9.46 & 84.9 & 113.87 \\
  100-1b & 9.60 & 87.0 & 115.27 \\
  100-2a & 9.41 & 86.5 & 116.28 \\
  100-2b & 9.43 & 84.4 & 115.42 \\
  100-3a & 9.42 & 84.4 & 116.77 \\
  100-3b & 9.47 & 84.4 & 116.77 \\
  100-4a & 9.43 & 84.9 & 117.79 \\
  100-4b & 9.45 & 84.9 & 117.79 \\
  143-1a & 6.91 & 120.8 & 161.77 \\
  143-1b & 6.99 & 120.3 & 162.78 \\
  143-2a & 6.78 & 119.8 & 162.26 \\
  143-2b & 6.80 & 119.3 & 163.28 \\
  143-3a & 6.91 & 120.3 & 158.73 \\
  143-3b & 6.86 & 120.3 & 160.75 \\
  143-4a & 7.01 & 118.8 & 167.83 \\
  143-4b & 7.01 & 119.3 & 161.26 \\
  143-5 & 7.45 & 120.3 & 166.31 \\
  143-6 & 7.08 & 120.3 & 165.81 \\
  143-7 & 7.18 & 120.8 & 167.83 \\
  143-8 & 7.20 & 120.8 & 165.3 \\
  217-5a & 4.73 & 184.0 & 249.72 \\
  217-5b & 4.75 & 183.9 & 249.12 \\
  217-6a & 4.66 & 182.5 & 253.26 \\
  217-6b & 4.64 & 189.6 & 252.76 \\
  217-7a & 4.63 & 188.6 & 253.77 \\
  217-7b & 4.68 & 189.6 & 250.74 \\
  217-8a & 4.69 & 182.5 & 253.26 \\
  217-8b & 4.73 & 182.0 & 252.76 \\
  217-1 & 4.68 & 189.6 & 249.72 \\
  217-2 & 4.61 & 189.1 & 253.26 \\
  217-3 & 4.59 & 191.1 & 252.76 \\
  217-4 & 4.61 & 193.1 & 252.76 \\
  353-3a & 4.47 & 310.9 & 403.91 \\
  353-3b & 4.46 & 310.4 & 405.93 \\
  353-4a & 4.40 & 323.5 & 400.88 \\
  353-4b & 4.39 & 313.9 & 406.94 \\
  353-5a & 4.41 & 302.3 & 405.43 \\
  353-5b & 4.42 & 299.8 & 405.93 \\
  353-6a & 4.47 & 300.3 & 406.94 \\
  353-6b & 4.45 & 314.4 & 397.84 \\
  353-1 & 4.57 & 310.4 & 401.38 \\
  353-2 & 4.46 & 312.9 & 407.45 \\
  353-7 & 4.44 & 326.1 & 404.4 \\
  353-8 & 4.53 & 318.5 & 405.92 \\
  545-1 & 3.94 & 466.1 & 638.93 \\
  545-2 & 3.63 & 464.5 & 633.87 \\
  545-3 & 3.79 & 467.6 & 633.87 \\
  545-4 & 4.17 & 479.2 & 635.89 \\
  857-1 & 3.73 & 748.1 & 986.59 \\
  857-2 & 3.66 & 736.5 & 982.65 \\
  857-3 & 3.76 & 747.1 & 984.21 \\
  857-4 & 3.67 & 744.1 & 970.02 \\
\hline
\end{tabular}
\end{table}

Table~\ref{PlanckIPF1.5:tab:SummaryBeams} reports our best knowledge of the FWHM
of the optical beams for each channel.  We stress that this table does
not provide parameters of the scanning beam of the processed data,
which accounts for the additional effects of the instrument time
response and of the time domain filtering in the data processing
\citep{planck2011-1.7}.

\begin{figure}
\begin{center}
\includegraphics[width=1\columnwidth,keepaspectratio,trim=0 0 0 70,clip]%
{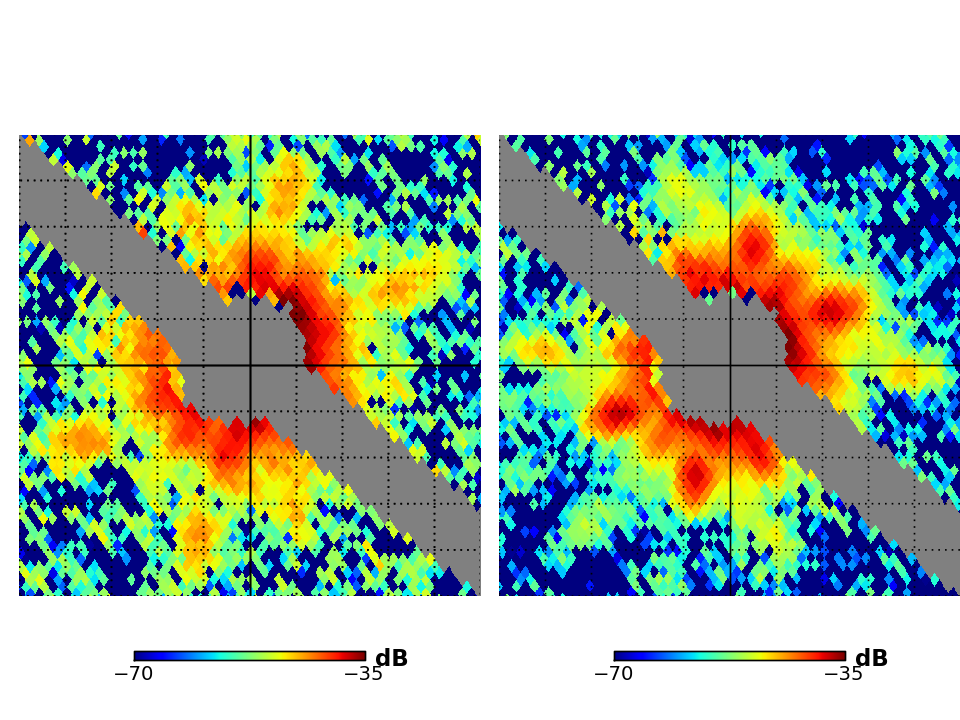}
\caption{The ``dimpling effect'' as seen at 545\,GHz (left panel) and
  857\,GHz (right panel).  The grid spacing is 10\arcm.  The color
  scale is in dB normalized to the peak signal of Jupiter.}
\label{PlanckIPF1.5:fig:dimpling}
\end{center}

\end{figure}

As reported in \citet{Maffei2010}, the 545\,GHz and 857\,GHz channels
are multimoded (more than one electromagnetic mode propagating through
the horn antennas) and their optical beams are markedly
non-Gaussian. The understanding of these channels through simulations
has progressed since \Planck\ was launched, especially in the
characterization of their modal content \citep{Murphy2010}.

In Table \ref{PlanckIPF1.5:tab:optical_beam_parameters} we compare pre-launch
calculations of the beams with the beams measured with Mars.
Differences in FWHM are less than 7\%.  We stress that this
discrepancy does not impact the scientific products of the \Planck\
mission since the scanning beams are the ones to be used for data
analysis purposes. From an instrumental point of view, the inflight
measurements must obviously be considered as the reference for the
performance of these channels.

The development of the HFI multimoded channels necessitated the novel
extension of previously existing modelling techniques for the analysis
of the corrugated horn antennas and waveguides, as well as for the
propagation of partially coherent fields (modes) through the telescope
onto the sky \citep{Murphy2001}. Extensive pre-launch measurement
campaigns were conducted for all the HFI horn antenna/filter
assemblies \citep{Ade2010}. The HFI multimoded channels are suitable
for the scientific goals of \Planck. Nevertheless, for future
instruments, more research can be envisaged in this field. The
characterization of the modal filtering in the horn-waveguide assembly
and the understanding of the coupling of the waveguide modes to the
detector need further theoretical and experimental study.

The similarity of the pre-launch expectations to our current knowledge
of the HFI focal plane (beams and their positions on the sky) tells us
that the overall structural integrity of the focal plane has been
preserved after launch. Furthermore, the optical beams as measured on
Mars are shown in Fig.~\ref{PlanckIPF1.5:fig:focal_plane_layout} and can be
compared with the equivalent representations of the focal plane layout
based on calculations in earlier papers
\citep{Maffei2010,tauber2010b}. A detailed account of the full focal
plane reconstruction can be found in \cite{planck2011-1.7}.

There is a ``dimpling'' of the reflector surfaces from the irregular
print-through of the honeycomb support structures on the reflector
surfaces themselves \citep{tauber2010b}.  GRASP calculations predict
that this will generate a series of rings of narrow bright grating
lobes around the main lobe. Since the small-scale details of the
dimpling structure of the \Planck\ reflectors are irregular, these
grating lobes tend to merge with the overall power scattered by the
reflector surfaces (Ruze scattering \citep{Ruze66}).
Fig.~\ref{PlanckIPF1.5:fig:dimpling} shows a HEALPIX
\citep{gorski2005} map of the first survey observation of Jupiter
minus the second survey observation of the same sky region to remove
the sky background.  We see the first ring of grating lobes as
expected in the map from all 545 and 857\,GHz channels, where the
signal-to-noise ratio on the planets is highest.  The inner
15\arcm\ of the beam is saturated and does not appear in the map.  At
857\,GHz, the discrete grating lobes appear at level below $-35$\,dB
with respect to the peak ($\sim 30$\,dB), and represent a negligible
fraction of the total beam throughput.  The shoulder of the beam,
extending radially to $\sim 15\arcm$, represents a larger contribution
to the throughput, ranging from 0.5\% to a few percent for the CMB and
sub-mm channels, respectively.

\section{Noise properties}
\label{PlanckIPF1.5:Sect5}

The \Planck\ HFI is the first example of space-based bolometers,
continuously cooled to 100\,mK for several years. Although the
detectors were thoroughly tested on the ground
\citep{lamarre2010,pajot2010}, it remained then to be seen how they would
behave in the L2 space environment. We describe here the noise
properties of the HFI in the first year of operation, focusing on the
differences between space and ground performance.

This section deals with the Gaussian part of the noise.  Section
(\ref{PlanckIPF1.5:Sect6}) describes the systematic effects that have
been analyzed in the data so far.

\begin{figure*}
\includegraphics[angle=180,width=1\textwidth]{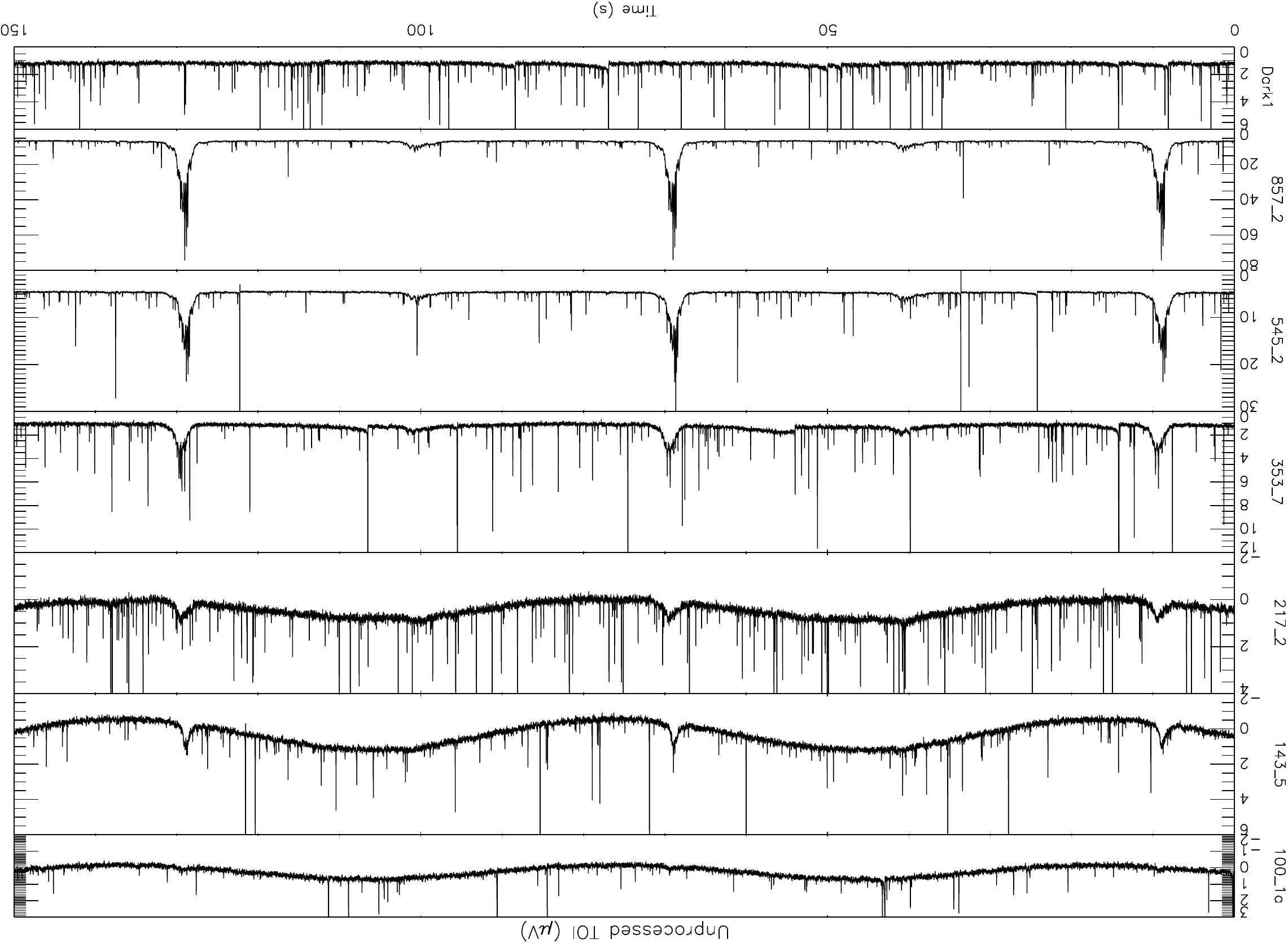}
\caption{Examples of raw (unprocessed) TOI for one bolometer at each
  of six HFI frequencies and one dark bolometer.  Slightly more than
  two scan circles are shown. The TOI is dominated by the CMB dipole,
  the Galactic dust emission, point sources, and glitches. The relative part 
  of glitches is over represented on these plots due to the thickness of the 
  lines that is larger than the real glitch duration.}
\label{PlanckIPF1.5:fig:TOIexample}
\end{figure*}

An example of raw time ordered information (TOI) is shown in
Fig.~\ref{PlanckIPF1.5:fig:TOIexample}. The TOI is dominated by the
signal from the CMB dipole, Galactic emission, point sources, and
glitches. Therefore, the noise properties cannot be directly deduced
from the \hbox{TOI}.  We first describe the general method used to
evaluate the noise, then we give general statements on the noise
properties.

\subsection{Noise estimation}

\begin{figure*}
\includegraphics[width=1\textwidth]{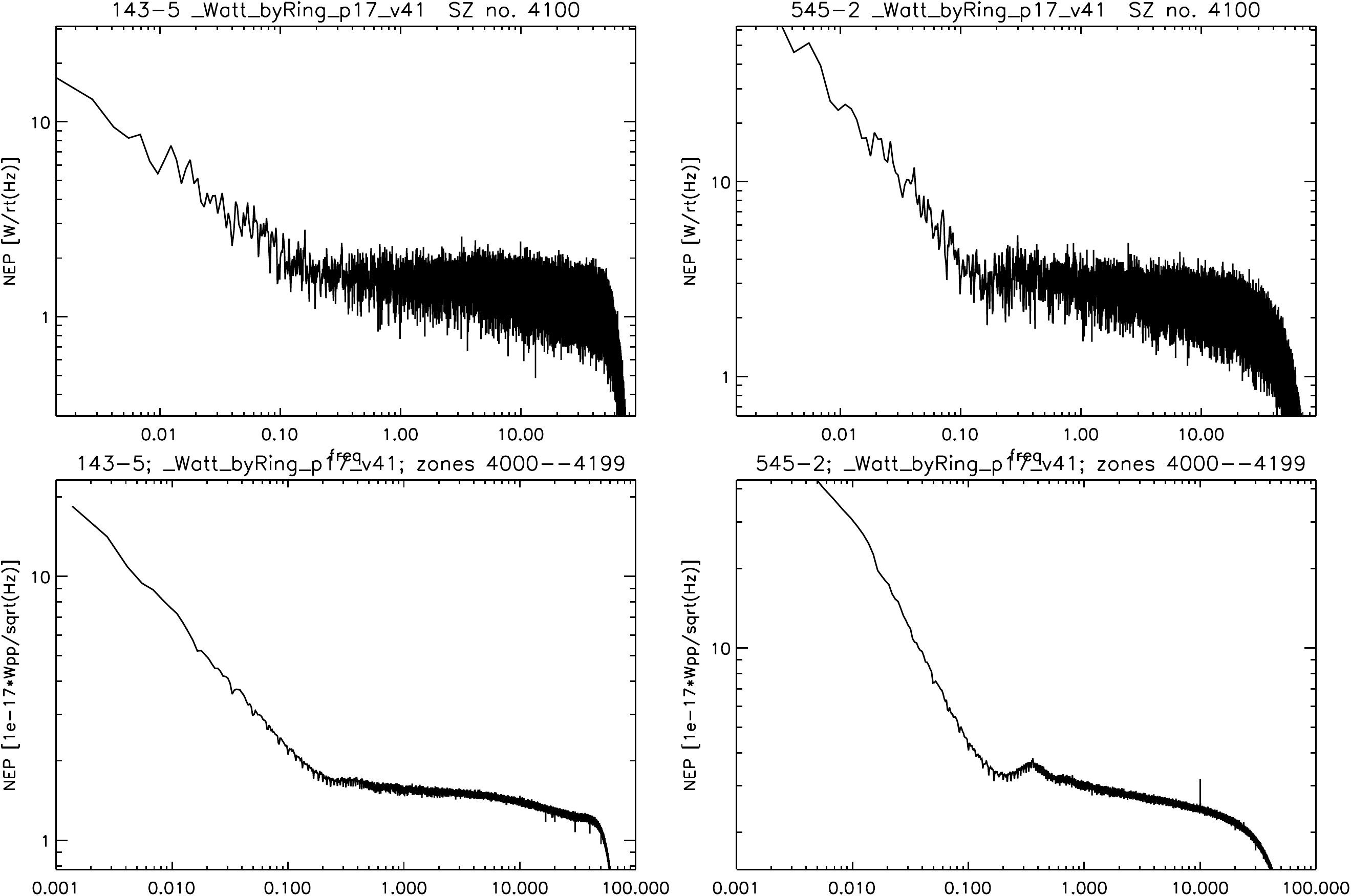}
\caption{Typical power spectrum amplitude of bolometers 143-5 and
  545-2. For the upper panels, this is the power spectrum density of
  valid samples, after an average ring (the sky signal) has been
  subtracted from the \hbox{TOI}.  Stacking of the result for 200
  rings is shown in the lower panel. Here, the instrument time
  response is not deconvolved from the data.}
\label{PlanckIPF1.5:fig:NEP}
\end{figure*}

The Level-2 \textit{detnoise} pipeline \citep{planck2011-1.7} is used
to determine a noise power spectrum, from which one extracts the noise
equivalent power (NEP) of the detectors (see \cite{planck2011-1.7} for
a full description).  The pipeline uses redundancies in the
observations to determine an estimate of the sky signal, which is then
subtracted from the full TOI to produce a pure noise timeline.  The
signal estimates are the integration of typically 40 circles of data
at a constant spin axis pointing.  The average signal, binned in spin
phase, provides an accurate estimate of the signal.  This signal as a
function of spin phase is then subtracted from the \hbox{TOI}.  The
residual is an estimate of the instantaneous noise.  Power spectra of
this residual timeline are then obtained for each pointing period (see
Fig.~\ref{PlanckIPF1.5:fig:NEP}) and fit for the white noise level,
i.e., the NEP, in the spectral region between 0.6 and 2.5\,Hz. The
lower limit of 0.6\,Hz is high enough that the low frequency excess
noise can be neglected and the upper limit small enough to keep the
time response near to its value at low frequencies (16\,mHz) at which
the instrument is calibrated.

The noise is stable at a level better than 10\% in the majority of
detectors.  Exceptions are: (1)~a few rings with unusual events that
contaminate the measurement, e.g., poorly corrected/flagged glitches
or passage over very strong sources such as the Galactic centre,
especially at 857\,GHz; (2)~a weak trend, smaller than 1\% in
amplitude, that correlates with the duration of the pointing period
(an expected bias due to the ring average signal removal);
(3)~bolometers affected by random telegraph signals (RTS) (see next
Section); and (4)~some uncorrelated jumps in the noise levels for
about ten bolometers at the 30\% level for isolated periods of a few
days.  The overall result is that a very clear \textit{baseline} value
can be identified and can be used to determine the NEP of each
bolometer.  This is then converted to NE$\Delta$T with the help of the
flux calibration.  The NE$\Delta$Ts thus measured are given in
Table~\ref{PlanckIPF1.5:tab:noiseJML}.  The quoted uncertainties are
derived from the rms of the NEPs in a band around the baseline.

\subsection{The noise components}

The detector noise is described by the combination of several
components:
\begin{itemize}
\item Photon and bolometer noise, which appear as white noise filtered
  by the time response of the bolometer, the readout electronics, and
  the TOI processing.
\item Electronics and Johnson noise, which produce noise that is
  nearly white across the frequency band, but with a sharp decrease at
  the high frequency end due to the on-board data handling and the TOI
  filtering.
\item The 4\,K lines (Sect~\ref{PlanckIPF1.5:Sect6}), appearing as
  residuals in the spectra.
\item The energy deposited by cosmic rays on the bolometers, which
  appears as "glitches", i.e., positive peaks in the signal, which are
  removed by the TOI processing (Sect.~\ref{PlanckIPF1.5:Sect6}, and
  \cite{planck2011-1.7}). Residuals from glitches appear in the noise
  spectrum as a bump between 0.1 and 1\,Hz.
\item Low frequency excess (LFE) noise, which is present below about
100\,mHz.
\end{itemize}

The last three sources of noise are detailed in
Section~\ref{PlanckIPF1.5:Sect6}.

There is additional noise (of the order of 0.5\% or less) due to the
on-board quantization of the data before transmission.  In general,
the noise level, as measured by the NEP, is between 10 and
20\,aW\,Hz$^{-1/2}$ for the 100 to 353\,GHz channels, and between 20
and 40\,aW\,Hz$^{-1/2}$ for the 545 and 857\,GHz channels. It is in
line with the ground-based expectations and the lower estimate of the
background load with a detector-to-detector variability of less than
$20\,\%$ (see Sect.~\ref{PlanckIPF1.5:Sect8}).

Due to the AC bias modulation scheme, the $1/f$ noise from the
electronics is aliased near the modulation frequency where it is
heavily filtered out. The benefit of this scheme is visible on the
noise power spectrum of the 10\,M$\Omega$ resistor which shows a flat
spectrum at the Johnson value down to 1\,mHz, a tribute to the
electronic chain stability.

At the present time, we assume that the LFE noise, not observed in
ground-based measurements, is mostly due to the 100\,mK bolometer
plate fluctuations.  While drifts in the 100\,mK stage that are
correlated between bolometers are removed, there are likely local
temperature fluctuations due to particle energy deposited close to
each detector.

\section{First assessment of systematic effects}
\label{PlanckIPF1.5:Sect6}

\subsection{4\,K lines}

\begin{figure}
\includegraphics[width=1\columnwidth]{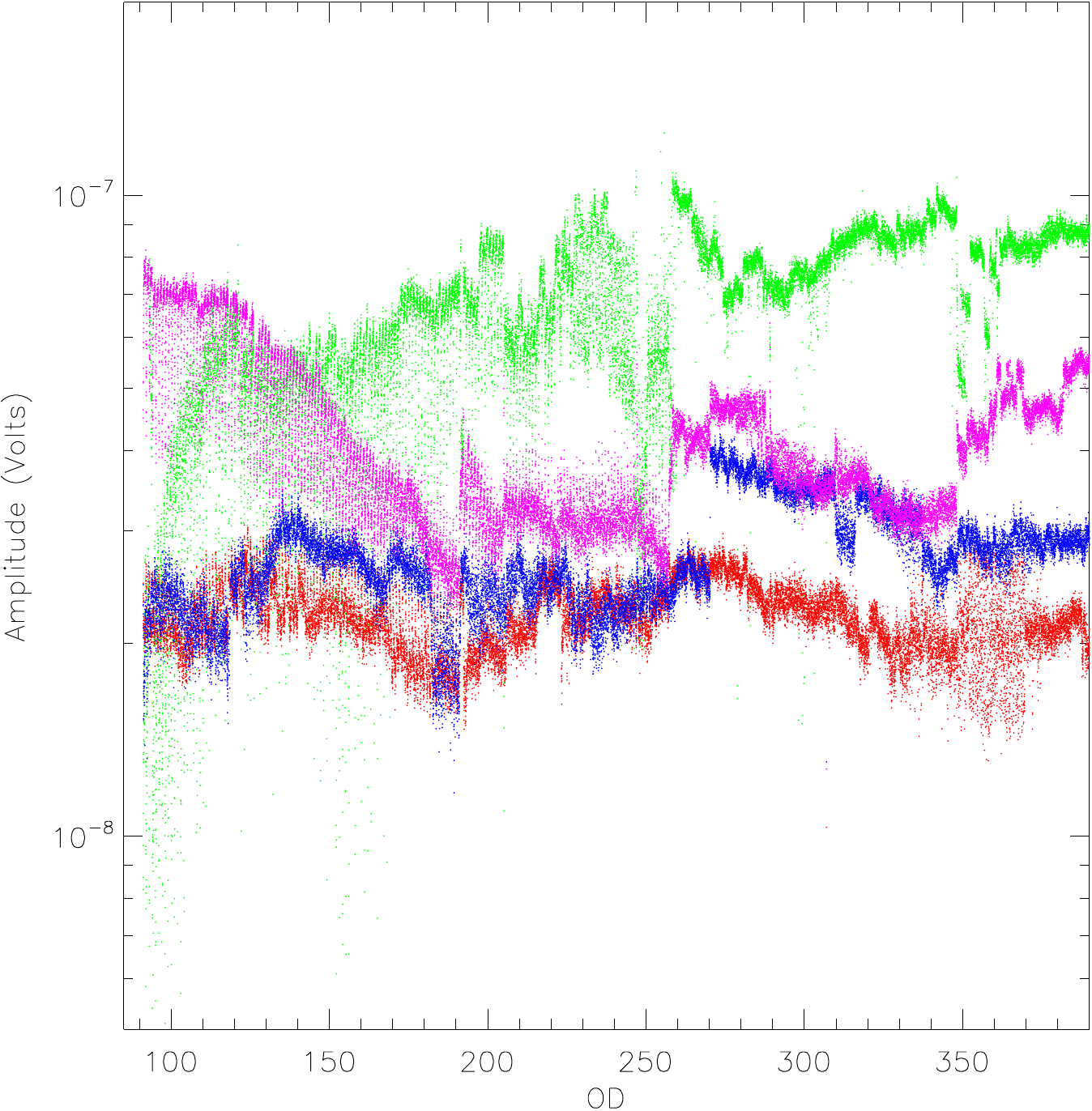}
\caption{Typical trend of the cosine and sine coefficient variation of
  the four main 4\,K lines measured in the TOI processing on the test
  resistance at 10, 30, 50, and 70\,Hz}
\label{PlanckIPF1.5:fig:A0_4K_all} 
\end{figure}

\begin{figure*}
\includegraphics[angle=0,width=0.5\textwidth]{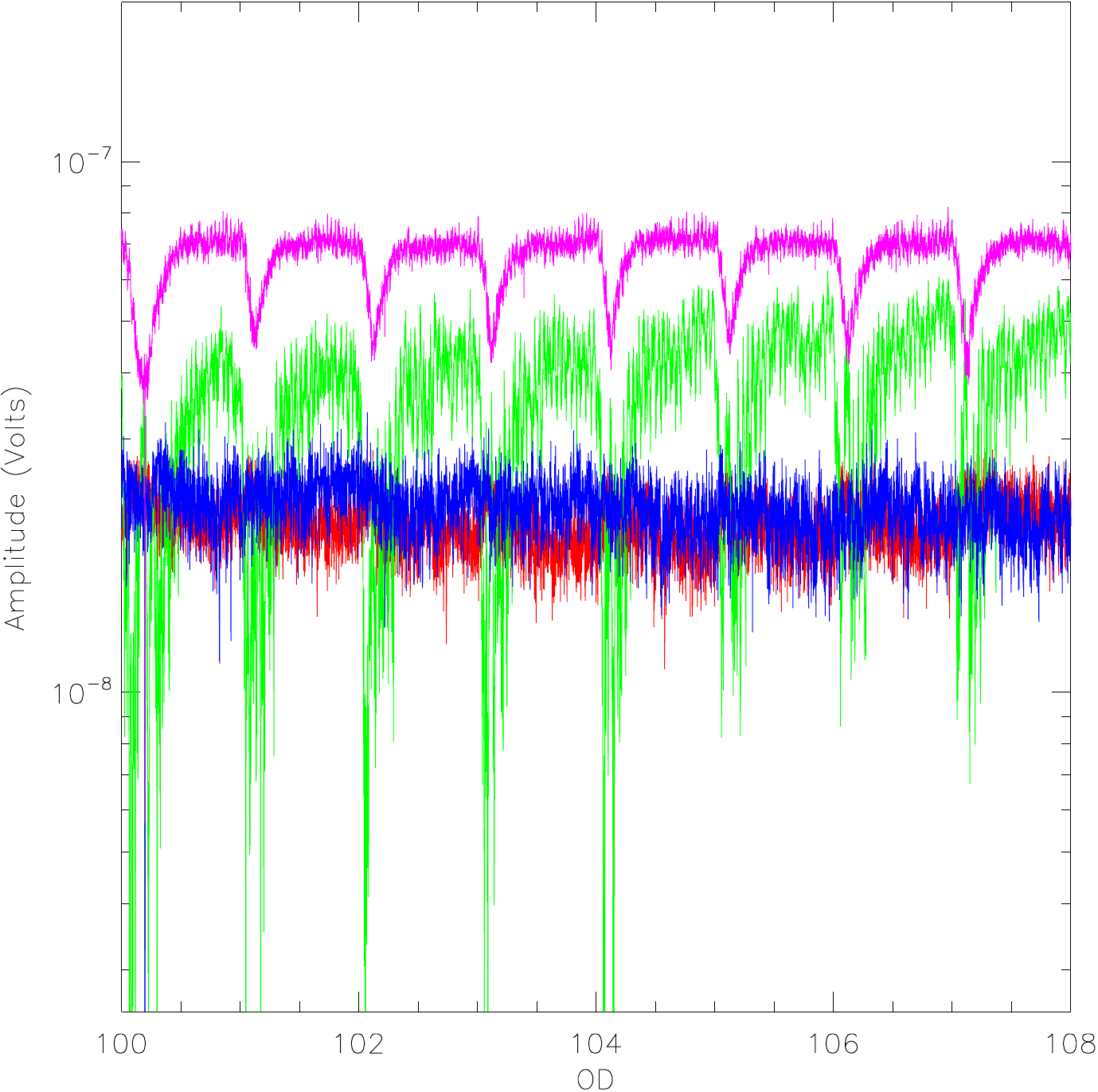}
\includegraphics[angle=0,width=0.5\textwidth]{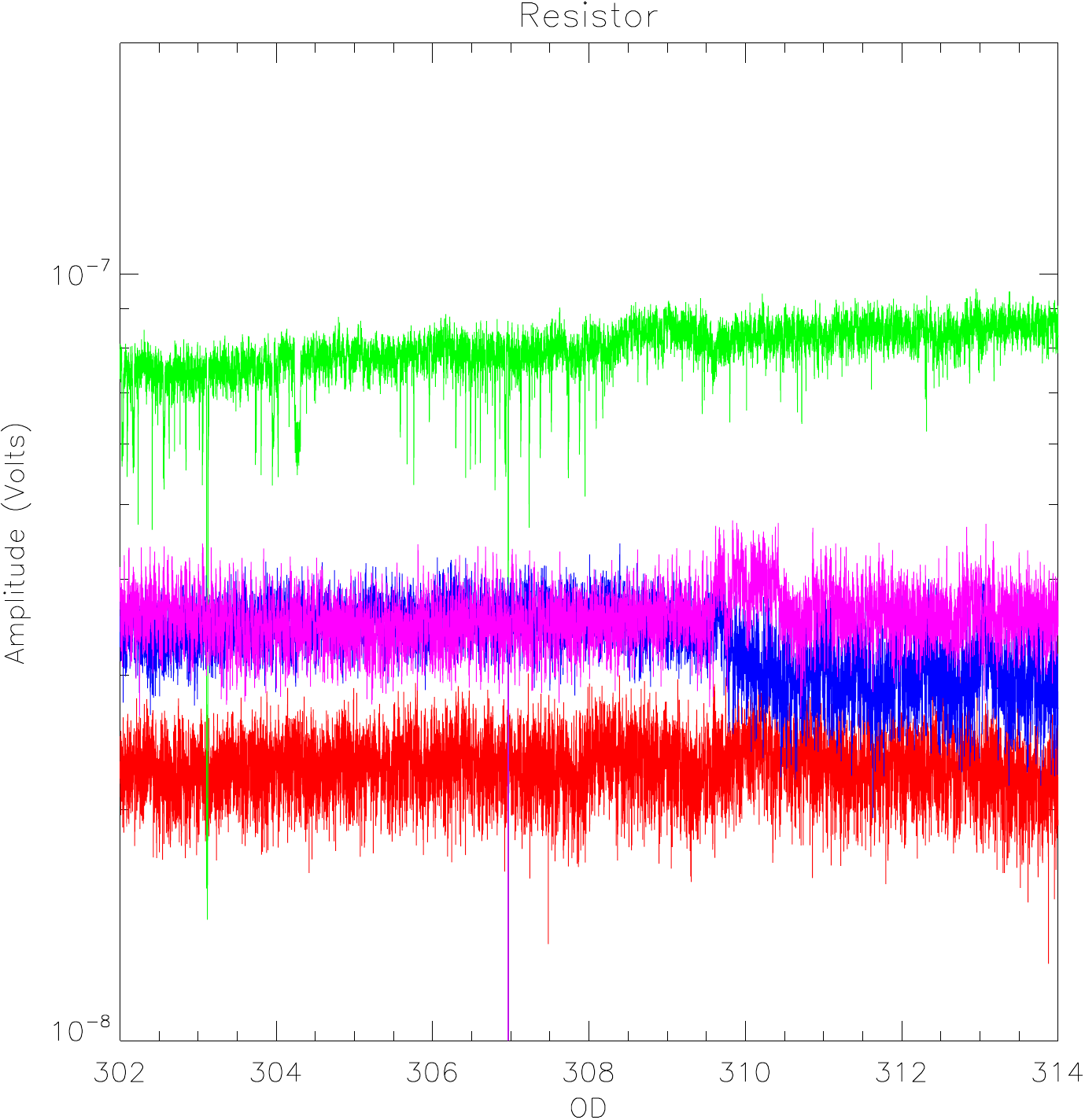}
\caption{Zoom on the four main 4\,K line systematics amplitude
  measured in the TOI processing on the test resistance. (\emph{Left})
  Period when the transponder was switched on once per day for the
  3-hours Daily Tele-Communication Period (DTCP). (\emph{Right})
  Period when the transponder was kept on at all times, later in the
  mission.}
\label{PlanckIPF1.5:fig:A0_4K_OnOff} 
\end{figure*}

The fundamental frequency of the \HeJT\ cooler (40.083373\,Hz) is
phase-locked to the frequency of the data acquisition (180.37518\,Hz)
in a ratio of 2 to 9. EMI/EMC impacts the TOI only as very narrow
lines.  Unfortunately, in flight, unlike in ground-based measurements,
these lines are not stable. The 4\,K line variations are illustrated
in Fig.~\ref{PlanckIPF1.5:fig:A0_4K_all}.  The variability of the lines is in part
due to temperature fluctuations in the service module of the Planck
spacecraft. Indeed, some of the variability was related to the power
cycling of the data transponder which, for stability reasons, has been
kept on continuously since 25 January 2010 (OD 258,
\citealt{planck2011-1.3}, see Fig.~\ref{PlanckIPF1.5:fig:A0_4K_OnOff}).

\subsection{Abnormal noise in the electronics}

\begin{figure}
\includegraphics[angle=180,width=0.5\textwidth]{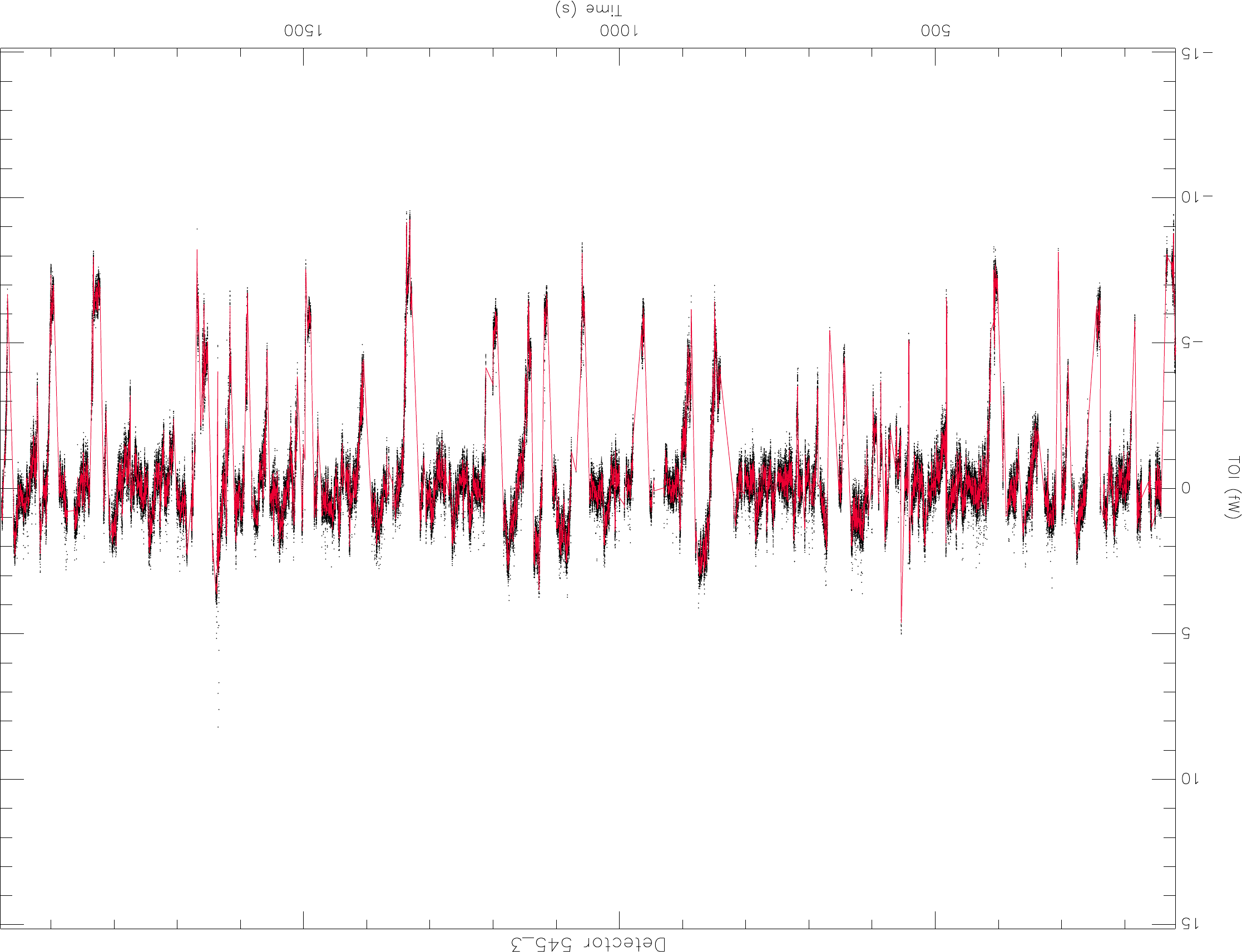}
\caption{Random telegraphic signal in the noise timeline (fW) of
  bolometer 545-3 plotted vs.~time. RTS is here a two-level signal
  with random occurrences of flipping. Full sampling is in black
  dots. A smoothed version (by 41 samples) is plotted with a red
  line.}
\label{PlanckIPF1.5:fig:RTS} 
\end{figure}

Of the 54 bolometers on HFI, three show a significant RTS, also known
as ``popcorn noise.''  These are 143-8, 545-3, and
857-4. Fig~\ref{PlanckIPF1.5:fig:RTS} illustrates their behaviour. The noise
timeline clearly exhibits a two-level system. The three RTS bolometers
in flight are the ones where RTS occurred most frequently in ground
measurements. However, in flight: 1)~the level difference is well
above the noise (at least ten times the rms); 2)~the two-level system
can be a three-level system or even larger; and 3)~the RTS is
intermittent.  For large duration, it can be unnoticeable, especially
for the 857-4 bolometer.

In an unrelated fashion, we see uncorrelated jumps in the noise 
TOI of many bolometers at a rate of just a few every year.

\subsection{Cosmic rays and their effects}

Energy is deposited by cosmic rays in various parts of the HFI
instrument.  We observe these events in the TOIs of all detectors as a
signal peak characterised by a very short rise time (less than
1.5\,ms) and an exponential decay.  These events are called glitches.
The other effect of the cosmic rays is a thermal input to the
bolometer plate, which induces low frequency noise on the
bolometers. Thermal effects are described in \cite{planck2011-1.7}
and their very long term consequences detailed in
Sect.~\ref{PlanckIPF1.5:Sect7}.

\subsection{Cosmic ray-induced glitch spectrum seen by Planck}

Cosmic rays consist of two main components at the L2 location of
\Planck: the Solar component and the Galactic component. The Solar
component is at low energy (a few keV) except during flares, when
energies can reach \hbox{GeV}. HFI is immune to the low energy
component and no major flares have yet been recorded.  The Galactic
component \citep{Bess2007} with a maximum between roughly 300\,MeV and
1\,GeV, is modulated by Solar activity. The \Planck\ mission began
during the weakest solar activity for a century
\citep{McDonald2010}. Hence, the Galactic cosmic ray component is
expected to be at its highest level.

\begin{figure}
\includegraphics[angle=90,width=.5\textwidth,totalheight=.15\textheight]{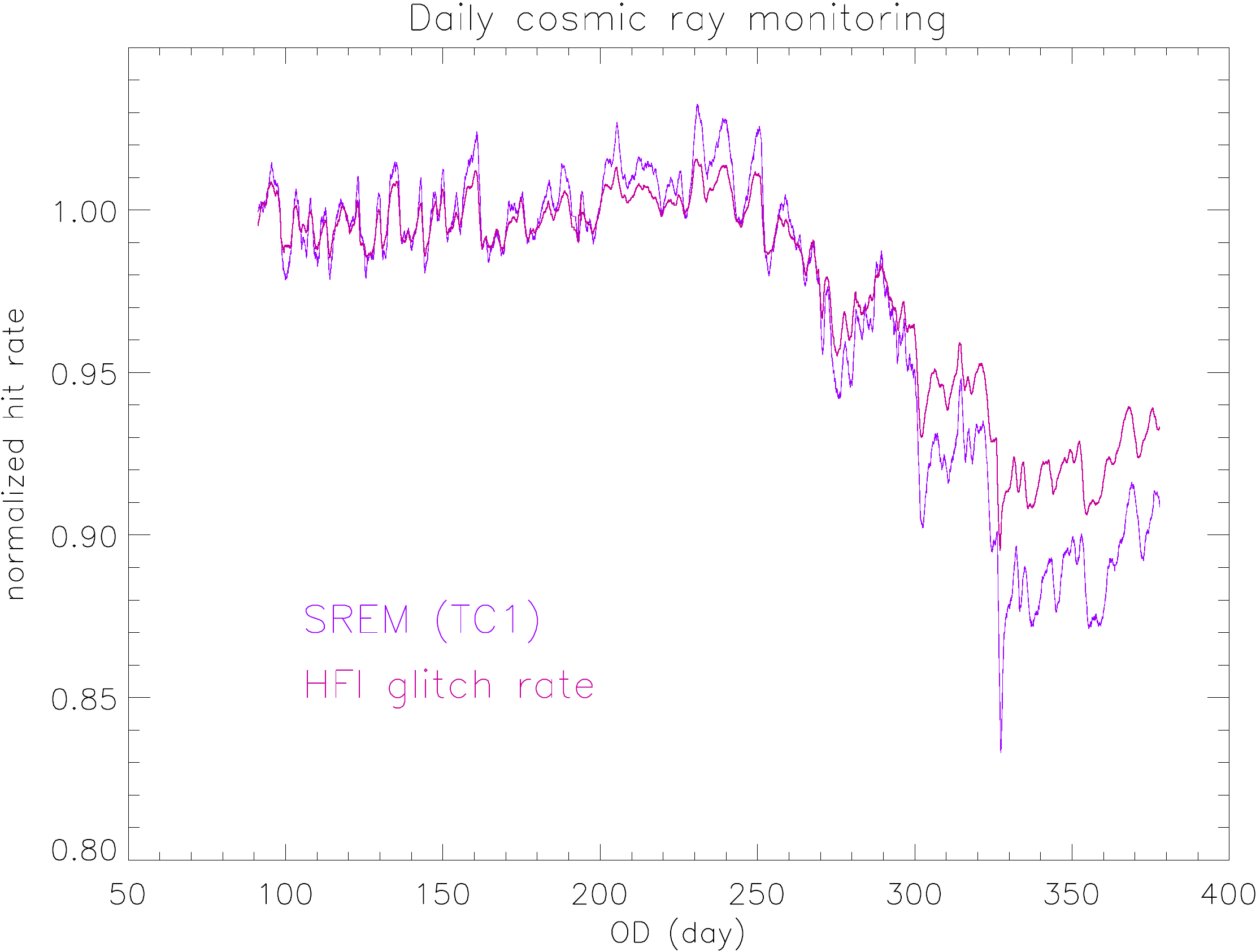}
\caption{SREM hit count and HFI bolometer average glitch rate
  evolution. SREM TC1 hit counts measure the protons with
  a deposited energy larger than 0.085\,MeV.}
\label{PlanckIPF1.5:fig:SremGlRate}
\end{figure}

The glitch rate evolution (Fig.~\ref{PlanckIPF1.5:fig:SremGlRate})
follows closely the proton monitoring by the space radiation
environnement monitor (SREM), a piggy-back experiment mounted on the
\Planck\ spacecraft.  This figure shows that cosmic rays are the main
source of HFI glitches. The glitch rate tended to decrease during the
mission since January 2010 due to the slow increase of the Solar
activity.

This glitch rate can be understood as the sum of two interaction
modes, depending on whether the cosmic ray has a direct or indirect
interaction with the bolometer.  High energy cosmic rays can also
interact with the bolometer plate and induce thermal effects and 
correlated glitches on the bolometers. These
are dealt with in Sect.~\ref{PlanckIPF1.5:Sect7}. The glitch
characteristics also depend on the location of the energy deposit
within the bolometer: the thermistor, the absorbing grid, or the
bolometer housing.

\subsubsection{Direct interaction} 

Cosmic ray particles can deposit energy directly on the thermistor or
the absorbing grid. This is observed at well-defined deposited
energies corresponding to the thickness of the element. Typically
thermistor hits are about 20\,keV, whereas grid hits are about
2\,keV. These events occur at a rate of a few per minute.

\subsubsection{Indirect interaction}

\begin{figure*}
\includegraphics[angle=180,width=1\textwidth,totalheight=.4\textheight]{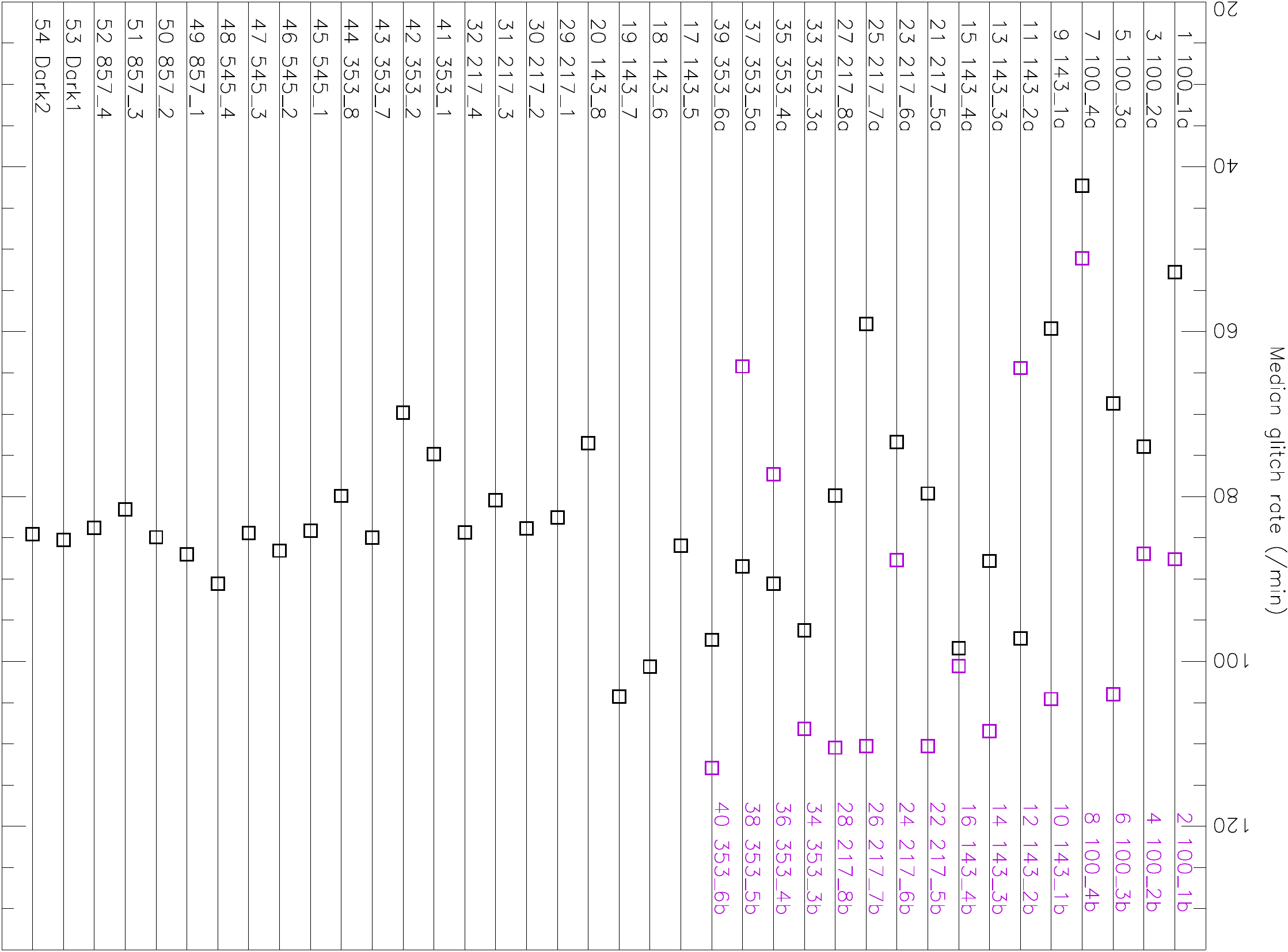}
\caption{Glitch rate of all HFI bolometers. An average over the first
  sky survey has been performed. The asymmetry between PSB bolometers
  sharing the same horn is an effect of detection threshold and
  asymmetric time constant properties between PSB \emph{a} and
  \emph{b}. }
\label{PlanckIPF1.5:fig:glrate}
\end{figure*}

Cosmic ray particles can also deposit energy indirectly. All particles
crossing some matter produce a shower of secondary electrons, through
ionization, that are mostly absorbed in the matter nearby. However,
interactions occurring within microns of the internal surface of
the bolometer box produce a shower of free secondary particles. A
fraction of these particles is absorbed by the thermistor and the grid
of the bolometer. This explains the large coincidence rate of gliches
between PSB bolometers \emph{a} and \emph{b} sharing the same mounting
structure. The energy of those glitches follows a power law
distribution spanning the whole range, from the detection threshold to
the saturation level.  This spectrum is expected for the delta and
secondary electrons produced via the ionization process. The total
rate of these events is typically one per second, and thus dominates
the total counts shown in Fig.~\ref{PlanckIPF1.5:fig:glrate}.

A more detailed description of the effect of cosmic rays on HFI
detectors is postponed to a dedicated paper, and glitch handling in
the data processing is described by \cite{planck2011-1.7}.

\section{Instrument stability}
\label{PlanckIPF1.5:Sect7}

The radiative power reaching each bolometer is the co-addition of the
flux from the sky and of the thermal emission of all optical elements
"seen" from the detector: filters, horns, telescope reflectors,
shields, and mechanical parts visible in the side-lobes. In addition,
fluctuations of the heat sink temperature (the bolometer plate) appear
like an optical signal. Any change in any of the parameters
(temperature, emissivity, geometrical coefficient) driving these
sources may be visible in the bolometer signal as a "DC level",
i.e. a stable or very slowly varying (days) component in the signal.

\begin{figure}
\includegraphics[width=1\columnwidth]{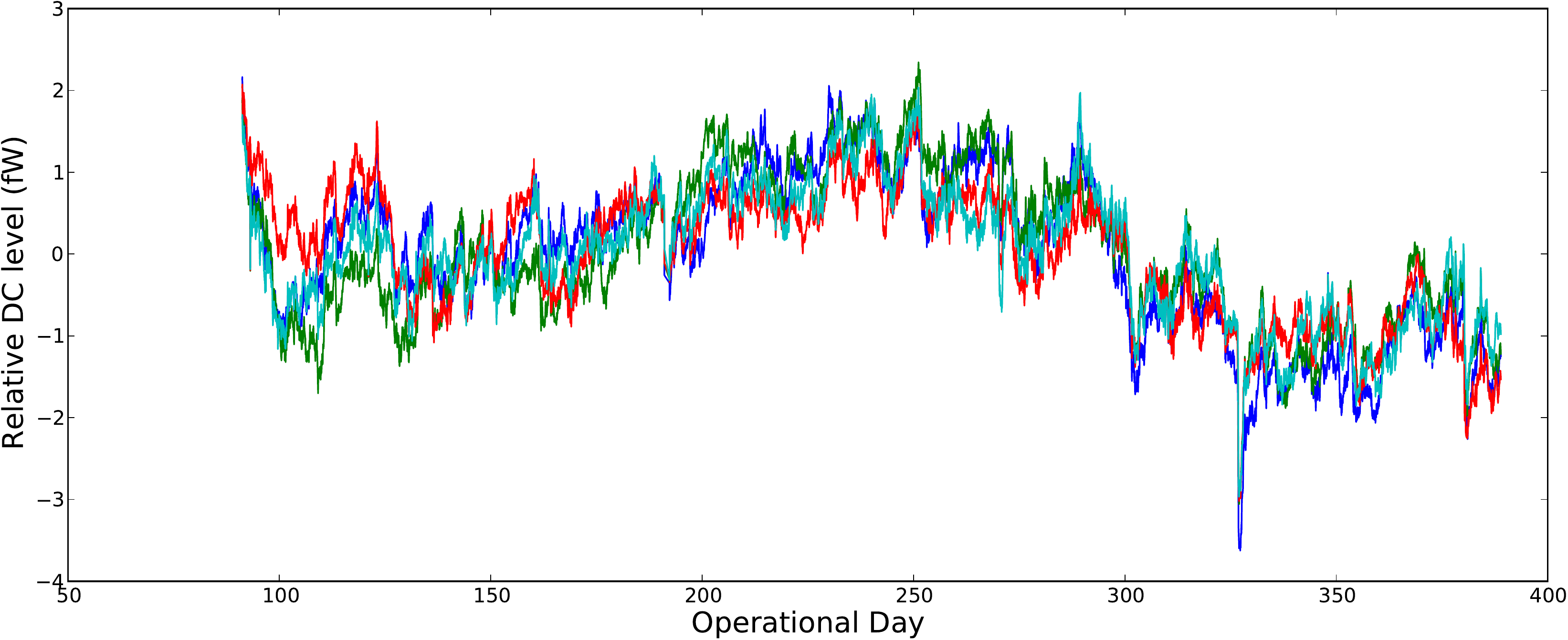}
\caption{The drift in the DC level of the 217 GHz SWB bolometers, in
  femtoWatts of equivalent power in the detector, for the first year
  of operation.}
\label{PlanckIPF1.5:fig:bolo_dc_levels}
\end{figure}

Monitoring the "DC level" supposes that one is able to remove the
varying sky signal from the stable sources. This is done in the
map-making process \citep{planck2011-1.7} by using the redundancy of
the scanning strategy.  Fig.~\ref{PlanckIPF1.5:fig:bolo_dc_levels}
shows for the 217\,GHz bolometers the history of the DC level during
nearly one year.  All follow a pattern similar to that of the cosmic
ray activity measured by the SREM (see Sect.~\ref{PlanckIPF1.5:Sect6}
and Fig.~\ref{PlanckIPF1.5:fig:SremGlRate}), which indicates that
cosmic rays are at the origin of the measured signal. One can check on
families of bolometers with non-uniform heat leaks $G$ that this
signal is directly related to temperature variations of the bolometer
plate and not to external optical sources. In fact, we can see here
residual fluctuations that the PID of the bolometer plate fails to
compensate because its efficiency is far from one. The similarity of
Fig.~\ref{PlanckIPF1.5:fig:bolo_dc_levels} and
Fig.~\ref{PlanckIPF1.5:fig:SremGlRate}, also shows that the effect of
gain variations and of DC level drifts of the readout electronics is
small with respect to other sources of signal drifts.

It should be noted that the "DC level" variation of 217\,GHz
bolometers is equivalent to an optical power of a couple of
femtowatts, while the total background power on these bolometers is
about 1\,p\hbox{W}. This fluctuation is mainly due to the
energy deposited by cosmic rays on the bolometer plate, which means
that the ``equivalent power'' of the other sources of temperature
fluctuation and of optical background fluctuations are no more than a
fraction of femtowatt, i.e. less than one part per thousand of the
background.

The change of gain induced by the DC level variations can be estimated
from the non-linearity measurements (see
Sect.~\ref{PlanckIPF1.5:SSS:Linearity}). In the case considered in
Fig.~\ref{PlanckIPF1.5:fig:bolo_dc_levels}, the relative gain change
is of the order of a few 10$^{-4}$.

During the CPV phase, the readout electronics was "balanced", i.e. the
offset parameter was tuned to get a signal near to zero. During the
first year of operation, and for all bolometers, deviations from this
point remained small with respect to the total range of acceptable
values. In consequence, no re-tuning of the readout electronics was
needed during this period and it is expected to be the same up to the
end of the mission.

\section{Main performance parameters}
\label{PlanckIPF1.5:Sect8}

The primary difference between the in-flight and pre-launch
performance of the HFI derives from the relatively high rate of cosmic
rays in the L2 environment.  At the energies of interest, the low
level of solar activity results in an elevated cosmic ray flux.  The
glitches that result from cosmic ray events must be identified and
removed from the time ordered information prior to processing the data
into maps. The TOI processing also removes a significant fraction of
the common mode component that appears in the bolometer TOIs at low
frequencies.  A residual low-frequency component is removed during the
map-making process \citep{planck2011-1.7}.

Table \ref{PlanckIPF1.5:tab:noiseJML} summarizes the noise properties of the
processed TOI \citep{planck2011-1.7}, by the following parameters:
\begin{itemize}
\item A white noise model. NEP$_1$ is the average of the Noise
Equivalent Power spectrum in the 0.6--2.5\,Hz range.
\item A model with a white noise NEP$_2$ plus a low frequency component:
NEP = NEP$_2$ [1 + (f$_{\rm knee}$/f)$^\alpha$]
\item The sensitivity NE$\Delta$T$_{CMB}$ to temperature differences
of the CMB.  Note that this quantity is not particularly relevant for
the channels at 545 and 857\,GHz for which it takes large values that
are highly dependent on the details of the spectral transmission
for each detector.
\item The sensitivity NE$\Delta$T$_{RJ}$ to temperature differences
for sources observed in the Rayleigh-Jeans regime.
\end{itemize}

\begin{figure}
\includegraphics[width=\columnwidth, keepaspectratio]{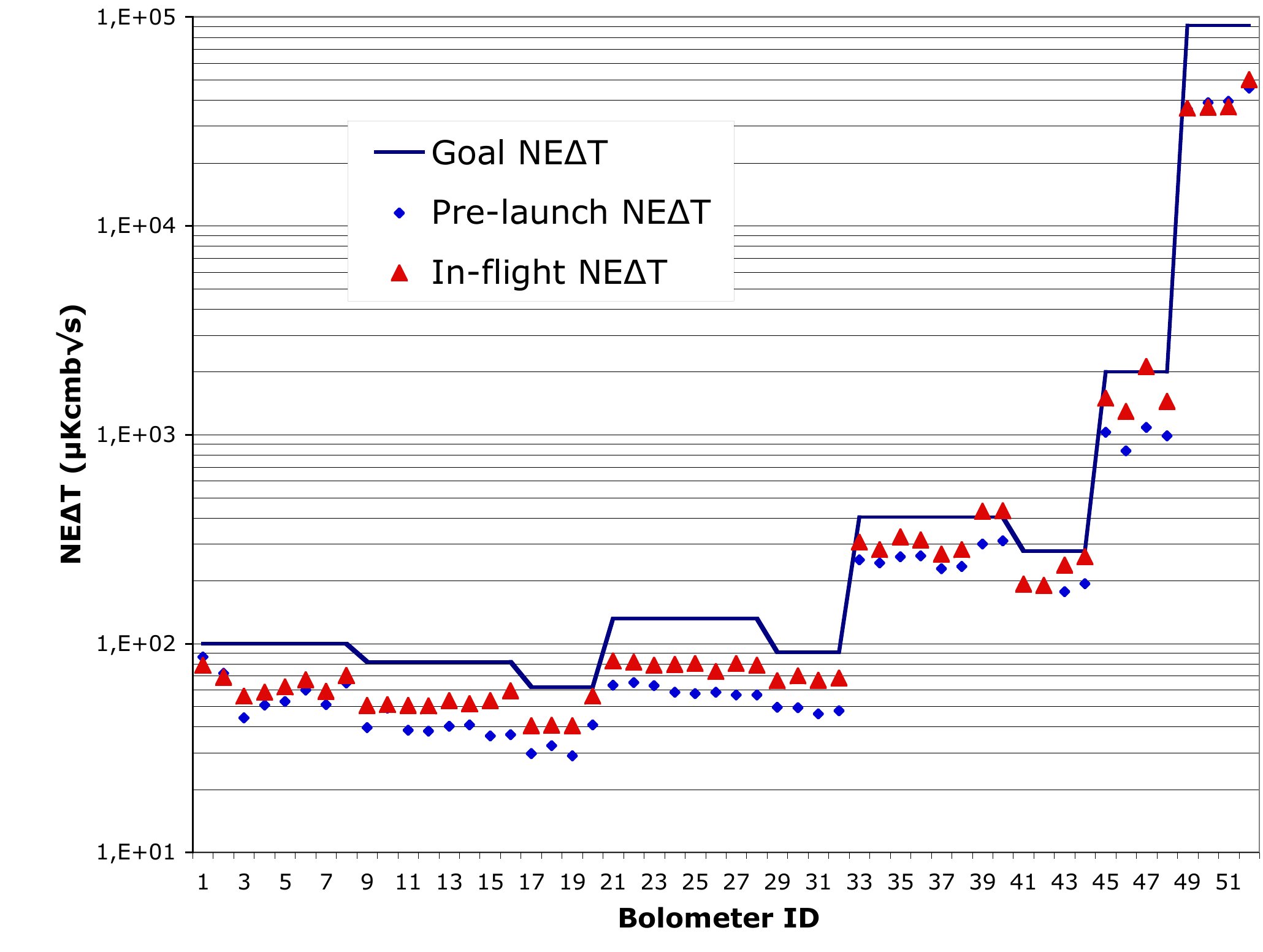}
\caption{Noise Equivalent Delta Temperature measured on the ground and
    in-flight with slightly different tools}
\label{PlanckIPF1.5:fig:NET}
\end{figure}

Figure \ref{PlanckIPF1.5:fig:NET} compares the goal, pre-launch and in-flight
NE$\Delta$Ts. The average in-flight NE$\Delta$Ts are 27\% higher 
than the pre-launch NE$\Delta$Ts.  While the pre-launch and in-flight 
values are not directly
comparable due to differences in the processing, these differences
can account for less than half of the observed variation.  The
remaining part is attributed to residual contamination from cosmic
rays that are not completely removed in the current TOI
processing.

The sensitivity goals are taken from Table~1 of
\citet{lamarre2010}, and are consistent with Table~1.3 of
\citet{planck2005-bluebook} corrected for the use of PSBs at 100\,GHz.
Note that Fig.~\ref{PlanckIPF1.5:fig:NET} supersedes Figure~11 of
\citet{lamarre2010} in which requirements and goals were improperly
plotted. The in-flight sensitivities estimated from NEP$_1$ exceed the
goals, which are defined by a total noise level equal to twice the
expected contribution of photon noise.  The average measured NEP$_1$
is typically 70\% of the initial goal.  The improvement in the NEP
over the design goals is primarily the result of having reduced the
photon background through careful design of the optical system.

\begin{table*}[tmb]
\caption{ NEP$_1$ is the average of the NEP of
processed signal in the band 0.6--2.5\,Hz. NEP$_2$ is the white noise
component of the NEP (see text). The NE$\Delta$Ts are derived from the
NEP$_2$. $\dag$ is for the bolometers suffering from RTS.} 
\label{PlanckIPF1.5:tab:noiseJML}
\centering
\begin{tabular}{l r c c c c c c}
\hline\hline

\multicolumn{2}{l}{Bolometer} & Noise & \multicolumn{3}{c}{Two component fit} & CMB & RJ \\
 &  & NEP$_1$ & NEP$_2$ & f$_{\rm knee}$ & \ \ \ \ $\alpha$\ \ \ \ & NE$\Delta$T$_{CMB}$ & NE$\Delta$T$_{RJ}$ \\
 Name & ID\# & 10$^{-17}$ W/$\sqrt{Hz}$ & 10$^{-17}$ W/$\sqrt{Hz}$ & \ mHz\  & & $\mu$K$_{CMB} \sqrt{s}$ & $\mu$K$_{CMB} \sqrt{s}$ \\
\hline
 100-1a &  1 & 1.13 & 1.04 & 218 & 0.93 & 78 &  60.7 \\
 100-1b &  2 & 1.21 & 1.14 & 166 & 1.02 & 69 &  53.0 \\
 100-2a &  3 & 1.22 & 1.16 & 126 & 0.96 & 56 &  42.9 \\
 100-2b &  4 & 1.31 & 1.22 & 182 & 0.95 & 58 &  44.7 \\
 100-3a &  5 & 1.22 & 1.16 & 117 & 1.01 & 61 &  47.4 \\
 100-3b &  6 & 1.09 & 1.01 & 173 & 1.02 & 66 &  51.4 \\
 100-4a &  7 & 1.23 & 1.18 & 109 & 0.97 & 59 &  45.2 \\
 100-4b &  8 & 1.18 & 1.08 & 212 & 0.95 & 70 &  53.6 \\
 143-1a &  9 & 1.35 & 1.31 & 91 & 1.11  & 50 &   30.4 \\
 143-1b & 10 & 1.18 & 1.09 & 197 & 1.01 & 51 &  30.6 \\
 143-2a & 11 & 1.28 & 1.20 & 161 & 0.97 & 50 &  30.3 \\
 143-2b & 12 & 1.30 & 1.27 & 106 & 1.18 & 50 &  30.1 \\
 143-3a & 13 & 1.35 & 1.26 & 202 & 1.01 & 53 &  32.2 \\
 143-3b & 14 & 1.18 & 1.09 & 190 & 1.02 & 51 &  30.9 \\
 143-4a & 15 & 1.27 & 1.18 & 185 & 0.99 & 53 &  31.7 \\
 143-4b & 16 & 1.32 & 1.24 & 161 & 1.07 & 59 &  35.5 \\
 143-5 &  17 & 1.53 & 1.46 & 138 & 1.10 & 40 &  23.9 \\
 143-6 &  18 & 1.37 & 1.25 & 230 & 1.03 & 40 &  24.1 \\
 143-7 &  19 & 1.49 & 1.40 & 154 & 1.09 & 40 &  23.8 \\
 $^\dag$143-8 &  20 & 2.2 & 1.60 & 1244 & 0.90 & 55 &  33.1 \\
 217-5a & 21 & 1.35 & 1.30 & 117 & 1.10 & 82 &  26.4 \\
 217-5b & 22 & 1.33 & 1.22 & 219 & 1.06 & 81 &  25.9 \\
 217-6a & 23 & 1.30 & 1.25 & 107 & 1.07 & 78 &  25.1 \\
 217-6b & 24 & 1.31 & 1.26 & 118 & 1.08 & 79 &  25.2 \\
 217-7a & 25 & 1.41 & 1.36 & 98 & 1.07 & 80 &   25.4 \\
 217-7b & 26 & 1.25 & 1.17 & 157 & 1.05 & 73 &  23.4 \\
 217-8a & 27 & 1.37 & 1.31 & 148 & 1.05 & 80 &  25.5 \\
 217-8b & 28 & 1.27 & 1.17 & 206 & 1.03 & 78 &  24.9 \\
 217-1 &  29 & 1.59 & 1.49 & 187 & 1.14 & 66 &  20.7 \\
 217-2 &  30 & 1.61 & 1.48 & 229 & 1.10 & 69 &  21.7 \\
 217-3 &  31 & 1.63 & 1.54 & 165 & 1.12 & 66 &  20.8 \\
 217-4 &  32 & 1.62 & 1.53 & 173 & 1.14 & 68 &  21.3 \\
 353-3a & 33 & 1.53 & 1.43 & 174 & 0.98 & 305 & 21.9 \\
 353-3b & 34 & 1.39 & 1.31 & 162 & 1.06 & 282 & 20.3 \\
 353-4a & 35 & 1.34 & 1.28 & 124 & 1.04 & 324 & 22.6 \\
 353-4b & 36 & 1.30 & 1.25 & 127 & 1.12 & 313 & 21.8 \\
 353-5a & 37 & 1.26 & 1.21 & 121 & 1.05 & 268 & 19.4 \\
 353-5b & 38 & 1.33 & 1.27 & 125 & 1.09 & 281 & 20.3 \\
 353-6a & 39 & 1.47 & 1.38 & 208 & 1.08 & 429 & 30.7 \\
 353-6b & 40 & 1.33 & 1.26 & 179 & 1.20 & 432 & 32.4 \\
 353-1 &  41 & 1.59 & 1.52 & 100 & 1.04 & 192 & 13.7 \\
 353-2 &  42 & 1.72 & 1.66 &  98 & 1.07 & 189 & 13.4 \\
 353-7 &  43 & 1.62 & 1.54 & 155 & 1.18 & 237 & 16.4 \\
 353-8 &  44 & 1.67 & 1.59 & 159 & 1.15 & 260 & 17.6 \\
 545-1 &  45 & 3.50 & 3.19 & 295 & 1.20 & 1490 & 8.7 \\
 545-2 &  46 & 2.93 & 2.66 & 322 & 1.20 & 1293 & 7.9 \\
 $^\dag$545-3 &  47 & 4.48 & 3.70 & 431 & 1.23 & 2116 & 12.7 \\
 545-4 &  48 & 2.76 & 2.51 & 297 & 1.19 & 1446 &  8.7 \\
 857-1 &  49 & 3.59 & 3.31 & 222 & 1.20 & 36566 & 3.4 \\
 857-2 &  50 & 4.10 & 3.75 & 265 & 1.15 & 36923 & 3.8 \\
 857-3 &  51 & 3.47 & 3.21 & 236 & 1.20 & 37037 & 3.5 \\
$^\dag$857-4 &  52 & 3.64 & 3.00 & 622 & 1.09 & 50180 & 5.4 \\
Dark1 &  53 & 1.17 & 1.14 & 136 & 1.42 & 16496 & -- \\
Dark2 &  54 & 1.39 & 1.35 & 148 & 1.40 & 19462 & -- \\

\hline
\end{tabular}
\end{table*}

\section{Conclusions}
\label{PlanckIPF1.5:Sect9}

We report on the in-flight performance of the High Frequency
Instrument on board the \Planck\ satellite.  These results are derived
from the data obtained during a dedicated period of diagnostic testing
prior to the initiation of the scientific survey, as well as an
analysis of the survey data that form the basis of the early release
scientific products.

With the exception of a single anomaly in the operation of the \HeJT\
cooler, the HFI has operated nominally since launch.  The settings of
the readout electronics determined during pre-launch testing were
found to be very near the optimal value in flight and were applied
without any modification.  A random telegraphic signal is observed in
the same three channels that exhibited this behaviour during the final
pre-launch testing. These channels are currently excluded from the scientific
analysis.  The instrument operation has been extremely
stable during the first year of operation, requiring no adjustment of
the readout electronics.

The optical design, and the alignment of the optical assembly, relied
on both theoretical analysis and testing at the subsystem level.  The
beams of the 545 and 857\,GHz channels, which employ multimoded
corrugated horns and waveguide, could not be measured on the ground.
The actual beam widths of these channels measured on planets are in
general smaller than the design goals and estimated values. The
optical properties of the single mode channels are in excellent
agreement with the design expectations.

A higher than expected cosmic ray flux, related to the level of Solar
activity, results in a manageable loss of signal and degradation of
thermal stability.  Discrete cosmic ray events result in glitches in
the scientific signal that are flagged and removed by an algorithm
making use of the signal redundancy in the timeline.  In addition to
these single events, the cosmic ray flux contributes a significant
thermal load on the sub-kelvin stage.  Variations in this flux produce
low-frequency fluctuations in the bolometer plate that induce a common
mode component to the scientific signal and the focal plane
thermometry. Although a component correlated with the dark bolometer
outputs is removed during the TOI processing, a residual low frequency
contribution is removed at the map-making stage. With the exception of
the three detectors affected by telegraph noise, the sensitivity
measured above 0.6\,Hz exceeds the design goals of all HFI
channels. After the removal of the residual low frequency noise, the
final sensitivity of the frequency maps exceeds the mission
requirements, and approaches the goals, as described in the companion
paper \cite{planck2011-1.7}.

\begin{acknowledgement}

The Planck HFI instrument (\url{http://hfi.planck.fr/}) was designed and
built by an international consortium of laboratories, universities and
institutes, with important contributions from the industry, under the
leadership of the PI institute, IAS at Orsay, France. It was funded in
particular by CNES, CNRS, NASA, STFC and ASI. The authors extend their
gratitude to the numerous engineers and scientists, who have
contributed to the design, development, construction or evaluation of
the HFI instrument. A description of the Planck Collaboration and a list 
of its members, indicating which technical or scientific activities they have 
been involved in, can be found at
\url{www.rssd.esa.int/index.php?project=PLANCK\&page=Planck\_Collaboration}.

\end{acknowledgement}


\end{document}